\documentclass[a4paper,fleqn,usenatbib]{mnras}

\usepackage{newtxtext,newtxmath}
\usepackage[T1]{fontenc}
\usepackage{ae,aecompl}

\usepackage{graphicx}	% Including figure files
\usepackage{amsmath}	% Advanced maths commands
\usepackage{amssymb}	% Extra maths symbols
\usepackage{mathtools}
\usepackage{txfonts}
\usepackage{pdflscape}
\usepackage{lscape}
\usepackage{natbib}
\usepackage{threeparttable, tablefootnote}
\usepackage{aas_macros}
\title[Morphology of the Magellanic Clouds]{The VMC Survey - XXXIV. Morphology of Stellar Populations in the Magellanic Clouds}

\author[D. El Youssoufi et al.]{
Dalal El Youssoufi,$^{1}$\thanks{E-mail: delyoussoufi@aip.de (DE)} Maria-Rosa L. Cioni,$^{1}$ Cameron P. M. Bell,$^{1}$ Stefano Rubele,$^{2}$
\newauthor Kenji Bekki,$^{4}$ Richard de Grijs,$^{5,6,7}$ L\'{e}o Girardi,$^{2}$ Valentin D. Ivanov,$^{8,9}$ Gal Matijevic,$^{1}$
\newauthor Florian Niederhofer,$^{1}$ Joana M. Oliveira,$^{10}$ Vincenzo Ripepi,$^{11}$ Smitha Subramanian$^{12}$,
\newauthor Jacco Th. van Loon$^{10}$
\\
$^{1}$Leibniz-Institut f\"ur Astrophysik Potsdam (AIP), An der Sternwarte 16, D-14482 Potsdam, Germany\\
$^{2}$Osservatorio Astronomico di Padova, INAF Vicolo dell'Osservatorio 5, I-35122 Padova, Italy\\
$^{3}$Department of Physics and Astronomy, Macquarie University, Balaclava Road, Sydney NSW 2109, Australia\\
$^{4}$INAF -- Osservatorio Astronomico di Padova, Vicolo dell'Osservatorio 5, I-35122 Padova, Italy\\
$^{5}$ICRAR, M468, University of Western Australia, 35 Stirling Hwy, 6009 Crawley, Western Australia, Australia\\
$^{6}$Centre for Astronomy, Astrophysics and Astrophotonics, Macquarie University, Balaclava Road, Sydney NSW 2109, Australia\\
$^{7}$International Space Science Institute--Beijing, 1 Nanertiao, Zhongguancun, Hai Dian District, Beijing 100190, China\\
$^{8}$European Southern Observatory, Karl-Schwarzschild-Str. 2, D-85748 Garching bei M\"unchen, Germany\\
$^{9}$European Southern Observatory, Ave. Alonso de Cordova 3107, Vitacura, Santiago, Chile\\ 
$^{10}$Lennard-Jones Laboratories, School of Chemical and Physical Sciences, Keele University, ST5 5BG, UK\\
$^{11}$INAF -- Osservatorio Astronomico di Capodimonte, via Moiariello 16, I-80131, Naples, Italy\\
$^{12}$Indian Institute of Astrophysics, Koramangala II Block, Bangalore-34, India\\
}

\date{Accepted XXX. Received YYY; in original form ZZZ}

\pubyear{2019}

\begin{document}
\label{firstpage}
\pagerange{\pageref{firstpage}--\pageref{lastpage}}
\maketitle

\begin{abstract}
%The Magellanic Clouds are nearby dwarf irregular galaxies that represent a unique laboratory for studying galaxy morphologies. Traced by different stellar populations, their morphology shows different properties. 
The Magellanic Clouds are nearby dwarf irregular galaxies whose morphologies show different properties when traced by different stellar populations, making them an important laboratory for studying galaxy morphologies. We study the morphology of the Magellanic Clouds using data from the VISTA survey of the Magellanic Clouds system (VMC). We used about $10$ and $2.5$ million sources across an area of $\sim105$~deg$^2$ and $\sim42$~deg$^2$ towards the Large and Small Magellanic Cloud (LMC and SMC), respectively. We estimated median ages of stellar populations occupying different regions of the near-infrared ($J-K_\mathrm{s}, K_\mathrm{s}$) colour-magnitude diagram. Morphological maps were produced and detailed features in the central regions were characterised for the first time with bins corresponding to a spatial resolution of $0.13$~kpc (LMC) and $0.16$~kpc (SMC). In the LMC, we find that main sequence stars show coherent structures that grow with age and trace the multiple spiral arms of the galaxy, star forming regions become dimmer as we progress in age, while supergiant stars are centrally concentrated. Intermediate-age stars, despite tracing a regular and symmetrical morphology, show central clumps and hints of spiral arms. In the SMC, young main sequence stars depict a broken bar. Intermediate-age populations show signatures of elongation towards the Magellanic Bridge that can be attributed to the LMC-SMC interaction $\sim200$~Myr ago. They also show irregular central features suggesting that the inner SMC has also been influenced by tidal interactions.
\end{abstract}


%%%%%%%%%%%%%%%%%%%%%%%%%%%%%%%%%%%%%%%%%%%%%%%%%%

%%%%%%%%%%%%%%%%% BODY OF PAPER %%%%%%%%%%%%%%%%%%

\section{Introduction}\label{section1}

Galaxies have diverse shapes, components and structural properties. Their morphology is considered a fossil record of their history and carries fundamental information about galaxy formation and evolution.
Located at approximately $50$ kpc (e.g.~\citealp{DeGrijs2014}) and $60$ kpc (e.g.~\citealp{DeGrijs2015}), the Large and Small Magellanic Clouds represent the nearest interacting pair of dwarf irregular (dIrr) galaxies. dIrr galaxies are characterised by a gas-rich environment, low metallicity levels and an unstructured shape.
The LMC is also known as a prototype of barred Magellanic spirals due to its asymmetric stellar bar with no bulge, a large star-forming region at one end of the bar, one prominent spiral arm as well as other spiral features. Its total mass is estimated at $1.7\times10^{10}$ M{$_\odot$} from the rotational velocities of mostly carbon stars measured to radii of $8.7$ kpc \citep{VanderMarel2014} while the total mass of the SMC is estimated at $2.4\times10^9$ M$_\odot$ from the \ion{H}{i} rotation curve \mbox{\citep{Stanimirovic2003}}.
The close encounters between the two Magellanic Clouds and between the Magellanic Clouds and the Milky Way have been corroborated by proper motion studies (e.g.~\citealp{Kallivayalil2006,Kallivayalil2013,Cioni2014,VanderMarel2014,Cioni2016}), star formation history studies (e.g.~\citealp{Harris2003,Noel2007,Harris2009,Rubele2012,Cignoni2013,Rubele2015,Hagen2016}) and dynamical modelling (e.g.~\citealp{Besla2007,Diaz2012, Salem2015}). These interactions provided the necessary forces (tidal and/or ram-pressure stripping) for the creation of the Magellanic Stream and Bridge. Due to their proximity to our own Galaxy, assuring the resolution of stellar populations into individual stars, and ongoing star formation, the Magellanic Clouds have been targets of intensive research for many years, making them  rather unparalleled laboratories for studying stellar evolution and galaxy interactions.
\subsection{LMC morphology}
The LMC is an almost face-on, gas-rich galaxy characterised by an inclined disc and an offset bar of which the origin is not well understood \citep{Zhao2000, Zaritsky2004}. The morphologies of the LMC traced by different stellar populations show different properties (e.g.~\citealp{DeVaucouleurs1972,Cioni2000,Nikolaev2000,Belcheva2011,Moretti2014}). While young stars exhibit a rather irregular structure characterised by spiral arms and tidal features, older stars dominating the mass of the LMC tend to be more smoothly and regularly distributed. The bar appears to be a luminous and prominent entity in both optical and near-infrared images, while it is not present in the distribution of \ion{H}{i} gas \citep{Stanimirovic2003}. The bar is also asymmetric and appears to be elliptical in the south east while it is flat in the north west \citep{Zaritsky2004}. Using Cepheids, \cite{Nikolaev2004} found that the bar lies at a different distance than the disc, concluding that it is closer to us by $\sim0.5$ kpc, while in intermediate-age populations like red clump (RC) stars the bar was found to be co-planar with the disk \citep{Subramanian2009}. The bar is also known for its centre being offset with respect to the disc. Recent numerical simulations of dwarf--dwarf galaxy interactions \citep{Pardy2016} with a $1$:$10$ mass ratio show that during the encounter of the LMC and SMC morphological structures shift in relation to the LMC's dynamical centre. The stellar disc of the LMC becomes displaced under the effect of the gravitational potential, and this can persist for up to $2$ Gyr, while the bar is never actually shifted, suggesting that the dynamical centre of the LMC is always coincident with the bar centre \citep{DeVaucouleurs1972} rather than with the \ion{H}{i} centre \citep{Stanimirovic2003} as previously assumed. Using Cepheids, \cite{Jacyszyn-Dobrzeniecka2016} redefined the classical LMC bar by including a western density, with the bar spanning the whole width of the galaxy, and found no offset from the plane of the LMC making the bar an integral part of the disc. After the redefinition of the bar, the dynamical centre is located at the bar centre.

The LMC is known for its non-planar structure \citep{VanderMarel2002,Nikolaev2004}. \cite{Salyk2002} found possible warps and twists in the south east of the LMC extending up to $2.5$ kpc. \cite{Choi2018A} detected a significant warp in the south west of the disc extending up to $4$ kpc in the direction of the SMC. The east and west sides of the bar are closer to us compared with its central region, indicating that the bar of the LMC is also warped \citep{Subramaniam2003}. Extra-planar features in front as well as behind the plane were also identified in both optical \citep{Subramanian2010} and near-infrared \citep{Subramanian2013} studies of RC stars.
The LMC disc is thick \citep{VanderMarel2002,Subramanian2009}, flared \citep{Alves2000} and intrinsically elongated \citep{VanderMarel2001a}. At large radii, the LMC disc appears strongly elliptical \citep{VanderMarel2001b}. Its inner and outer parts are situated at different inclination angles making its eastern part closer to us. The orientation angles of the disc have been measured using different tracers and methods such as the RC \citep{Salyk2002,Subramanian2010,Subramanian2013}, Cepheids \citep{Nikolaev2004,Haschke2012,Jacyszyn-Dobrzeniecka2016,Inno2016}, RR Lyrae stars \citep{Haschke2012,Jacyszyn-Dobrzeniecka2017}, carbon to oxygen rich asymptotic giant branch (AGB) stars \citep{VanderMarel2001, Cioni2001}, outer isophotes \citep{DeVaucouleurs1972}, as well as HI gas. Using these different tracers, the inclination angle is found to be between $22\pm6$ deg and $37.4\pm2.3$ deg while the position angle of the line of nodes is between $122.5\pm8.3$ deg and $170\pm5$ deg.

The prominent work of \cite{DeVaucouleurs1972} revealed a distinct multi-arm spiral structure in the LMC. \cite{Besla2016} followed up on this work and explored the stellar substructures in the outskirts of the LMC. They found that stellar arcs and spiral arms exist in the northern periphery with no southern counterpart. They examined numerical simulations of the outskirts of the LMC disc and found that these features can be reproduced in isolation of the Milky Way potential. This entails that the disturbed nature of the Magellanic Clouds is largely due to the LMC-SMC interactions rather than to the effect of the Milky Way. This is supported by high-precision proper motion measurements (e.g.~\citealp{Kallivayalil2013}) implying that the Magellanic Clouds are most likely on their first passage by the Milky Way.
\cite{Choi2018b} detected a ring-like structure in the outskirts of the LMC disc. This stellar overdensity is clearly visible in the distribution of RC stars while it is not visible in young  main sequence stars. This structure was first detected by \cite{DeVaucouleurs1955} who referred to it as a faint outer loop. It was also found in the distribution of intermediate-age star clusters \citep{Westerlund1964} as well as in a map of the number ratio of carbon rich to oxygen rich AGB stars \citep{Cioni2003}.
\subsection{SMC morphology}
The SMC is an elongated galaxy known for its less pronounced bar and its eastern extension, connecting the SMC to the Bridge, known as the Wing \citep{Shapley1940}. Young and old stellar populations display different spatial distributions. Young stars are concentrated in the central regions of the galaxy as well as in the Wing, following the large-scale irregular and asymmetric structure of the HI distribution \citep{Stanimirovic2003}. Older populations, however, are uniformly distributed and they can be characterised by circular and elliptical structures (e.g.~\citealp{Cioni2000}; \citealp{Zaritsky2000}; \citealp{Harris2003}; \citealp{Haschke2012}; \citealp{Rubele2015}). 
Variable stars have been used extensively to provide a thorough study of the three dimensional structure of the SMC. Able to trace both the young and old populations as well as being distance indicators, these stars are also used to study the depth of the galaxy. RR Lyrae stars indicate that the old population follows an ellipsoidal shape without showing any signs of substructures and/or asymmetries. It was found that the line of sight depth can range from $1$ to $14$ kpc (e.g.~\citealp{Subramanian2012}; \citealp{Jacyszyn-Dobrzeniecka2017}; \citealp{Muraveva2018}). Cepheids trace the younger population. Their age distribution in the SMC is bimodal, younger stars are located closer than older ones {\citep{Subramanian2015,Jacyszyn-Dobrzeniecka2016,Ripepi2017}}. \cite{Scowcroft2016} found an elongation along the NE-SW axis of up to $20$ kpc with the NE closer to us. These results are consistent with other studies (e.g.~\citealp{Haschke2012}; \citealp{Subramanian2012}; \citealp{Subramanian2015}; \citealp{Jacyszyn-Dobrzeniecka2016};   \citealp{Ripepi2017}). The SMC bar is elongated along the line-of-sight \citep{Gardiner1995}; line-of-sight depth in the eastern part is higher than in western parts and can reach up to $23$ kpc in some regions \citep{Nidever2013}. Using RC stars, \cite{Subramanian2012} found that the SMC is elongated along the NE-SW axis, with a tidal radius  between $7$ and $12$ kpc. \cite{Subramanian2017} found a foreground population highlighted by its distance bi-modality in the distribution of RC stars. This feature can be traced in the direction of the Magellanic Bridge and its origin probably involves material stripped from the SMC.\\
In the present work, we use data from the VISTA near-infrared $YJK_{\mathrm{s}}$ survey of the Magellanic Clouds system (VMC) to investigate the spatial distribution of stellar populations of different ages across the Magellanic Clouds and carry out a comprehensive study of their morphological properties. The paper is organised as follows. Section \ref{section2} gives a description of the dataset used in our investigation and our criteria for data selection. Section \ref{section3} focuses on the morphology of the Magellanic Clouds based on various stellar populations, Section \ref{section4} addresses the comparison of the morphological maps with previous studies followed by a summary and conclusions in Section \ref{section5}. Supporting information is given in the Appendices.
\section{Observations and Data Selection}\label{section2}
\begin{figure*}
	\begin{center}
		\includegraphics[scale=0.059]{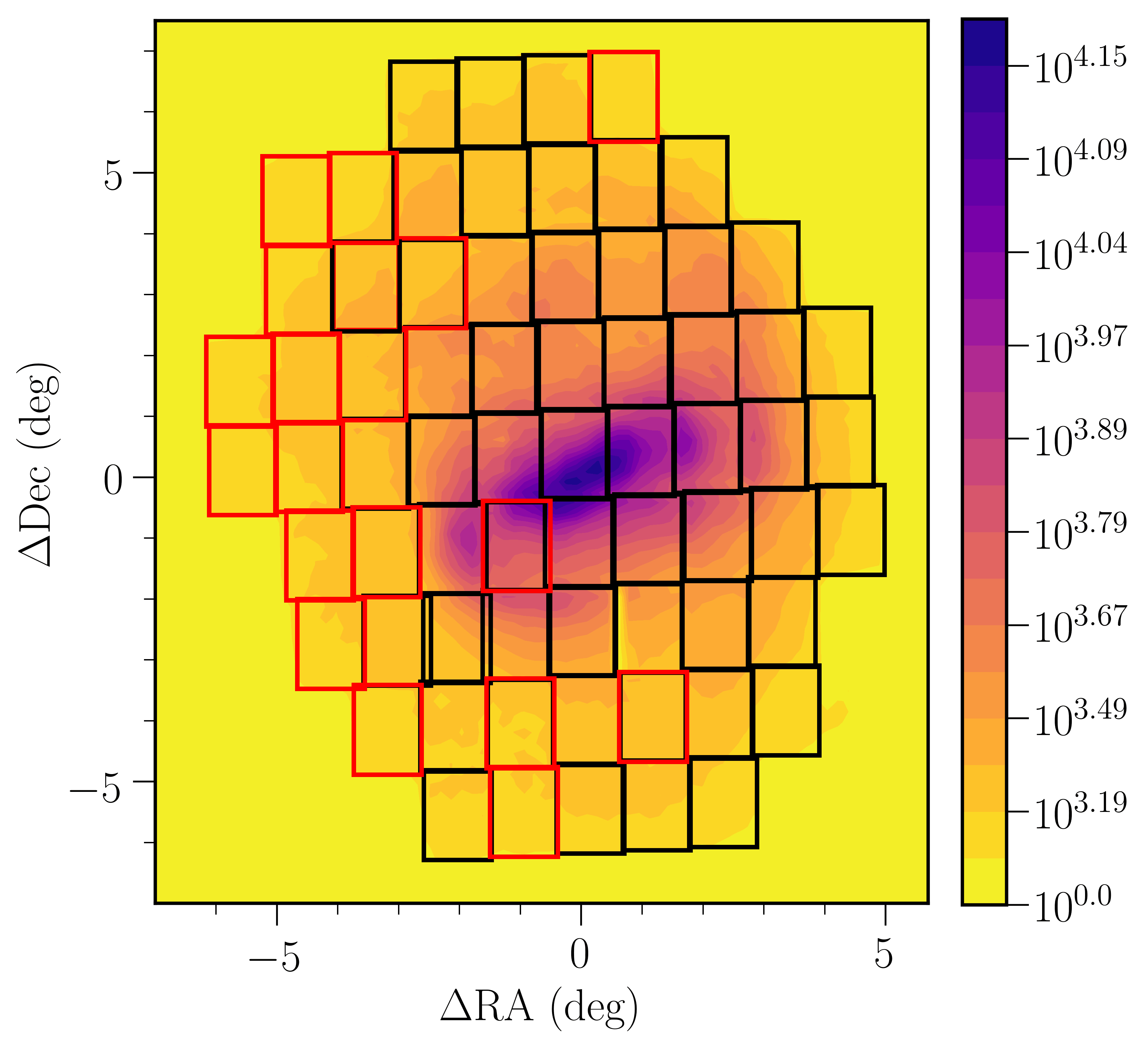}
		\includegraphics[scale=0.065]{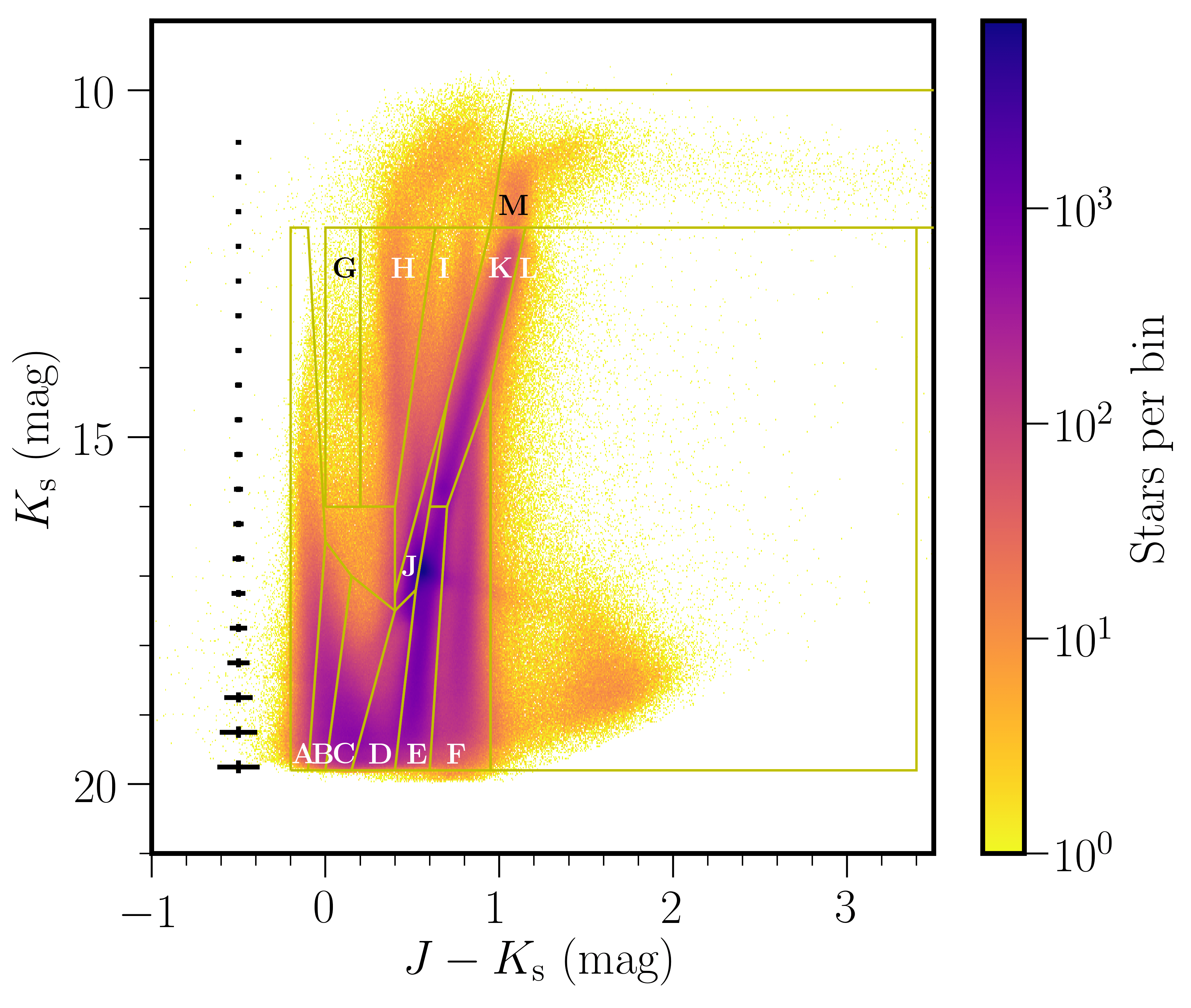}
		
	\end{center}
	\begin{center}
		\includegraphics[scale=0.059]{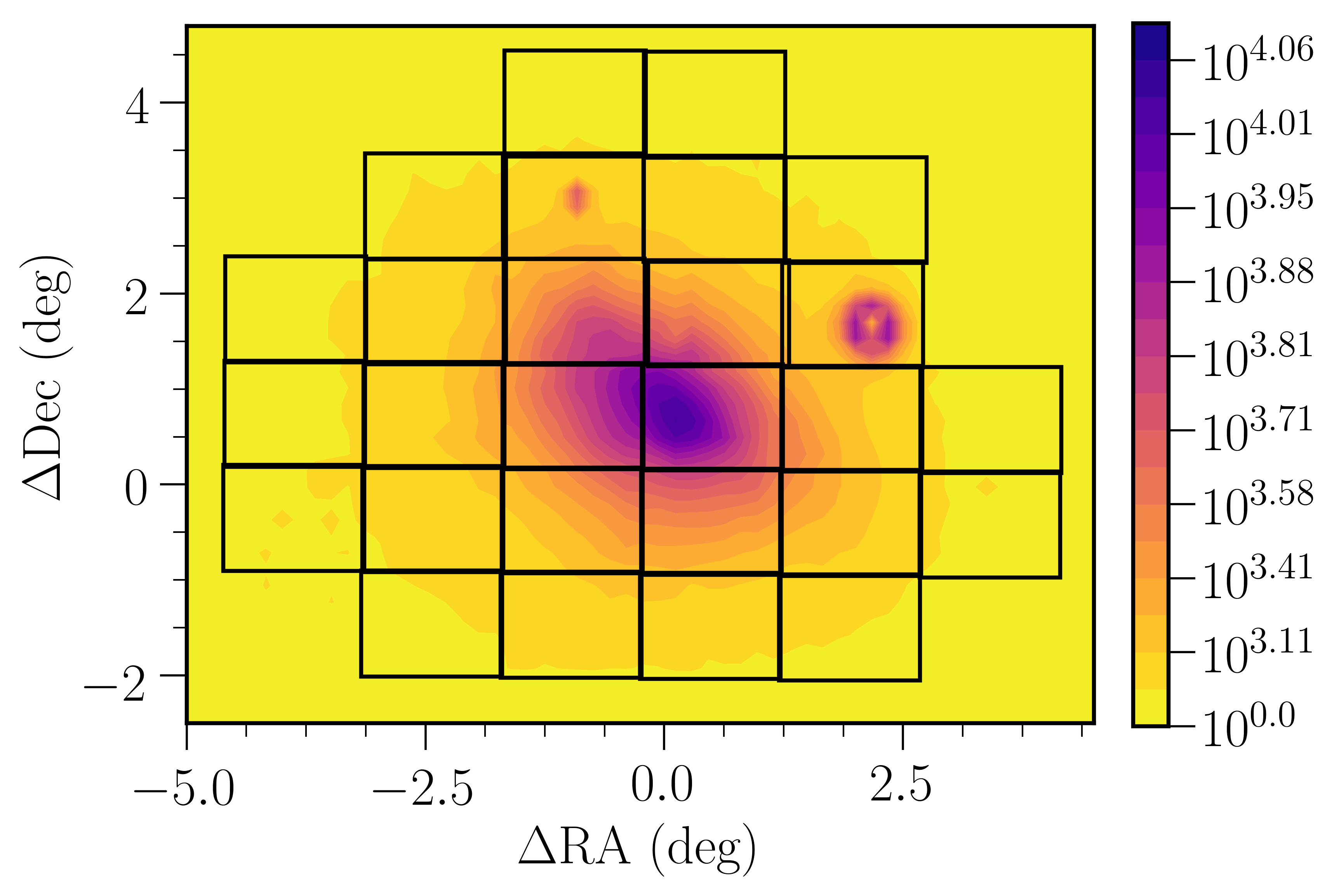}
		\includegraphics[scale=0.065]{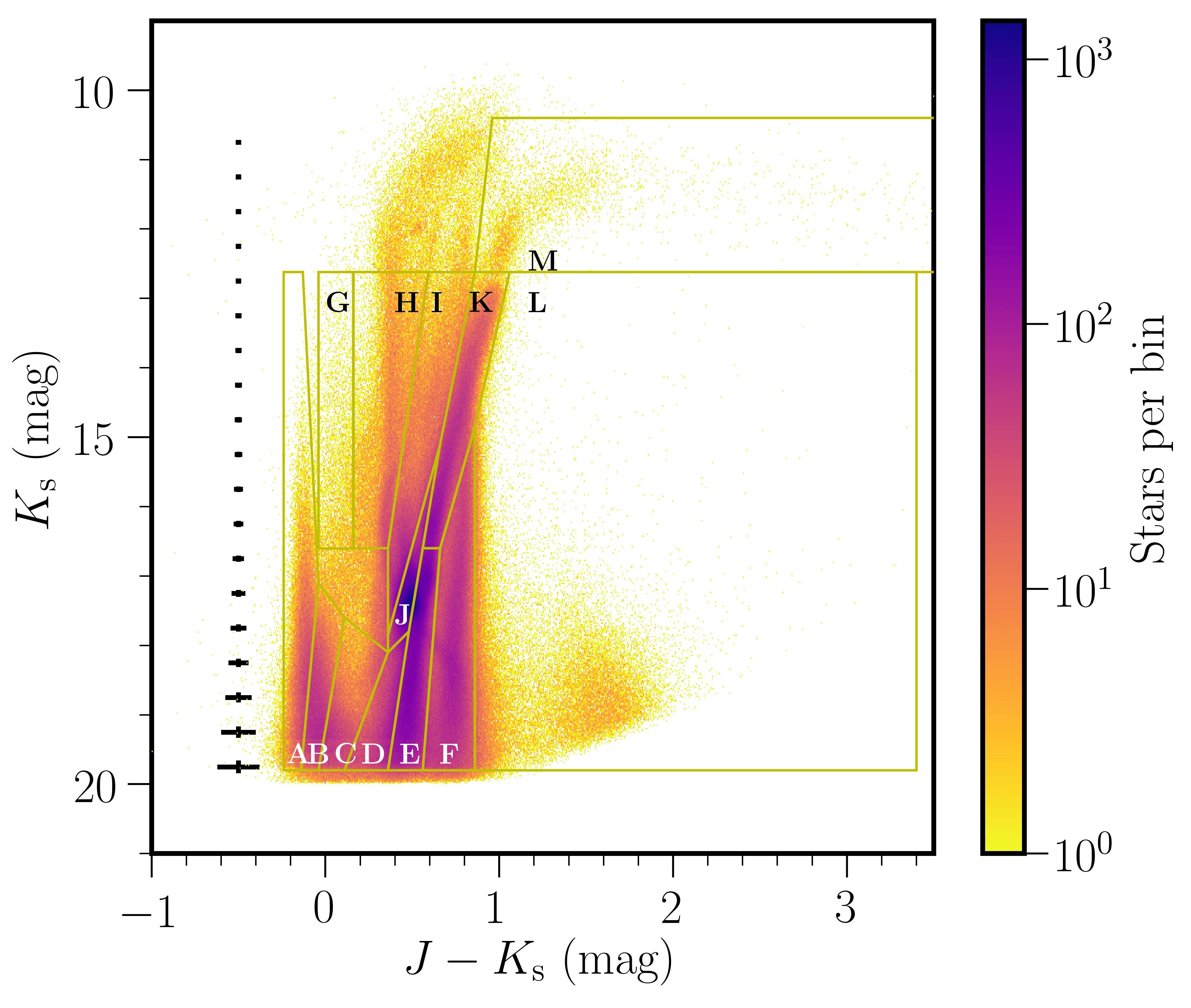}
	\end{center}
	
	\caption{(left) Distribution of VMC tiles in the LMC (top left) and SMC (bottom left). Tile boundaries are colour coded by completion of observations as follows: black (fully observed), and red (partly observed). Contours refer to the number density of stars per bin. Bins of $0.15\times0.15$ deg$^2$ are used and maps are centred at (RA$_0$, Dec$_0$)=($81.00^\circ$, $-69.73^\circ$) for the LMC and (RA$_0$, Dec$_0$)=($13.05^\circ$, $-73.82^\circ$) for the SMC. We used a zenithal equidistant projection, east is to the left and north to the top. The concentrations of stars west and north of the SMC are due to the Milky Way globular clusters 47 Tucanae (47 Tuc) and NGC 362, respectively. (right) Near-infrared ($J-K_\mathrm{s}$, $K_\mathrm{s}$) Hess diagram of the LMC (top right) and SMC (bottom right). The colour scale indicates the stellar density on a logarithmic scale while the yellow boxes, marked by letters, indicate the boundaries of different classes of objects. Region M is limited to $J-K_\mathrm{s}$=$3.5$~mag for clarity, but extends to $J-K_\mathrm{s}$=$6.5$~mag.} 
	\label{fig:realCMD}
	
\end{figure*}

The VMC survey \citep{Cioni2011} is a deep near-infrared photometric survey of the Magellanic Clouds carried out with the 4.1 m Visible and Infrared Survey Telescope for Astronomy (VISTA). It observed $\sim170$ deg$^2$ until its completion in October 2018. Covering $68$ tiles across the LMC ($105$ deg$^2$), $27$ across the SMC ($42$ deg$^2$), $13$ across the Bridge ($21$ deg$^2$) and $2$ within the Stream ($3$ deg$^2$). It provides data in three photometric bands reaching a $5\sigma$ limits of $Y$=$21.9$, $J$=$22$, and $K_\mathrm{s}$=$21.5$~mag in the Vega system.
The observing strategy involves multi-epoch observations. Each tile covers $\sim 1.77$ deg$^2$ and is observed in at least  three epochs in the $Y$ and $J$ bands and in twelve epochs in the $K_\mathrm{s}$ band. The total exposure time per tile in the $Y$, $J$ and $K_\mathrm{s}$ bands are $2400$ s, $2400$ s, and $9000$ s, respectively. Each tile is the result of stacking six pawprints in order to cover the gap between the $16$ detectors of the VISTA infrared camera (VIRCAM; \citealp{Sutherland2015}). The observations were acquired under homogeneous sky conditions with a uniform tile coverage in service mode. The median FWHM of the image seeing in each band is $Y$=$1.03^{\prime\prime}\pm0.13^{\prime\prime}$, $J$=$1.00^{\prime\prime}\pm0.10^{\prime\prime}$, and $K_\mathrm{s}$=$0.93^{\prime\prime}\pm0.08^{\prime\prime}$. The main science goals are to determine the spatially resolved star formation history of the Magellanic Clouds and derive their three dimensional structure. This study makes use of all VMC observations obtained until 2017 September~$30$. The data were extracted from the VISTA Science Archive (VSA\footnote{\url{http://horus.roe.ac.uk/vsa}}; \citealp{Cross2012}), reduced with the VISTA Data Flow System (VDFS; \citealp{Irwin2004}) pipeline versions 1.3/1.5, and calibrated according to \cite{Gonzalez-Fernandez2018}. 
Figure \ref{fig:realCMD} showcases the Magellanic Clouds observations used in this study. All $27$ SMC tiles and $50$ LMC tiles have been fully observed, while $18$ LMC tiles were partly observed. These tiles are generally located in the outskirts of the LMC where crowding is less significant. Therefore the reduced number of epochs should have little influence on their depth. All VISTA raw data and a fraction of the processed data used in this study are publicly available at the ESO and VISTA archives, while the VMC data set including the full SMC, Bridge, and Stream is going to be released as part of Data Release \#5 (expected in 2019).

\subsection{Selection of stellar populations} \label{selectionsp}
Our aim is to derive the morphology of the Magellanic Clouds using different stellar populations. We used the ($J-K_\mathrm{s}$ , $K_\mathrm{s}$) colour-magnitude diagram (CMD) in combination with stellar populations models of the LMC, SMC and the Milky Way to select different classes of objects. These models are produced with the TRILEGAL code \citep{Girardi2016} in its latest version \citep{Marigo2017}. We selected data from the \textit{vmcsource} table, which contains merged sources from individual VMC source detections. Deprecated data resulting from observations obtained outside the nominal VMC observing requirements were not used. Our selection criteria consisted of objects that were classified as stars with at least a $70$\% probability (flags mergedclass=$-1$ and mergedclass=$-2$). We only chose unique (priOrSec$\leqslant0$ or priOrSec=frameSetID) objects detected in the $J$ and $K_\mathrm{s}$ bands, with photometric uncertainties $<0.1$ mag in both bands. No selection criteria based on extraction quality flags (flag ppErrbits) were applied. We used aperMag3, which corresponds to the default point source aperture corrected mag ($2.0^{\prime\prime}$ aperture diameter). In order to express VISTA magnitudes in the Vega system, we added $0.011$~mag to the $K_\mathrm{s}$ band \citep{Gonzalez-Fernandez2018} while no adjustment is needed in the $J$ band. Our dataset contains $10,030,967$ and $2,429,550$ sources in the LMC and SMC, respectively, where about $96$\% of them have a $90$\% probability or greater being stars.

\begin{figure*}
	\centering
	\includegraphics[scale=0.1]{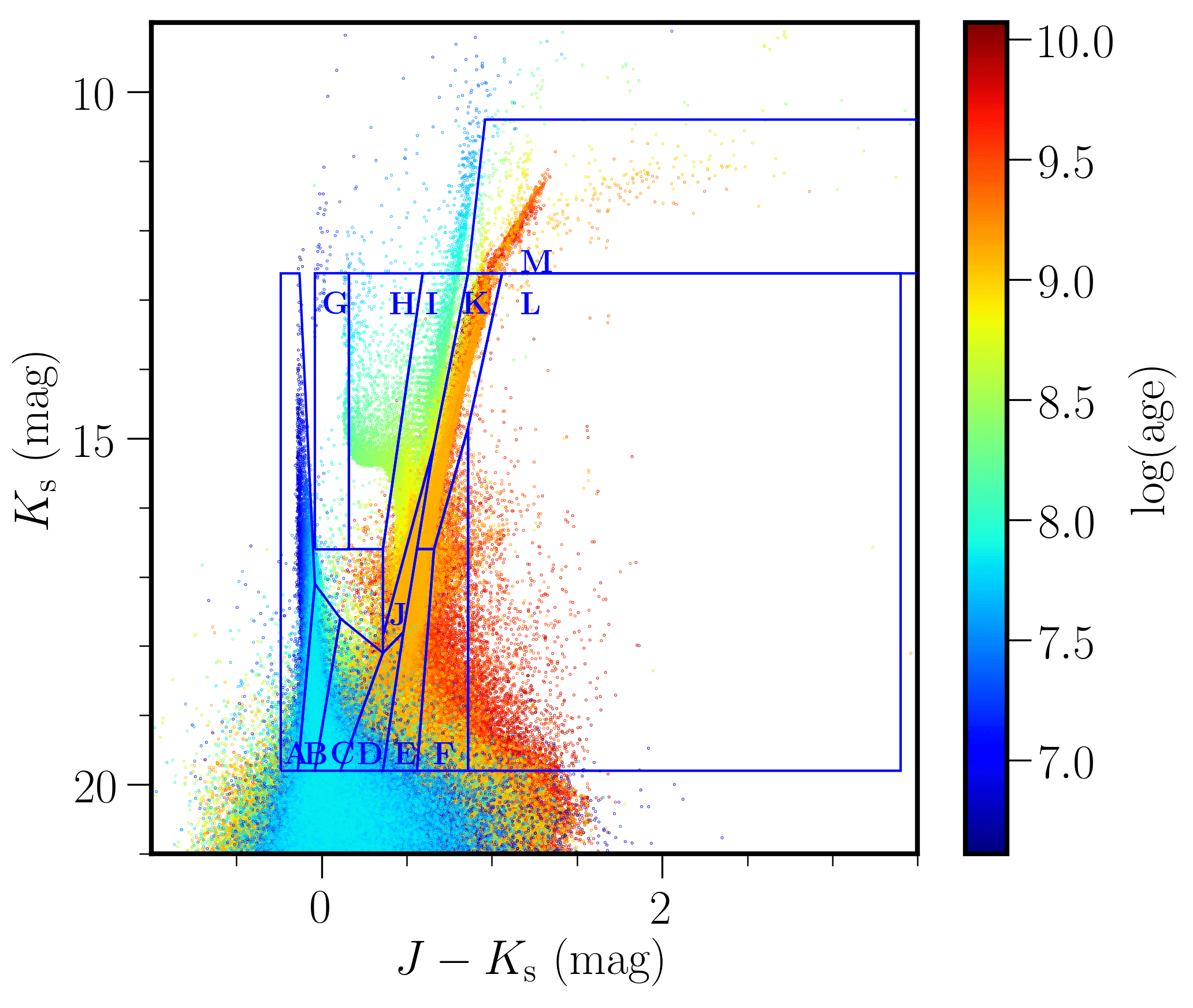}
	\includegraphics[scale=0.1]{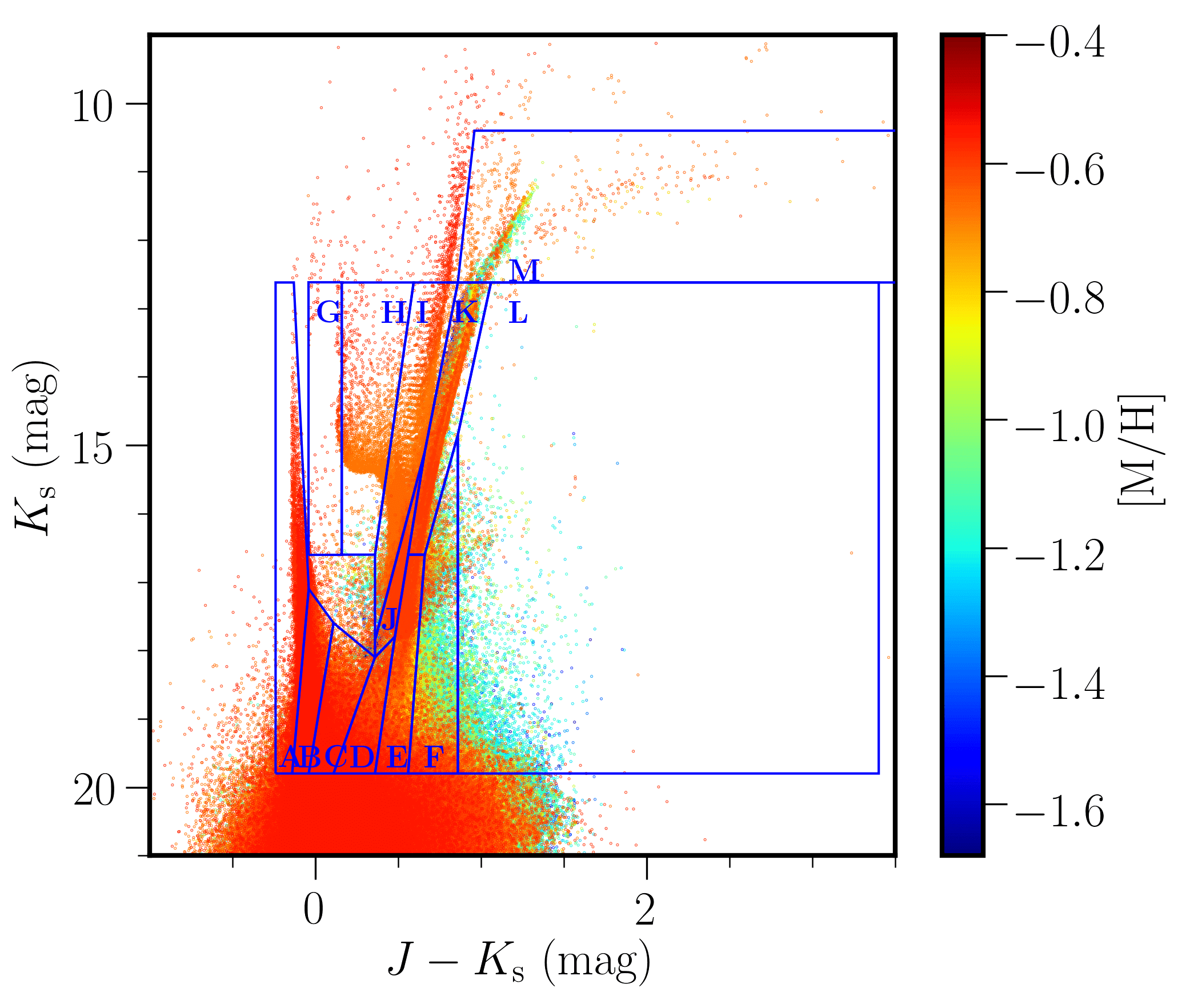}\\
	\includegraphics[scale=0.065]{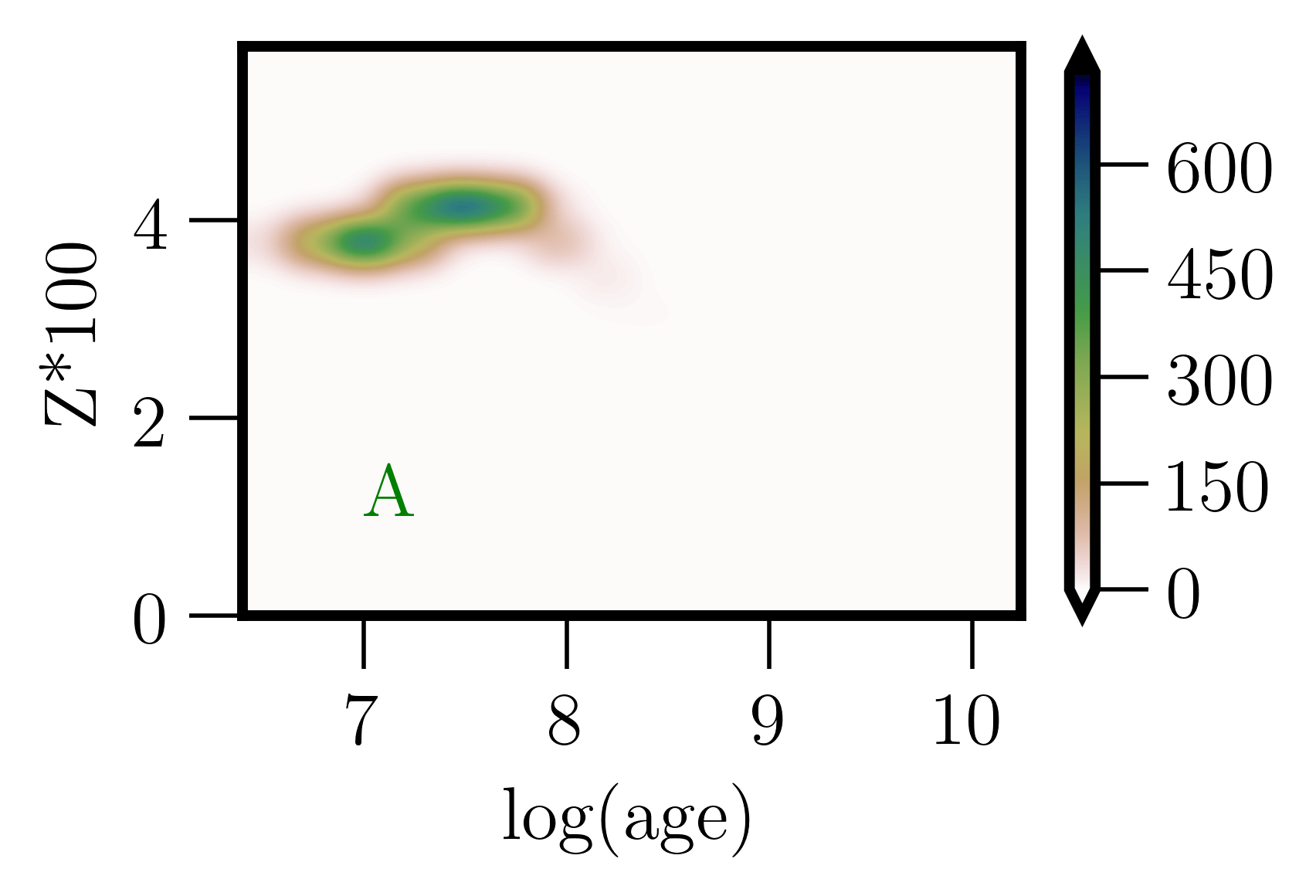}
	\includegraphics[scale=0.065]{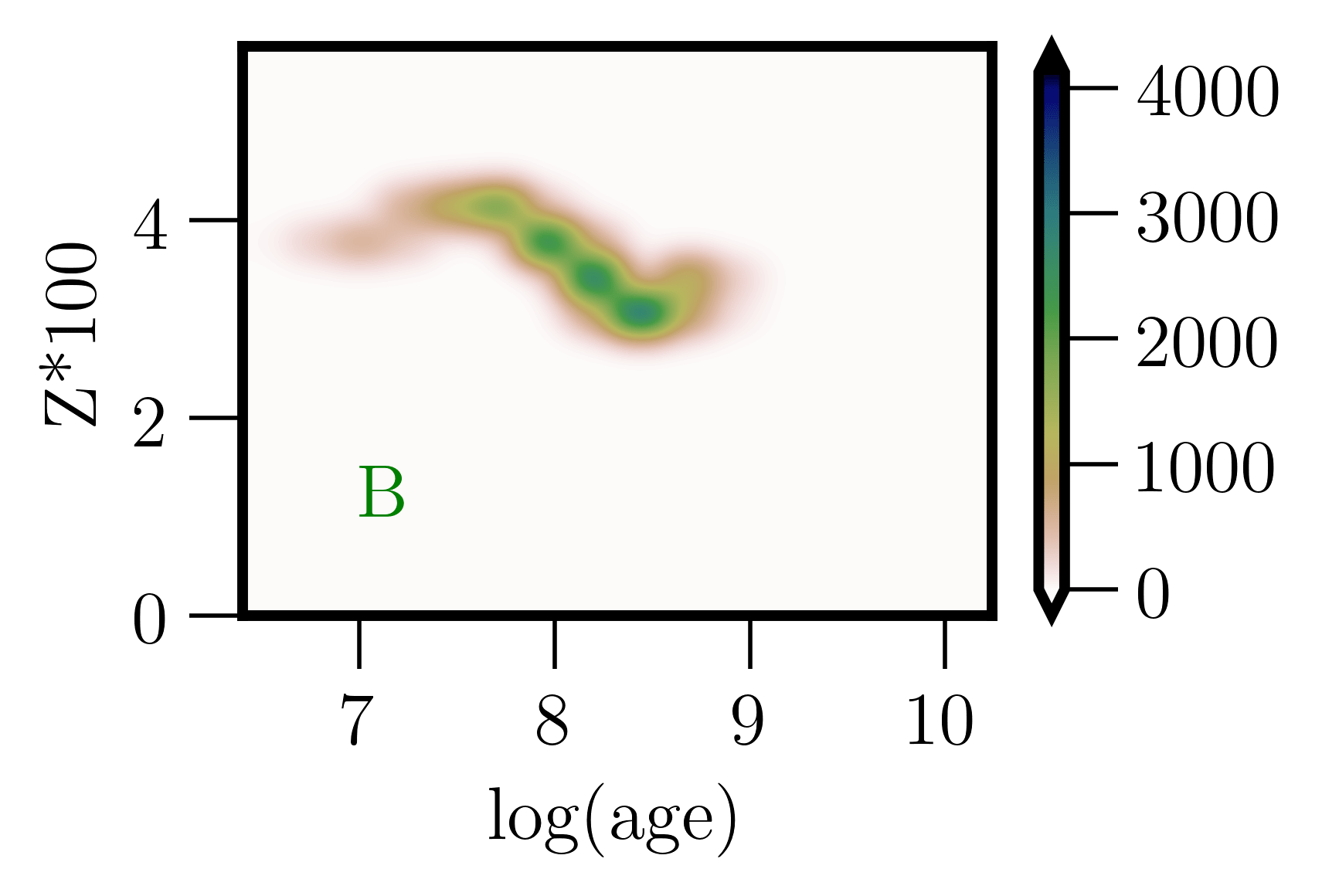}
	\includegraphics[scale=0.065]{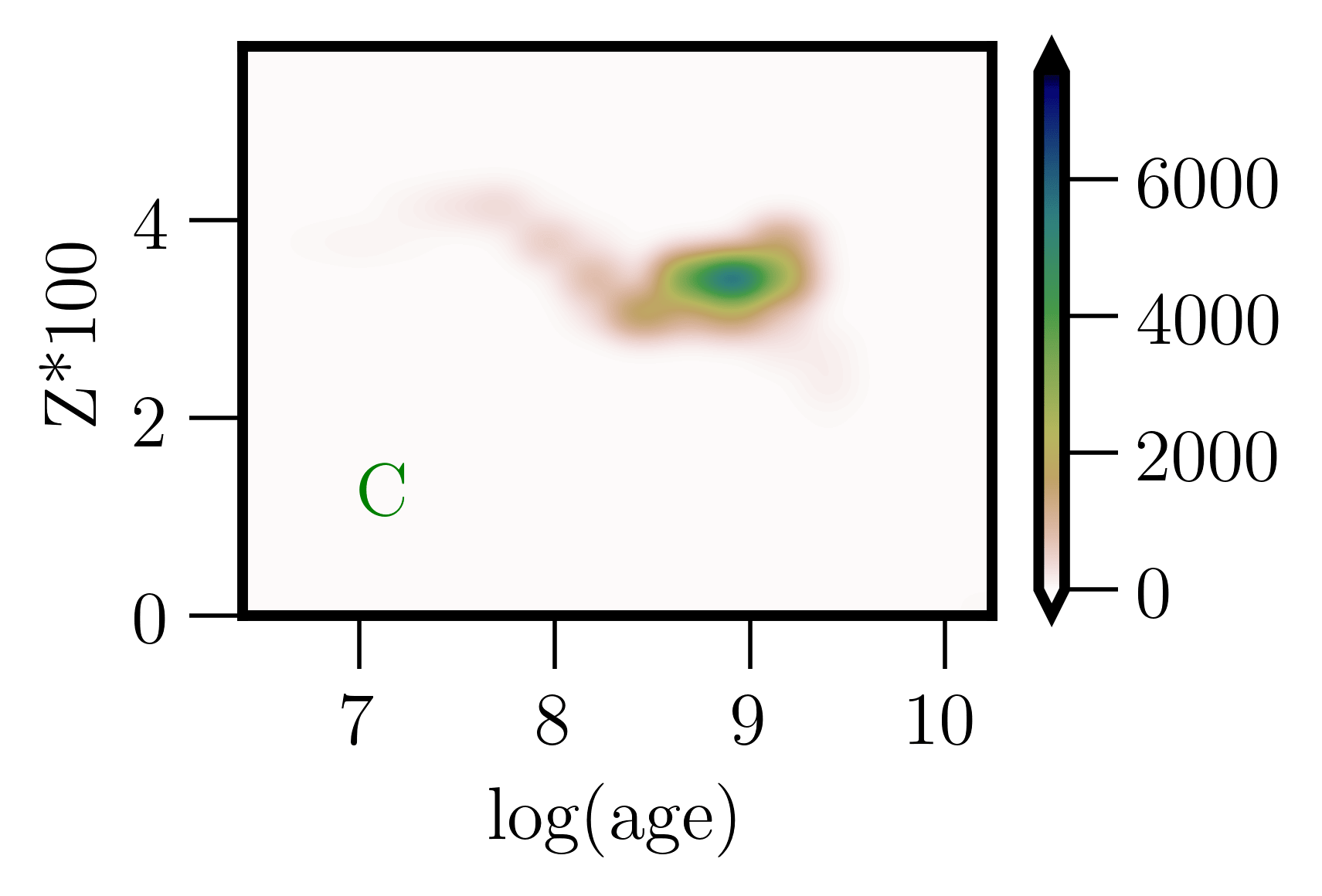}
	\includegraphics[scale=0.065]{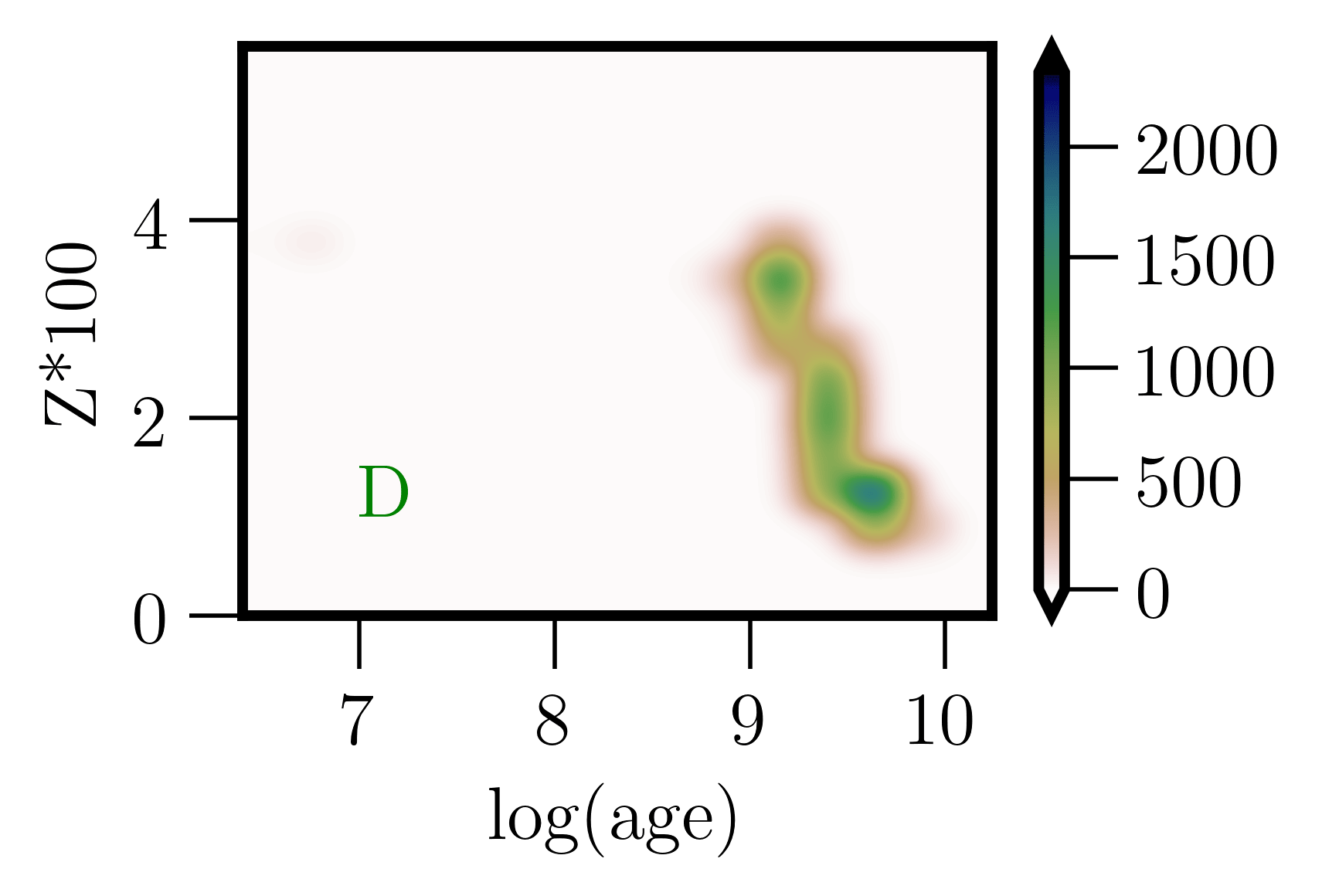}\\
	\includegraphics[scale=0.065]{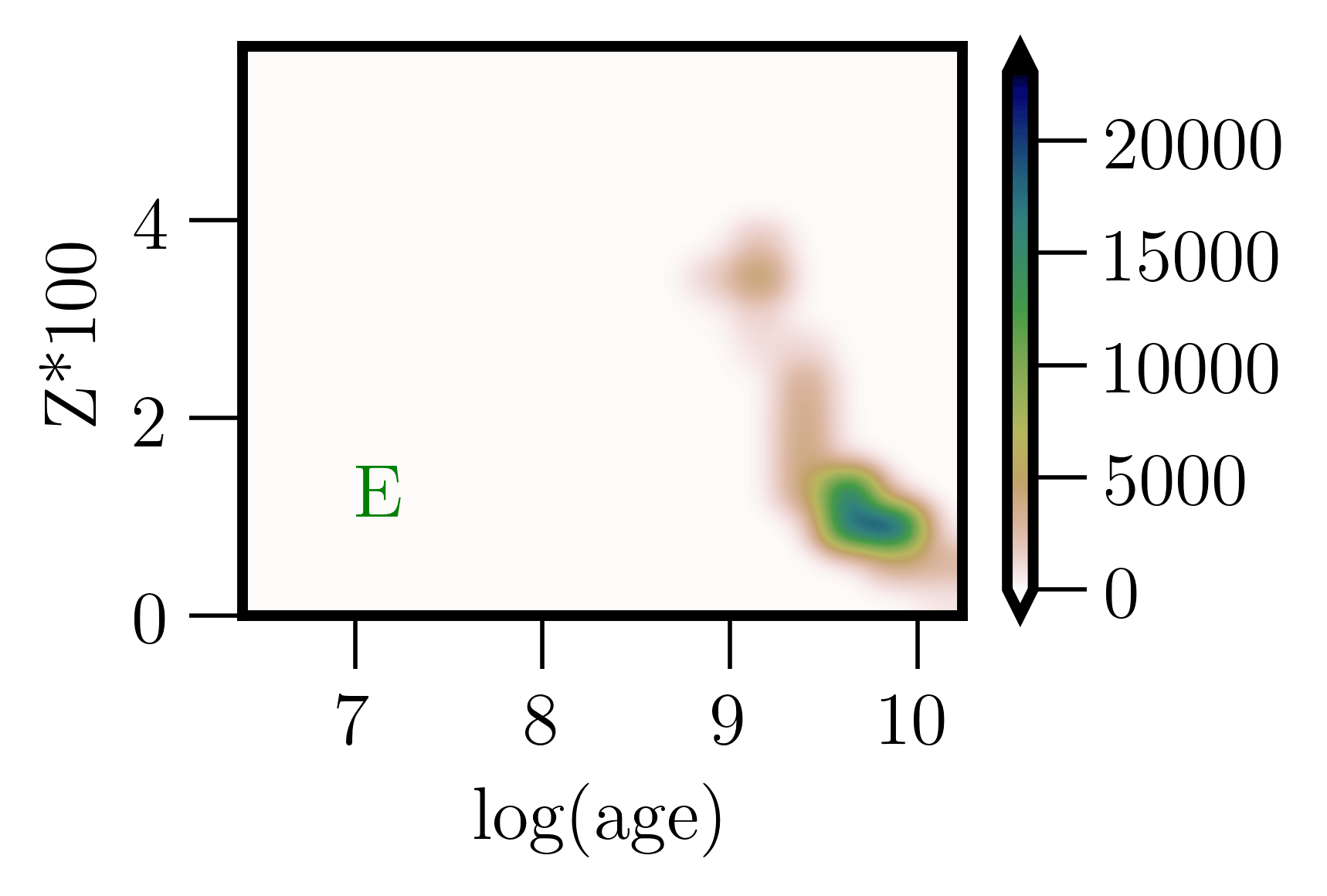}
	\includegraphics[scale=0.065]{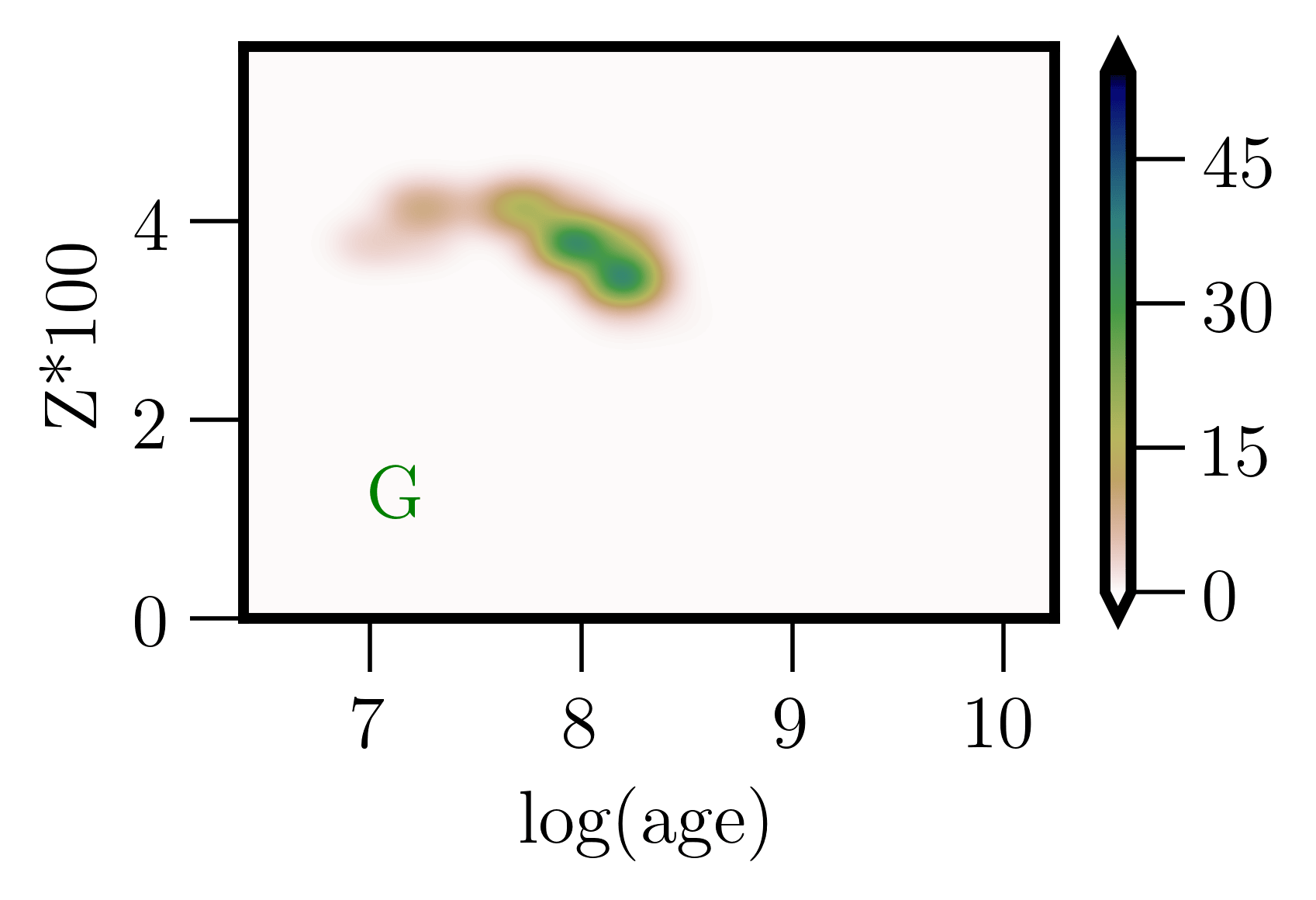}
	\includegraphics[scale=0.065]{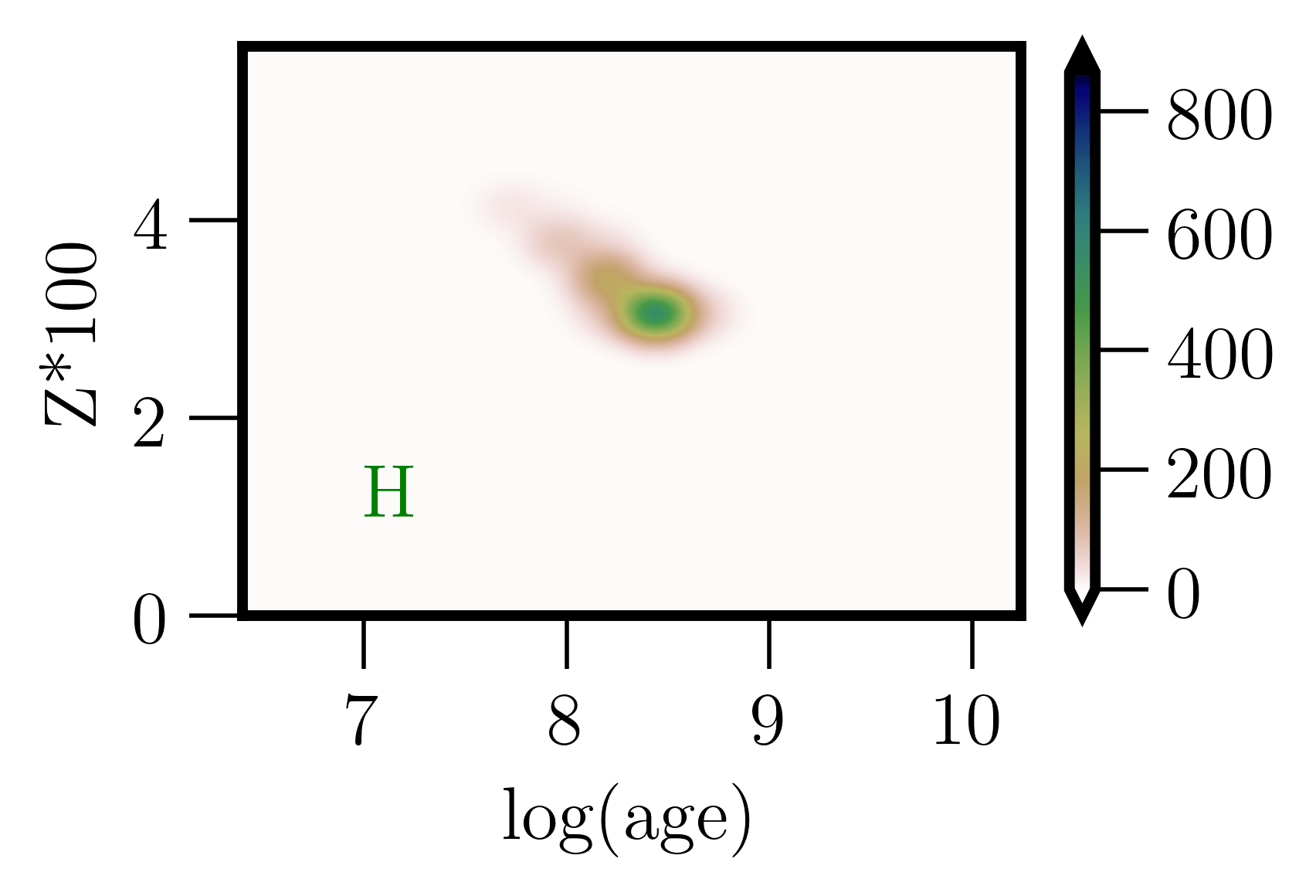}
	\includegraphics[scale=0.065]{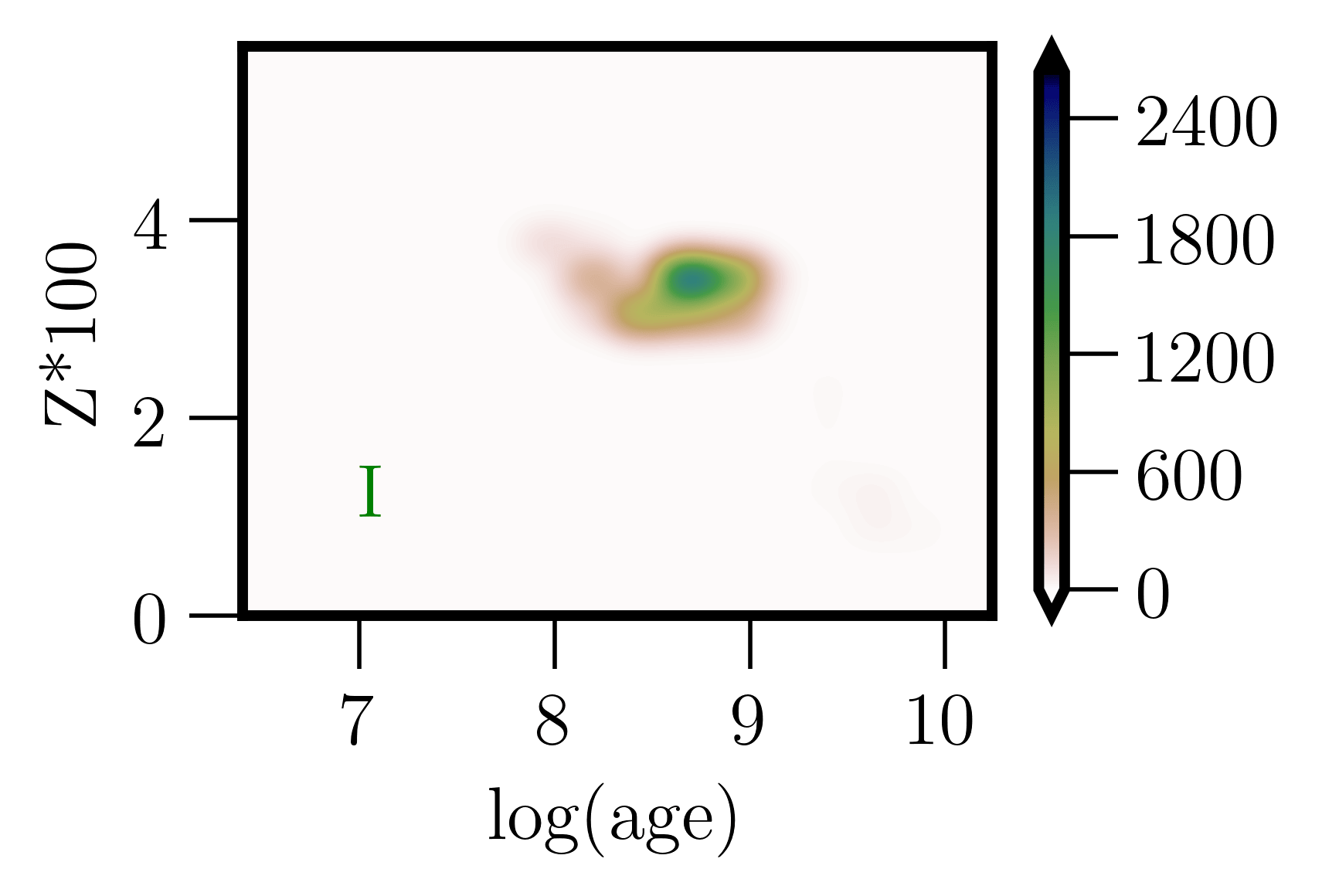}\\
	\includegraphics[scale=0.065]{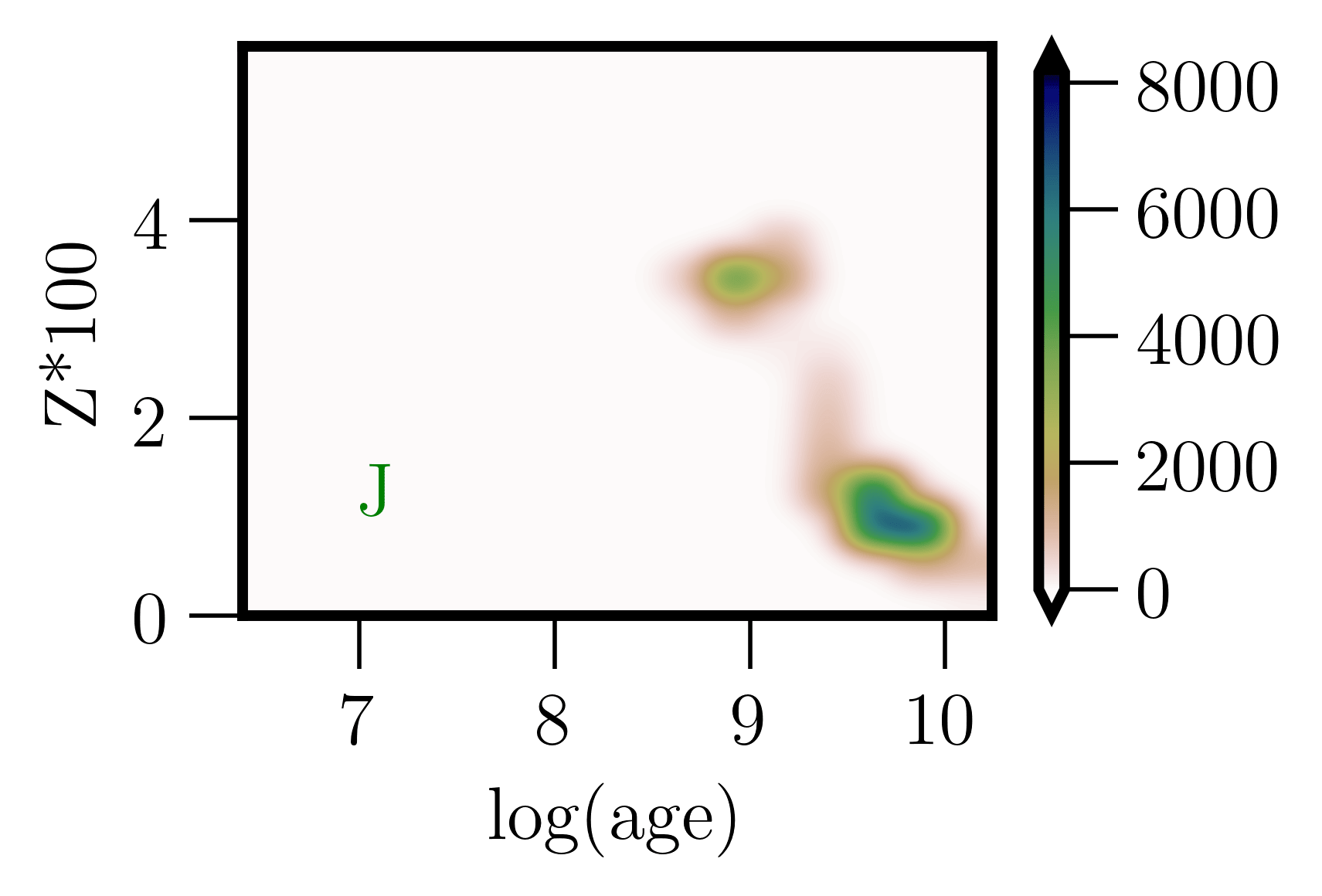}
	\includegraphics[scale=0.065]{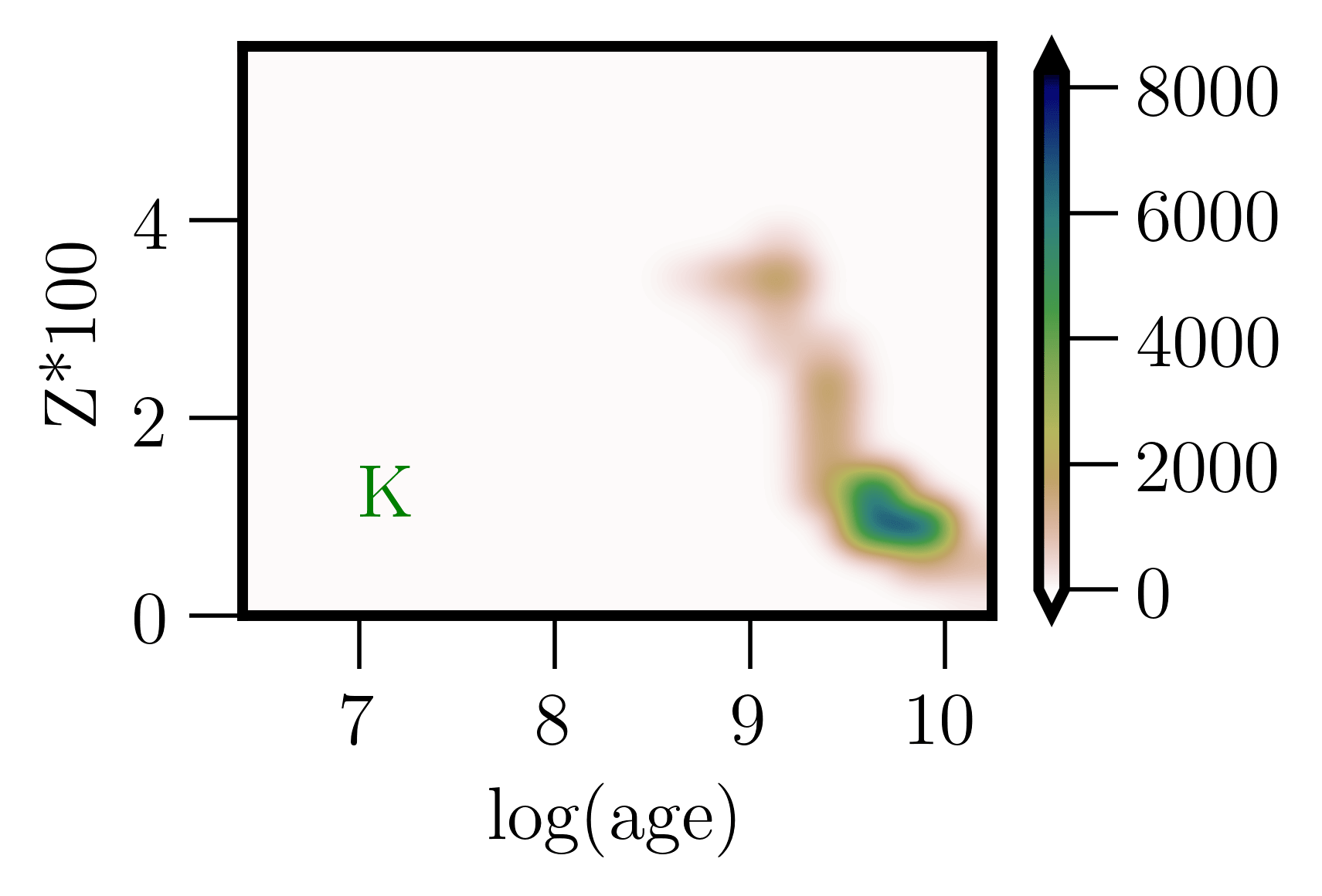}
	\includegraphics[scale=0.065]{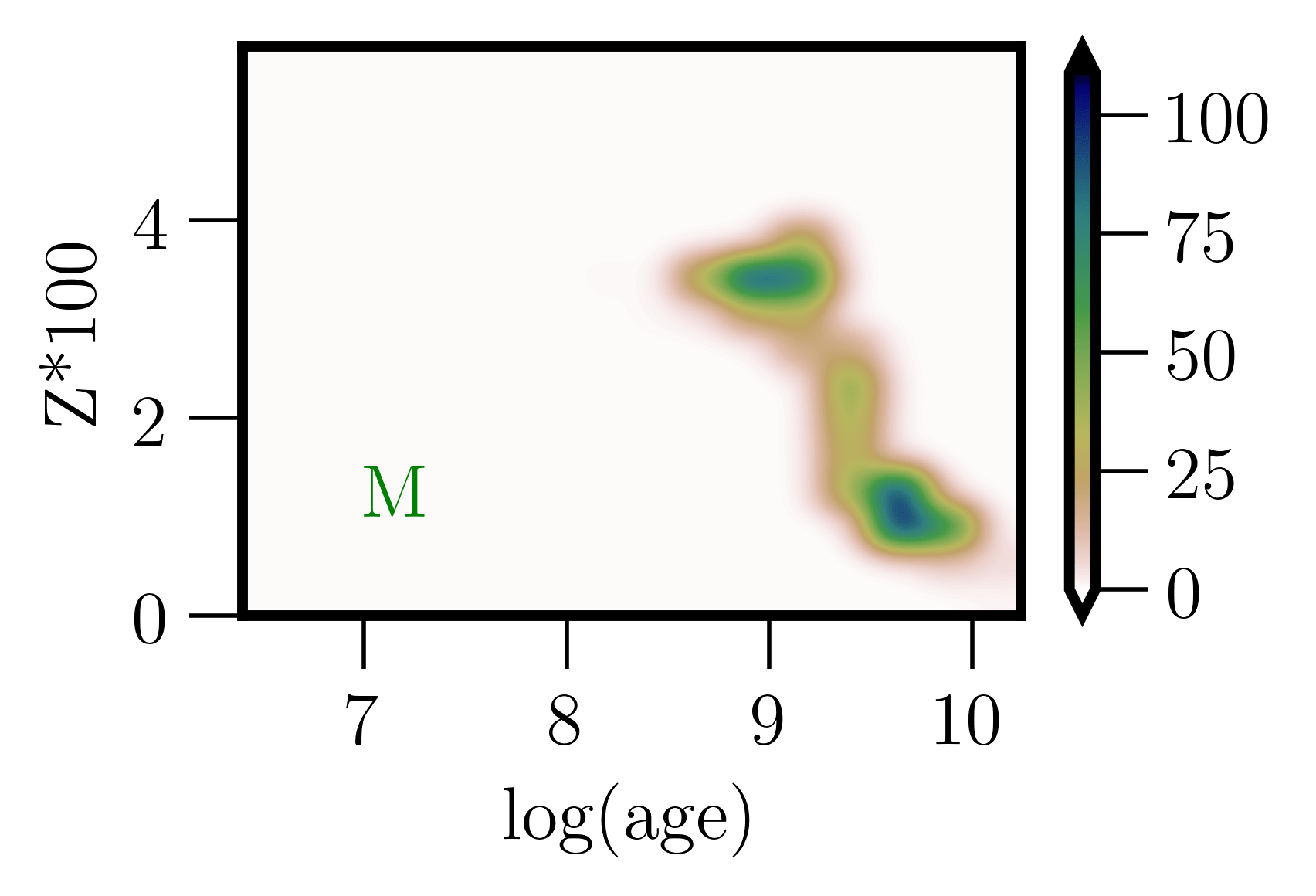}

	\caption{(top) Simulated ($J-K_\mathrm{s}$, $K_\mathrm{s}$) CMDs illustrating stellar populations in the SMC. The colours correspond to a range of ages (left) and metallicities (right). The boxes refer to the regions used to disentangle different stellar populations. (bottom) Age-metallicity diagrams showing the distribution of ages and metallicities for stars inside each CMD region. The bin size is $0.08$ dex$^2$. The colour bars reflect the number of objects per bin.} 
	\label{fig:SIMUCMD}
\end{figure*}

We used CMD regions defined by \cite{Cioni2014} for the LMC while for the SMC, we shifted these regions to take into account differences in mean distance and metallicity of the stellar populations, as in \cite{Cioni2016}. The exact locations of these regions are based on the analysis of the star formation history by \cite{Rubele2012}. The dimensions of the original regions were mainly conceived for proper motion studies, based on single-epoch pawprint images \citep{Cioni2016,Niederhofer2018}. Here we used stacked multi-epoch tile images and hence we updated those regions as follows. We extended the faint limit of regions A, B, C, D, E, F, and L to $K_\mathrm{s}$=$19.8$ mag to maximise the number of stars and the bright limit of regions A, G, H, I, K, and L to $K_\mathrm{s}$=$12.62$~mag for the SMC and $K_\mathrm{s}$=$11.98$ mag for the LMC coinciding with the location of the tip of the red giant branch \citep{Cioni2000a}, we also added region M encompassing thermally pulsing AGB stars.	
Figure \ref{fig:realCMD} outlines the regions that disentangle different stellar populations across the LMC and the SMC. Figure \ref{fig:SIMUCMD} (top panels) shows the same regions overplotted on CMDs produced from theoretical models \citep{Rubele2018}. These models represent synthetic stellar populations covering a grid of age and metallicity bins shifted to an established distance modulus and extinction, adapted to the conditions of our observations by applying the photometric errors and completeness obtained from artificial star tests as well as to the VISTA system zero points. The stellar models assume scaled-solar abundances of
metals with [M/H] $\equiv$ [Fe/H] as well as a present-day solar metal content of
Z = 0.0152. A model describing the Milky Way foreground is also used to assess the Milky Way contamination of the Magellanic Cloud stellar populations \citep{Girardi2016}. The \cite{Rubele2018} models refer to the SMC while new models of the LMC populations are currently being produced.  
The distributions of age and metallicity are colour-coded in the diagrams and can be appreciated from the associated panels depicting the age-metallicity relations within each region. The synthetic CMDs include only the stellar populations of the SMC. 
Tables \ref{table:lmc} and \ref{table:smc} indicate for each region the number of stars, their median age, and the percentage of Milky Way foreground stars using the model and data from the \textit{Gaia} data release \#2 (DR2, \citep{Brown2018}) (see Section~(\ref{sect22})) , as well as the type of the dominant stellar population. We omit region L because it is populated mostly by background galaxies.

\subsection{Completeness}
\begin{figure*}
	\begin{center}
	
		\includegraphics[scale=0.1]{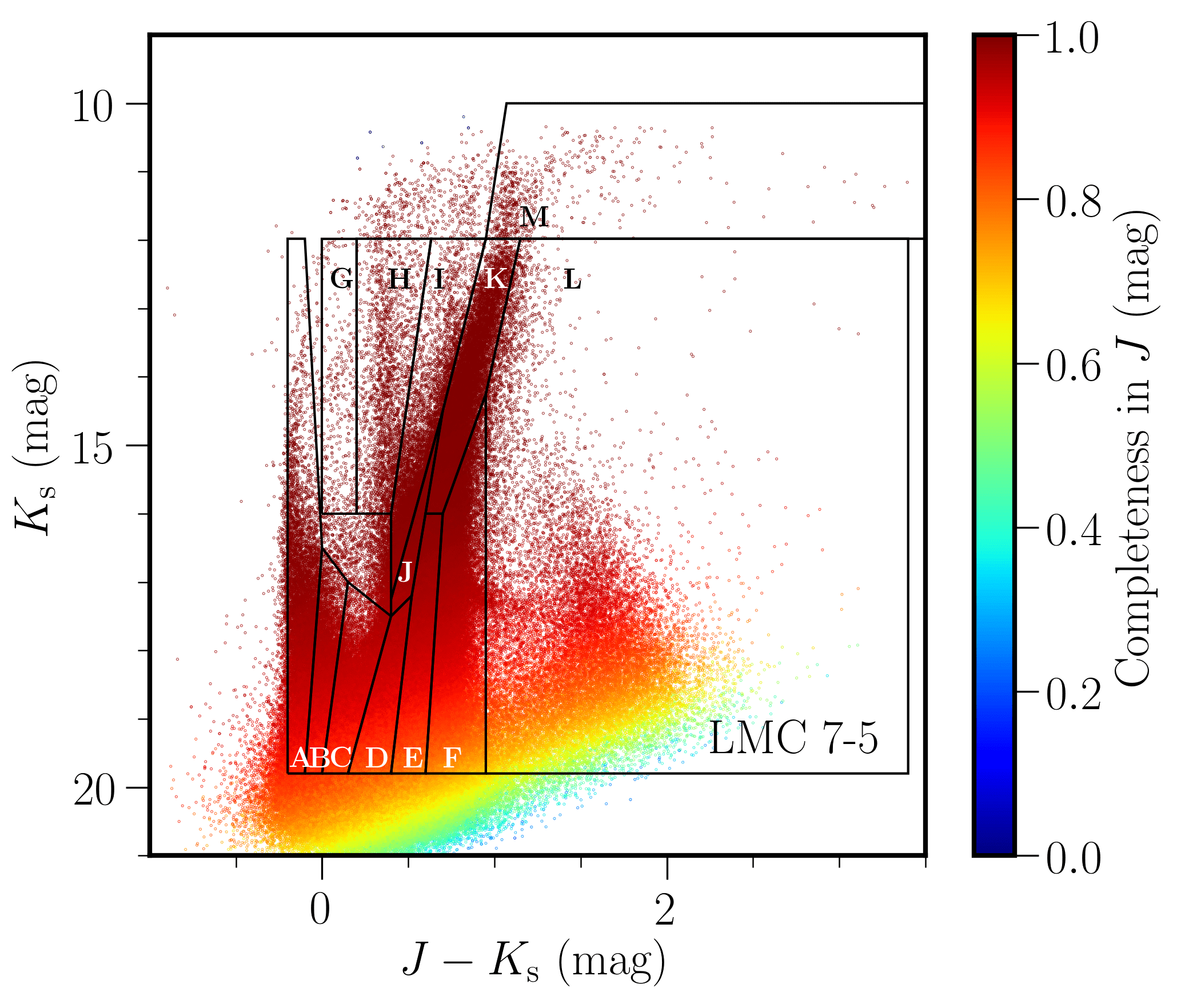}
		\includegraphics[scale=0.1]{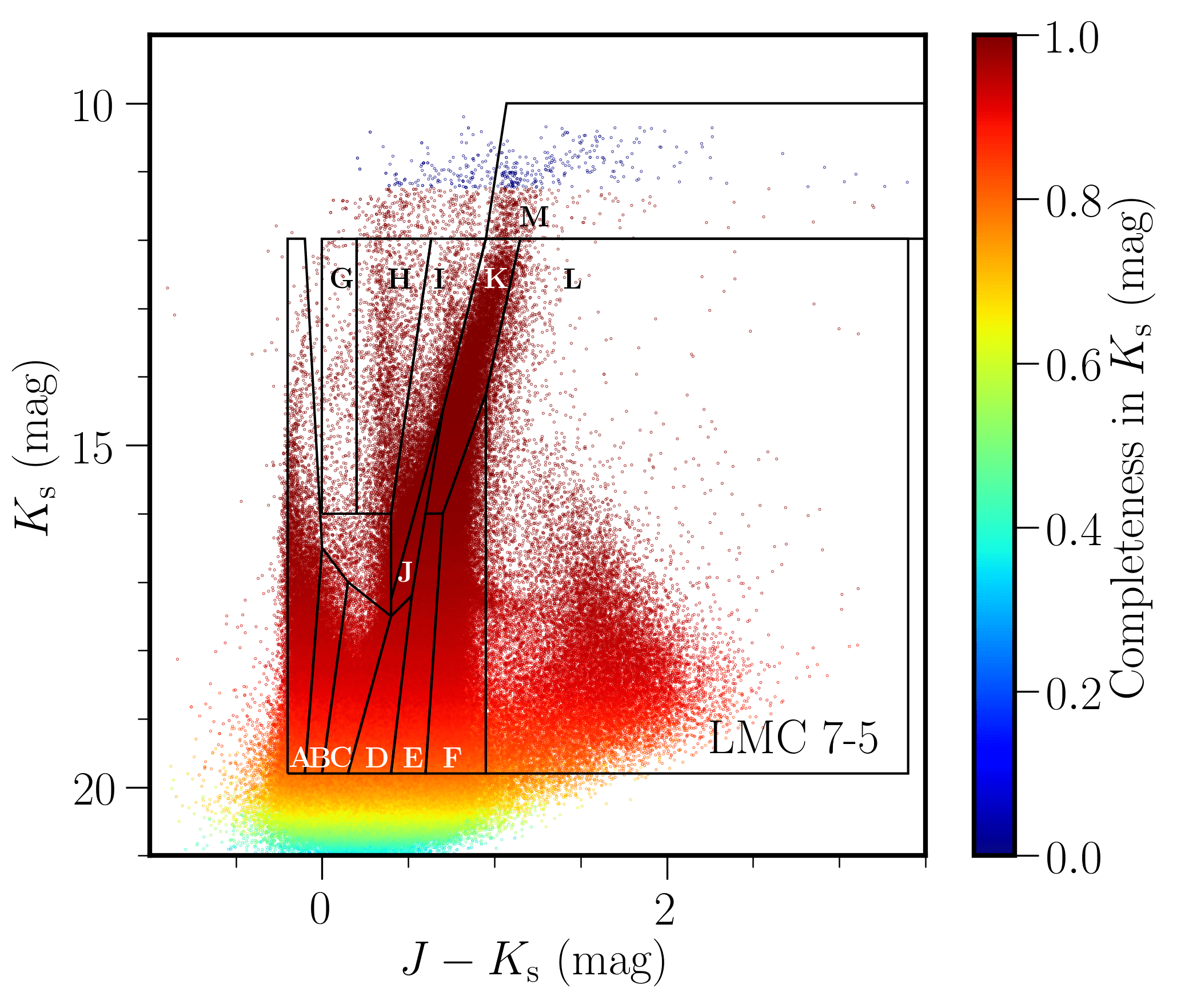}\\
		\includegraphics[scale=0.035]{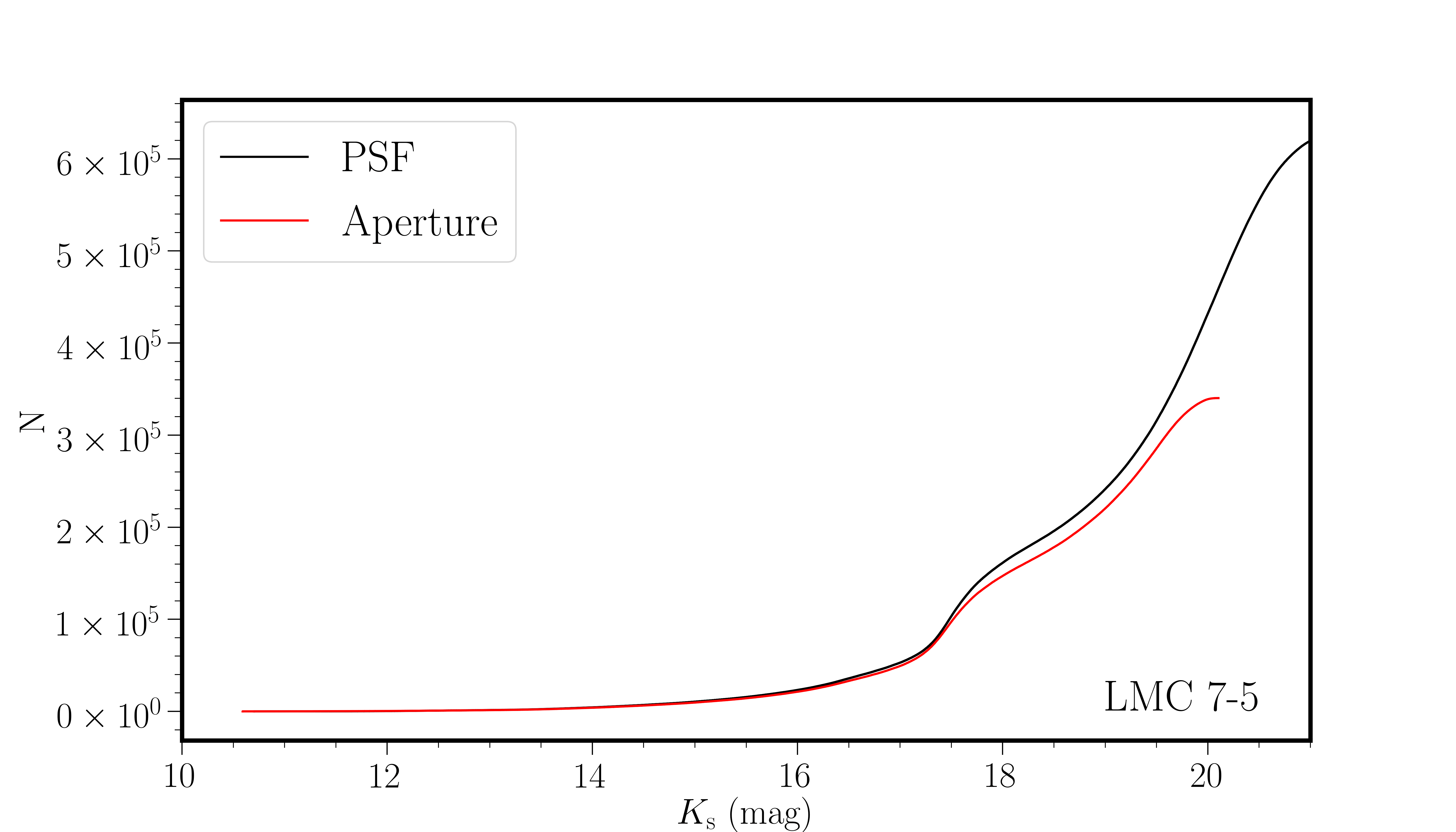}
		\includegraphics[scale=0.035]{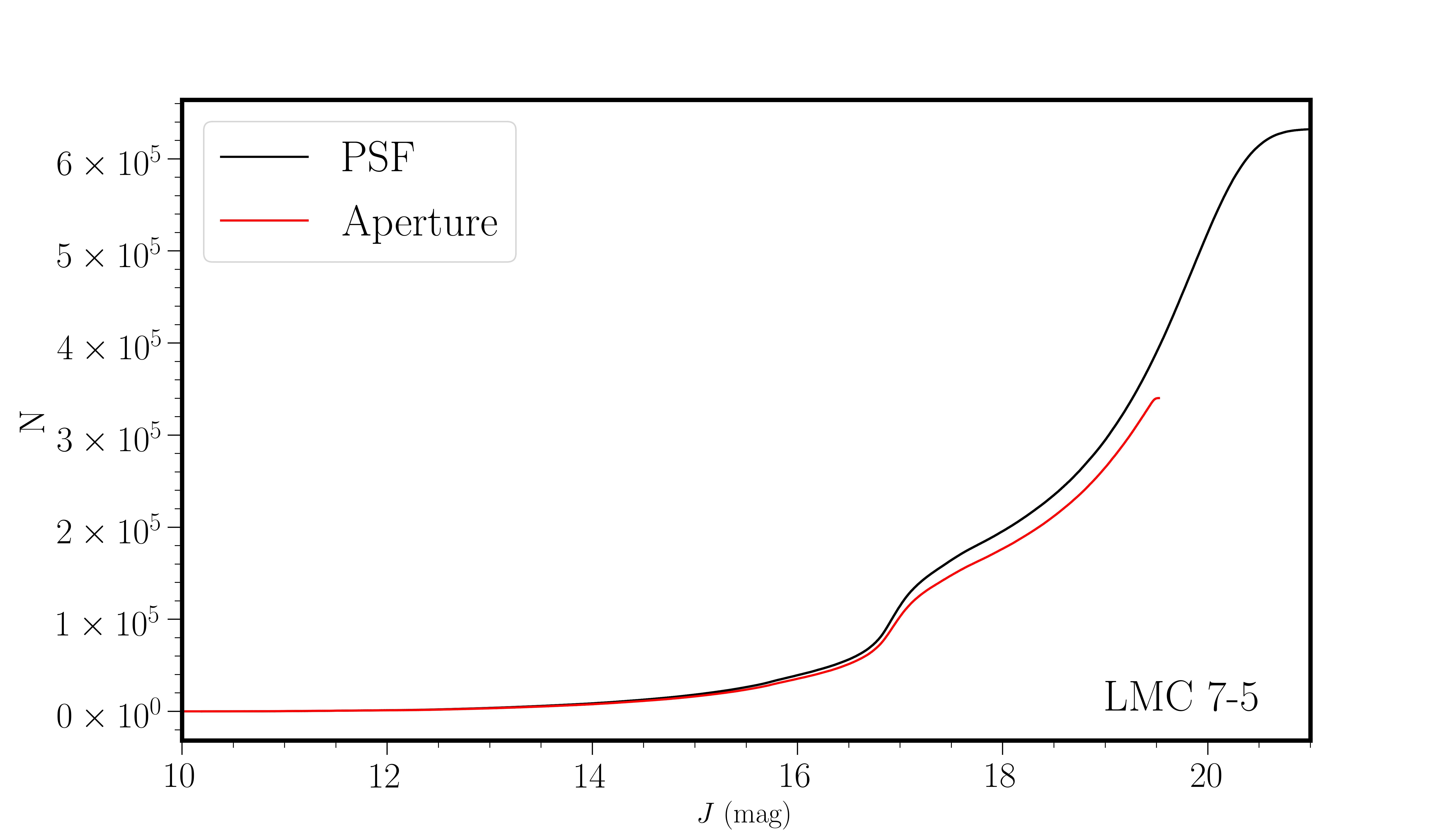}
	\end{center}
	\caption{(top) Examples of completeness diagrams in the $J$ (right) and $K_\mathrm{s}$ (left) bands derived from PSF photometry catalogues on a ($J-K_\mathrm{s}$, $K_\mathrm{s}$) CMD for the tile LMC $7\_5$. (bottom) Comparison between the number of sources extracted with PSF and aperture photometry in the $J$ (left) and $K_\mathrm{s}$ (right) bands. }
	\label{fig:completeness}
\end{figure*}

To assess the completeness of the VDFS aperture photometry catalogues, we used the completeness analysis performed on point-spread-function (PSF) photometry catalogues that are available for some tiles \citep{Rubele2012,Rubele2015}. This analysis consists of running large numbers of artificial star tests on tile images in order to trace the distributions of photometric errors and completeness, as a function of position, magnitude, and colour. Figure \ref{fig:completeness} shows the results of the completeness calculations for the $J$ and $K_\mathrm{s}$ bands, separately.
The region boundaries for the selection of stellar populations are influenced by the variation of the completeness across different VMC tiles where the PSF photometry was available. We focus on tile LMC $7\_5$ as it represents the average quality of our observations. It is located in the inner disc of the LMC where crowding is moderate and comparable to that in the central regions of the SMC. In the external regions of both the LMC and SMC the completeness is higher, while there are tiles in the central regions of the LMC where the completeness is lower, i.e.~in the tile containing the 30 Doradus star-forming region.
The completeness maps show a rather good recovery of stars even in the crowded central parts of the galaxies. 
In tile LMC $7\_5$ the completeness level at the faint end corresponds to $75$\% while it is only $50$\% in tile LMC $6\_4$. Tile LMC $6\_4$ is located close to the bar of the LMC and it is therefore more crowded than tile LMC  $7\_5$. The completeness of SMC tiles is higher than that of LMC tiles even in crowded regions. At the brightest magnitudes, outside the CMD regions, incompleteness is due to objects close to the saturation limit. Blue objects brighter than $K_\mathrm{s}$ = $11$~mag in the completeness diagram of the $K_\mathrm{s}$ band reflect saturated sources. These sources are included in region M because they are clearly detected in $J$ and for the morphology study we are only interested in their number.

Completeness established based on PSF photometry may however differ from completeness derived using aperture photometry. In order to test the completeness levels in the aperture photometry, we performed a comparison of the number of stars in the aperture and PSF catalogues (Fig.~\ref{fig:completeness}). No object classification criteria (mergedclass flag) was used in our selection, only objects detected in the $J$ and $K_\mathrm{s}$ bands with photometric uncertainties $<0.1$ mag in both bands. The figure shows that both PSF and aperture photometry extract the same number of sources until $K_\mathrm{s}\sim17$ and $J\sim17.5$~mag. Fainter than these magnitudes we notice a drop of approximately $10$\% in both bands. We conclude that the completeness in the aperture-photometry catalogues is the same as that derived from PSF photometry for sources brighter than $K_\mathrm{s}\sim17$ and $J\sim17.5$~mag while it is $10$\% worse for fainter sources in aperture-photometry. 

\subsection{Uncertainties}\label{sect22}
There are a few factors to consider that influence the position of stars in the CMD and their selection using boxes, such as the presence of dust, photometric uncertainties and distance variations. Additionally, not all stars in the CMD belong to the Magellanic Clouds, but may be part of the Milky Way. In this subsection, we use stellar population models to determine the influence of Milky Way stars. Tables \ref{table:lmc} and \ref{table:smc} showcase the percentage of Milky Way sources. For the LMC, These values were derived using models from an ongoing SFH study by S. Rubele et al. (in preparation) within several LMC tiles, while for the SMC we used stellar population models from \cite{Rubele2018}. Regions A, B, C, D, E, J, K and M have negligible fractions of Milky Way sources. Region F has the highest percentage of sources belonging to the Milky Way with $94\%$, followed by region H with $77$ per cent, regions G and I have $13$ and $15$ per cent of Milky Way stars, respectively. The SMC has similar contamination levels except for regions K, I and D which have percentages up to $24$ per cent higher than in similar regions of the LMC. We also used \textit{Gaia} data release \#2 (DR2) to test the fraction of Milky Way contamination using a cross-matched \textit{Gaia}-VMC catalogue. In order to obtain this catalogue, accounting for the time at which \textit{Gaia} and VMC observations were obtained was necessary. \textit{Gaia} (J2015) objects were moved to the epoch of the VMC (J2000) survey and therefore only sources with proper motions in the \textit{Gaia} catalogue were used. To query the cross-matched catalogue, we used the same selection criteria for the VMC data given in Sect.~\ref{selectionsp}. Additionally, several \textit{Gaia} selection criteria employed in previous papers were tested. The $\omega$/$\bar{\omega}$ (parallax / parallax error) $\leq10$ in addition to $G\leq19$~mag, as in \cite{Helmi2018A}, only disentangles Milky Way objects up to  $K_\mathrm{s}\sim15$~mag and hence was not used in this study. Choosing only objects with $\omega\leq0.2$~mas is problematic as parallaxes are significantly smaller than the typical measurement error. Therefore only objects with $\omega \leq 0.2$~mas, not consistent with zero at more than $3\sigma$  and have an \texttt{astrometric\char`_excess\char`_noise} $\leq0.2$~mas were chosen to be Milky Way objects, as in \cite{Vasiliev2018}. Using a simple parallax cut might lead to a greater potential of mistaking Magellanic Clouds stars for Milky Way stars and therefore overestimating the Milky Way contamination percentage (see Figs.~\ref{fig:ANNMW1} and \ref{fig:ANNMW2}). The limitations of \textit{Gaia} DR2 enables us to estimate the Milky Way contamination for only regions G H, I, K, M and F. We excluded region J because residual red clump stars are still present, and provide a lower limit of the percentage of Milky Way stars to $K_\mathrm{s}\sim15$~mag. Apart from region F, which is the one that is mostly influenced by the Gaia sensitivity, the percentages of Milky Way stars between the Milky Way model and the Gaia data are in very good agreement. The spatial distribution of Milky Way objects is expected to be homogeneous across the Magellanic Clouds. Therefore we have not attempted to correct our morphological maps for it. The percentages of Milky Way stars are reported in Tables \ref{table:lmc} and \ref{table:smc}. CMDs showing the Milky Way stars towards the Magellanic Clouds are shown in Figs.~\ref{fig:ANNMW1} and \ref{fig:ANNMW2}.

The dust content towards the Magellanic Clouds is generally quite low. \cite{Choi2018A} found an average LMC reddening corresponding to $E$($g-i$) = $0.15 \pm 0.05$~mag using RC stars. Using the same stellar population, \cite{Haschke2012} found a mean reddening of the LMC of $E$($V-I$) = $ 0.09 \pm 0.07$~mag, while $E$($V-I$) = $ 0.04 \pm 0.06$~mag is found for the SMC. The reddening values can be converted using the following equation $E$($J-K_\mathrm{s}$) = $ 0.43 \times$ $E$($V-I$) \citep{Ripepi2016} which results in $E$($J-K_\mathrm{s}$) = $0.02-0.04$ mag. We have not corrected the VMC data for reddening. It is however important to note that  common extinction values of $A_V$=$0.45$~mag and  $A_V$=$0.35$~mag were applied to the theoretical models of the LMC and SMC, respectively. Using the \cite{Cardelli1989} extinction curve for $R_V$=$3.1$, $A_V$ can be converted to $A_J$ and $A_{K_\mathrm{s}}$ which gives absorption coefficients of $A_J$=$0.1$~mag and $A_{K_\mathrm{s}}$=$0.04$~mag. Since the width of the boxes ranges from $0.1$~mag to $0.3$~mag, the reddening will have a minor effect on the possible displacement of sources outside the boxes. To quantify reddening effects on stellar populations, we examined differences in the number of sources in region K of the LMC. We chose this box because it is tilted in colour compared to the other boxes, which facilitates the displacement of sources outside it in the presence of reddening. If we correct for the reddening with an average value of $E$($J-K_\mathrm{s}$) = $0.03$~mag, 93 per cent of the sources are still present in the selection box.
	
\begin{figure}
\hspace{-1.4pt}	
\includegraphics[scale=0.36]{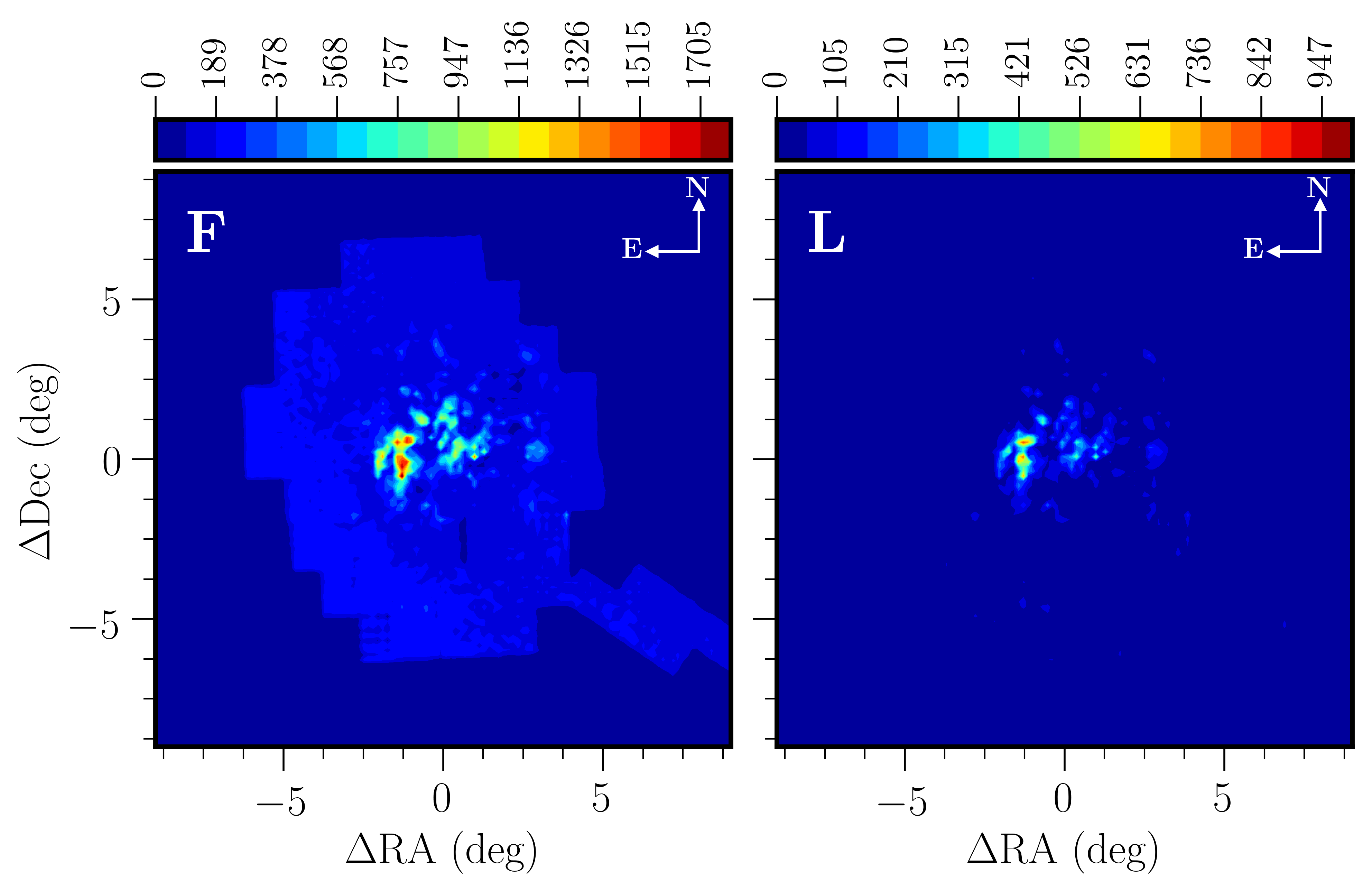}\\
	\includegraphics[scale=0.4]{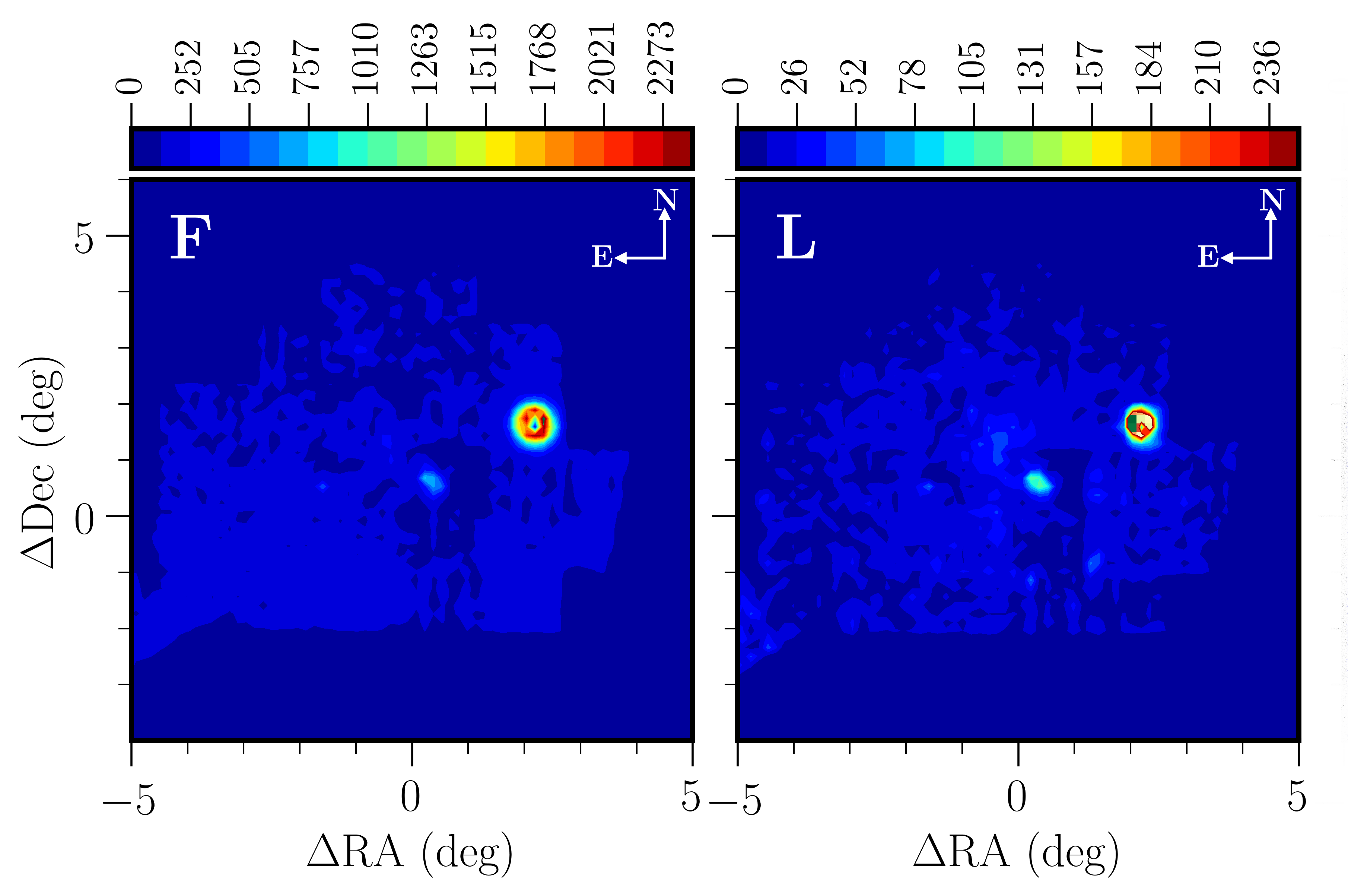}

	\caption{Stellar density/contour maps of regions F (dominated by Milky Way stars, but containing reddened Magellanic Clouds giants) and L (dominated by background galaxies, but containing Magellanic Clouds giants and stars of the 47 Tuc star cluster) for the LMC (top) and SMC (bottom). The bin size is $0.03$ deg$^2$. The colour bars show the number of stars per bin.}
	\label{fig:MW}
\end{figure}

Regions F and L represent regions dominated by Milky Way stars and background galaxies, respectively. Their spatial distributions are expected to be homogeneous across the Magellanic Clouds. The remaining overdensities (Fig.~\ref{fig:MW}) reflect reddened LMC and SMC sources that can be used as indicators of high extinction regions in the Magellanic Clouds, except for that associated with the 47 Tuc cluster.

Photometric uncertainties in the $J$ and $K_\mathrm{s}$ bands were limited to $<0.1$~mag, which corresponds to uncertainties in ${J-K_\mathrm{s}}$ colours  of $<0.14$~mag. These uncertainties are well contained within the widths of the boxes at a limiting magnitude of $K_\mathrm{s}\sim17.5$~mag (Figs.~\ref{fig:realCMD} and \ref{fig:errors}). Displacement of sources from one region to another for sources fainter than these magnitudes is therefore possible, but since our selection criteria were the same for the theoretical models and the data, we expect that the effect is in part accounted for by the age uncertainties (Tables \ref{table:lmc} and \ref{table:smc}).\\
Using RC stars, \cite{Subramanian2009} found that the observed dispersion due to the line-of-sight depth ranges from $0.023$~mag to $0.45$~mag (a depth from $500$ pc to $10.4$ kpc) for the LMC and, from $0.025$ to $0.34$~mag (a depth from $670$ pc to $9.53$ kpc) for the SMC. Using RR Lyrae stars, the line-of-sight depth is found to be in the range $1-10$~kpc, with an average depth of $4.3\pm1.0$ kpc \citep{Muraveva2018}, while in the LMC, the bar can be traced as a protruding overdensity with a line-of-sight depth of almost $5$ kpc (\citealp{Haschke2012}). \cite{Ripepi2017} found that the SMC is elongated by more than $25-30$ kpc in the east/north-east towards south-west direction. 
The vertical extension of the CMD boxes is larger than the average $0.2$~mag variation due to populations at different distances. However, regions that are tilted in colour may suffer from depth effects. A shift of $0.2$~mag of the sources in region K, for the LMC, leaves 86 per cent within the box.

\section{Morphology of the Magellanic Clouds}\label{section3}
Figures \ref{fig:morphlmc} and \ref{fig:morphsmc} outline morphological maps of stellar population regions for both the LMC and the SMC. They represent two dimensional density maps based on a set of $20$ evenly spaced filled contours from the lowest to the highest density. The colour bar depicts star distributions in bins of 0.15$\times$0.15 deg$^2$ .
The projection origins for the LMC and SMC are, set at, respectively (RA$_0$, Dec$_0$) = ($81.24^\circ$, $-69.73^\circ$) and (RA$_0$, Dec$_0$) = ($13.05^\circ$, $-72.83^\circ$), corresponding to the densest point in the LMC bar and the optical centre of the SMC \citep{DeVaucouleurs1972}. Coordinates were transformed from angular to Cartesian through a zenithal equidistant projection \citep{VanderMarel2001}.

\subsection{LMC}
In order to investigate the morphological features of the LMC, we selected LMC structures, some of which are well known, based on density contours combined with the age tomography of the various stellar populations. Figure \ref{fig:substructure} gives an overview of the main features we focus on while describing the morphology of the LMC.
	\subsubsection{Young stellar populations}
\begin{itemize}

	\item{\textbf{Region A} represents a population of young main sequence stars with a median age of $\sim20$ Myr ($\pm$ 15 Myr). The overall morphology traced by this population is clumpy and irregular. The bar is thin and spans a length of about $4^\circ$. It has no obvious central overdensity and clumps are distributed along it. Several star forming regions appear as prominent overdensities, Shapley constellation III ($\Delta$RA=$-1^\circ$, $\Delta$Dec=$2.5^\circ$), 30 Doradus ($\Delta$RA=$-1^\circ$, $\Delta$Dec=$0.5^\circ$), N79 ($\Delta$RA=$2.5^\circ$, $\Delta$Dec=$0^\circ$), and N11 ($\Delta$RA=$2.5^\circ$, $\Delta$Dec=$2.5^\circ$). We note the presence of spiral arms emerging from the ends of the bar to the south and north of the galaxy. The SE arm is faint and shows an overdensity towards its tip ($\Delta$RA=$-0.5^\circ$, $\Delta$Dec=$-2^\circ$) while the SW arm ($\Delta$RA=$2^\circ$, $\Delta$Dec=$1^\circ$) is more enhanced with a few overdensities. The NW arm represents the main spiral arm of the galaxy and has a break at ($\Delta$RA=$1^\circ$, $\Delta$Dec=$3^\circ$) which divides it in two parts. \cite{Harris2009} named the vertical part (from $\Delta$Dec=$2^\circ$ to $\Delta$Dec=$4^\circ$ and $\Delta$RA=$2^\circ$) the blue arm, and the horizontal part, extending to $\Delta$RA=$3.7^\circ$, the NW arm. In the following section, we will refer to the whole structure as the NW arm.}
	
	\item{\textbf{Region G} represents a population of blue supergiants and giant stars with a median age of  $\sim81$ Myr ($\pm$ 39 Myr). These stars are mainly distributed across a thin bar, similarly to the stars of region A. The SW arm has a clear overdensity at the location of the N79 star forming complex. The star forming region 30 Doradus represents the highest density feature in this map. The vertical and horizontal parts of the NW arm, embedding Shapley constellation III, are clearly visible, but the SE arm is feeble.}
	
	\item{\textbf{Region H} represents a population of red supergiants and giant stars with a median age of $\sim170$ Myr ($\pm$ 59 Myr) and contains most of the Cepheids. This region is heavily influenced by the presence of Milky Way stars, which contrary to those in region F are difficult to disentangle using CMD criteria. They are probably responsible for the increase in the number of stars in the outer regions of the galaxy leaning towards the north east, which is in the direction of the Galactic plane. However, the bar is clearly traced. It has a prominent central density and no obvious gaps. The NW arm is also visible, while the SE and the SW arms are not clear. The low density contours connecting the bar with external clumps encompass the known star forming complexes (30 Doradus, Shapley constellation III, N11, and N79.)}
	
	\item{\textbf{Region B} represents a population of main sequence stars with a median age of $\sim195$ Myr ($\pm$ 206 Myr). The spiral arm features are denser and broader than in Region A. Shapley's Constellation III no longer constitutes the densest area of the map. We note three major overdensities, one at the east and two at the west end of the bar from where the SE and the SW arms originate. The overdensity at ($\Delta$RA=$1.5^\circ$, $\Delta$Dec=$0.5^\circ$) connects the bar to the SW arm. The bar has no central overdensity, but is instead traced by a thin ($<1^\circ$) smooth structure. A protuberance emerging from the SW arm in the direction of the SE arm is also visible. The NW spiral arm is enhanced and it shows an additional faint arm-like feature protruding to the external regions. There is a significant lack of stars between the bar and the NW arm, referred to as the NW void by \cite{Harris2009}, and between the bar and the SW arm. The lack of stars in a linear vertical feature below the bar is due to an observational gap.}
	
	\item{\textbf{Region I} represents a population of yellow supergiants and giant stars, similar in median age ($\sim446$ $\pm$ 257 Myr) to those in region H. We identify the following features: a prominent bar, a major overdensity near the central region, a weak SE arm, as well as a weak NW arm. The bar appears disconnected from the SE arm.}
	
	\item{\textbf{Region C} represents a population of main sequence stars with a median age of $\sim891$ Myr ($\pm$ 739 Myr). We produced two morphological maps for this region in order to assess completeness variations among tiles. One map contains stars with $K_\mathrm{s}<19.8$ mag (Fig.~\ref{fig:mapscd}) as was done for the other regions, while the other contains stars with $K_\mathrm{s}<19.4$ mag (Fig.~\ref{fig:morphlmc}). In these figures, the SW arm is connected to the bar and to the SE arm rather than extending south to the outer regions. The bar appears disconnected from the overdensity to the west. The NW arm, although less dense than in the previous maps, clearly shows both the horizontal and vertical parts. The additional northern arm is clearly visible. Furthermore, another external arm-like feature is present in the SW and extends to $\Delta$Dec=$-5^\circ$.}

\end{itemize}

	\subsubsection{Intermediate-age and old stellar populations}
\begin{itemize}
	
	\item{\textbf{Region D} represents a population of main sequence and/or subgiant stars with a median age of $\sim2.45$ Gyr ($\pm$ 1.53 Gyr). We also produced two morphological maps for this region for the same reasons as stated above (Figs.~\ref{fig:morphlmc} and \ref{fig:mapscd}). The overall morphology of the galaxy appears more regular than in main sequence stars with no significant spiral arms, except for the SE inner arm and the SW outer arm. This stellar population has a prominent overdensity at the east end of the bar. The bar itself is inconspicuous. This map complements the previous maps by filling the gaps above and below the bar. The horizontal part of the NW arm is not easily disentangled from its vertical part.}
		
	\item{\textbf{Region M} represents the thermally pulsing AGB population with a median age of $\sim2.45$ Gyr ($\pm$ 1.58 Gyr). The AGB phase in the theoretical models is not robust to draw solid conclusions about the properties of the population. Therefore the age of the population should be considered carefully. The bar is well defined and it has a clear central density. The structure is rather regular and smooth, especially compared to that of region D traced by a population with a similar median age. The North-West and South-East arms are present but they are faint.}
	
	\item{\textbf{Region K} represents the upper RGB stars with a median age of $\sim3.23$ Gyr ($\pm$ 2.16 Gyr). This stellar population is located above the RC and shows that the galaxy has a regular structure with a prominent thick ($\sim2^\circ$) bar. The bar is twice as thick as those outlined by supergiant and main sequence stars. There is also a major overdensity at its centre, and a gap with respect to the SE arm. The outermost regions of the bar show a flaky distribution that is not found in the other maps and that cannot be explained by the tiling pattern of the VMC survey. The density of stars in the outer disc regions is asymmetric with protuberances that may or may not be associated with spiral features.}
	
	\item{\textbf{Region J} is dominated by the intermediate-age RC population, but also includes old horizontal and red giant branch stars, resulting in a median age of $\sim3.71$ Gyr ($\pm$ 3.42 Gyr), see Fig.~\ref{fig:SIMUCMDLMC}. The overall morphology of the stellar population resembles that of stars in region D, but with a denser and larger overdensity encompassing the bar region. The gap between the bar and the South-East arm is clearly visible. The North-West arm is faint.}
	
	\item{\textbf{Region E} represents the lower (below the RC) red giant branch (RGB) stars with a median age of $\sim3.72$ Gyr ($\pm$ 2.48 Gyr). The spatial distribution of these stars is leaning towards a radially symmetric structure characterised by inner clumps and irregular contours. The largest overdensity is located west of the centre, but this may be due to a lack of reddened stars that end up instead in region F (cf.~Fig.~\ref{fig:substructure}). On the other hand, this overdensity complements the low-density region produced by old main sequence and subgiant stars. The South-West arm is still visible, but this is not the case for the other inner and outer arms highlighted in the previous maps.}

\end{itemize}

In summary, main sequence stars exhibit coherent structures and the spatial extent of this population grows with age. Several star forming regions are outlined as overdensities (Shapley's constellation III, 30 Doradus, N11, and N79) which grow dimmer as we progress in age. The bar is traced by a relatively thin and clumpy structure without a prominent central feature. Overdensities are present at each end of the bar. They are stronger and appear to detach from the bar itself at older ages. Main sequence stars also trace the distinct multi-arm structure of the LMC. We found a clear connection between the SW spiral arm and the bar which becomes more enhanced with age. Although supergiant stars also represent a young population, they show less substructure than main sequence stars. Their spatial extent is more or less similar at the different ages, but the bar has a different length. Star forming regions are more evident as overdensities as traced by older supergiants. The spiral arms are not as clearly traced by supergiants as they are by young main sequence stars. Intermediate-age and old stars are represented by upper and lower RGB stars, red clump (RC) stars as well as thermally pulsing AGB stars. The oldest stars we detect are over 10 Gyr old, but their number is significantly lower than younger stars populating similar CMD regions. The distribution traced by lower RGB stars exhibits more irregularities in its inner parts, possibly due to the influence of RC stars, and the bar is wider than in the other maps. All four populations show hints of spiral arms, although they are located in the opposite direction compared with those traced by young stars.

\subsection{SMC}
\begin{figure*}
	\begin{center}
		\includegraphics[scale=0.235]{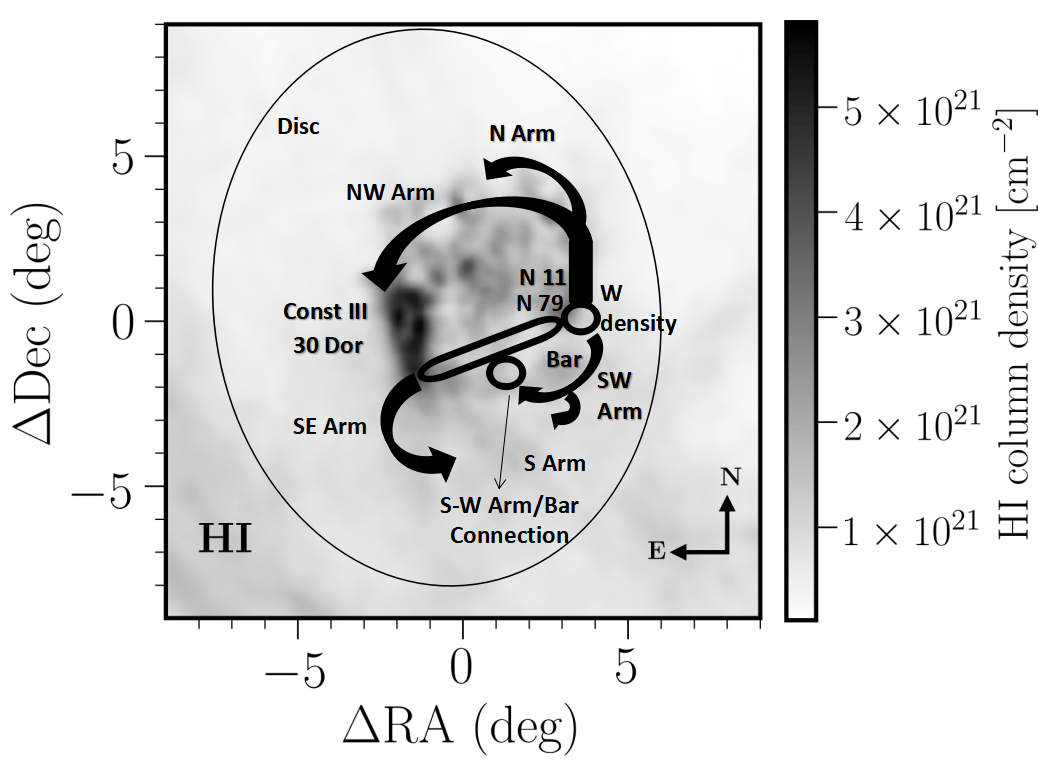}
		\includegraphics[scale=0.28]{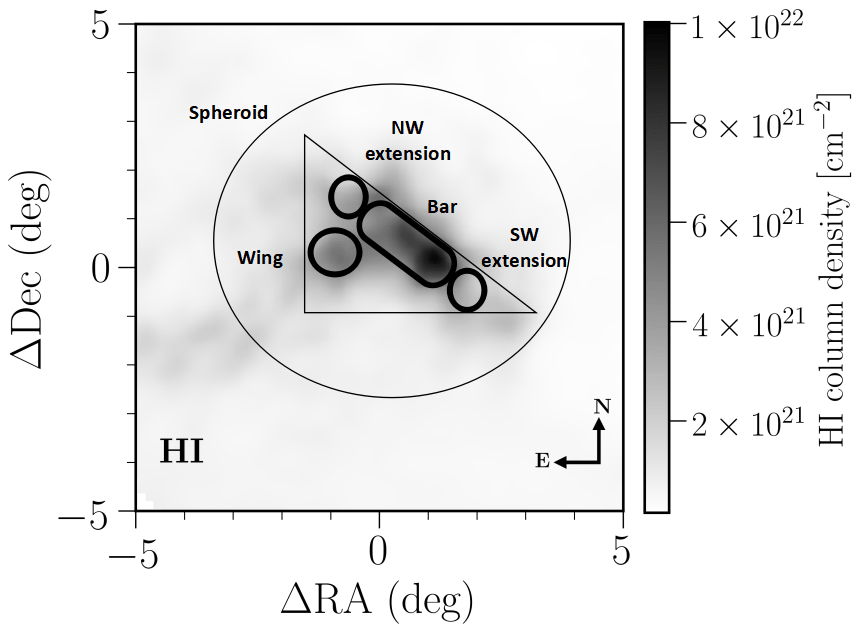}
	\end{center}
	\caption{Morphological features discussed for the LMC (left) and SMC (right) superimposed on the \ion{H}{i} gas column density obtained from \protect\cite{Bekhti2016}.}
	\label{fig:substructure}
\end{figure*}

Figure \ref{fig:morphsmc} reflects the spatial distributions of various stellar populations and highlights the transition of the SMC morphology as function of time, from that of a spherical system to an asymmetric and irregular one. Towards the SMC line-of-sight and within the area surveyed in this study there are two Milky Way star clusters (NGC 362 and 47 Tuc). They appear in the morphology maps as clear overdensities, but they are not discussed in detail. NGC 362 is located to the north of the main body of the galaxy and 47 Tuc is located to the west. Figure \ref{fig:substructure} gives an overview of the main features we focus on while describing the morphology of the SMC.

\subsubsection{Young stellar populations}
\begin{itemize}
	\item{\textbf{Region A} represents young main sequence stars with a median age of $\sim20$ Myr ($\pm$ 15 Myr). The presence of the Wing is most evident in this stellar  population ($\Delta$RA=$-2^\circ$, $\Delta$Dec=$0.5^\circ$). The asymmetric appearance of the young population is outlined by a broken bar. The bar density shows a discontinuity at $\Delta$RA=$-0.5^\circ$ after which the bar bends east by $\sim30^\circ$. The stellar densities within the two parts of the bar are rather homogeneous. Protuberances possibly associated with tidal interaction events are apparent: north of the upper bar (NE extension), south west of the lower bar (SW extension) and to the Wing. The faint structures around $\Delta$RA=$-4^\circ$ are located towards the direction of the Magellanic Bridge.}
	
	\item{\textbf{Region G} represents a population of blue supergiants stars with a median age of $\sim112$ Myr ($\pm$ 80 Myr). The bar is composed of a single density protruding and bending slightly to the north east. This is a mild effect compared with the broken bar shown in the map from region A. The NE and SW extensions are clearly visible, but the Wing is faint.}
	
	\item{\textbf{Region B} represents a population of main sequence stars with a median age of $\sim141$ Myr ($\pm$ 150 Myr). In this map the bar is not prominent as in Region A and is composed of only one southern triangular density. North of this structure, instead of the upper bar shown in the previous map of younger main sequence stars, there is a fuzzy and irregular density distribution of stars. The extension of the Wing is reduced in size and so are the NE and SW extensions.}
	
	\item{\textbf{Region H} represents a population of red supergiants stars with a median age of $\sim234$ Myr ($\pm$ 103 Myr) and contains most Cepheids. It's spatial distribution follows the one shown in region G. It has however a lower density and a significantly more irregular structure.}
	
	\item{\textbf{Region I} represents a population of yellow supergiants with a median age of $\sim512$ Myr ($\pm$ 886 Myr). The overall spatial distribution of this young population shows the asymmetry of the galaxy. The central region is comprised of a single overdensity that however does not correspond to the $\sim45^\circ$ inclination of the main body of the galaxy but appears instead rather vertical. The South-West extension and the Wing are inconspicuous. The inner population is more extended to the south and there appears to be a sharp edge north east of the highest density area, but these features are less pronounced than in the maps of other young stellar populations.}
		
	\item{\textbf{Region C} represents a population of main sequence stars with a median age of $\sim707$ Myr ($\pm$ 668 Myr). The body of the galaxy is more extended than in the previous main sequence stars maps. The outer contours outline a number of protuberances, in particular towards the south east, the Wing, the North- and South-extensions. The principal overdensity is located at the same position as in Region B, but the fuzzy distribution of stars has become denser and outlines a somewhat different structure.}

\end{itemize}

\subsubsection{Intermediate-age and old stellar populations}
\begin{itemize}	
	
	\item{\textbf{Region D} represents a population of main sequence and subgiant stars with a median age of $\sim2.57$ Gyr ($\pm$ 1.65 Gyr). The density distribution is still not completely smooth. Several clumps are visible, but the overall structure appears more regular and symmetric than in young stellar populations. The faint outer contours show protuberances which might be due to tidal effects. The main overdensities are located north, east and south of the highest density areas in the maps of regions B and C. The top overdensity corresponds to the break point of the bar as traced by stars in region A.}
	
	\item{\textbf{Region M} represents a population of thermally pulsing AGB stars with a median age of $\sim2.45$ Gyr ($\pm$ 1.86 Gyr). The AGB phase in our theoretical evolutionary models is not reliable enough to draw solid conclusions about the properties of the population, therefore the age of the population should be considered carefully. The overall spatial distribution shows a rather regular structure, but it is not as spatially extended as that shown in regions J and K. There is only one major overdensity region coincident with the southern component in region K. The shape of this high-density area appears tilted with respect to that at the same location in region K. There are several protuberances overall around the central area.}
	
	\item{\textbf{Region K} represents a population of upper RGB stars (above the RC) with a median age of $\sim4.26$ Gyr ($\pm$ 2.65 Gyr). The spatial distribution outlined by the outer and intermediate contours is similar to that of other regions, such as regions J and E, dominated by old stars. The features in the central area are similar to those shown in region E, except that here they are more regular and only the two densest regions are evident. Small localised overdensities are not obvious. The southernmost density is denser than its northern counterpart.}
	
	\item{\textbf{Region J} is dominated by the intermediate-age RC population, but also includes old horizontal and red giant branch stars resulting in a median age of $\sim4.07$ Gyr ($\pm$ 3.28 Gyr), see Fig.~\ref{fig:SIMUCMD}. The overall morphology of the galaxy is regular and lacks a significant elongation. Unlike red giant and subgiant stars, the central region of the RC population is not comprised of two overdensities but rather of an arc-like overdensity feature. The inner parts of this feature suggest a mild density gradient from east to west. The region enclosed by this arc corresponds to the high density region in the distribution of red giant branch stars.}
	
	\item{\textbf{Region E} represents a population of lower RGB stars (below the RC) with a median age of $\sim4.46$ Gyr ($\pm$ 2.57 Gyr). The outer structure of the galaxy is rounder than that shown above. The central regions are characterised by two main overdensities along the NE to SW axis that complement the overdensities in the distribution of stars from region D. A stripy density pattern, that cannot be explained by technical and/or observational effects, appears to the south of the galaxy.}

\end{itemize}

In summary, the bar traced by the youngest main sequence stars is denser than the main overdensities in the rest of the young populations including older main sequence and supergiant stars. The separation between the wing and the bar becomes less evident with age. The contour lines in region C are set in the opposite direction of the proper motion movement of the galaxy. The inner morphology as traced by red clump stars shows an arc-like structure open towards the south, while the distribution of subgiants and main sequence stars of a similar median age shows several clumps that appear to trace an arc-like structure open towards the west. RGB stars show two central overdensities that seem more populated by lower RGB stars than by the upper RGB stars. Thermally pulsating AGB stars have a central nucleus coincident with the southern density in RGB stars. The external morphology is represented by a circle in intermediate-age and old stars. The oldest stars we detect are over 10 Gyr old, but their number is significantly lower than younger stars populating similar CMD regions.

\section{Discussion}\label{section4}

\subsection{Morphology and interaction history}
Galaxy interactions play an important role in shaping galaxy morphology. Constraining the dynamical history of the Magellanic Clouds is crucial to understand how these interactions have reflected on their morphologies. Our current perception of the orbital history of the Magellanic Clouds indicates that they are either on their first passage of the Milky Way or on an eccentric long period orbit (e.g.~ \citealp{Bekki2005,Bekki2007,Besla2007,Diaz2012}).
Star formation history (SFH) peaks might correspond to interactions between the LMC and SMC and between the Magellanic Clouds and the Milky Way. These peaks occurred at $100-200$ Myr, $1-3$ Gyr, $4-6$ Gyr, and $7-10$ Gyr (e.g.~ \citealp{Harris2003,Noel2007,Harris2009,Indu2011,Rubele2012,Cignoni2013,Rubele2015}). The age uncertainty in CMD regions is quite significant and regions encompass one or more SFH peaks.\\
\cite{Besla2016} examined the impact of tidal interactions on the periphery of the LMC and found that LMC-SMC interactions in isolation are sufficient for the creation of the asymmetric spiral arms and arcs in the outskirts of the LMC disc similar to those observed. These structures are produced independently of the Milky Way tidal field. After $6.3$ Gyr of evolution the SMC has just passed through the disc of the LMC ($b < 10$ kpc).
In the LMC, the northern arm feature appears at $100$ and $160$ Myr in \cite{Harris2009} SFH maps. In our morphology maps, the northern spiral arm is most prominent in main sequence stars at $\sim195$ Myr, becomes clearly present at $\sim891$ Myr, which hints of its vertical part are present at $\sim 2.45$ Gyr. There is no clear presence of the SW arm before $\sim195$ Myr, this feature is connected to the bar at $\sim891$ Myr, while they are completely combined at $\sim2.45$ Gyr. The SE arm is prominent in red clump stars and main sequence/subgiant stars, but it is also present in RGB and thermally pulsing AGB stars. It agrees well with the star formation bursts since one of the strongest SFH signatures in this substructure peaks at $600$ Myr and at $2.5$ Gyr. The ages of these populations range from $\sim1.4$ to $\sim8$~Gyr so it is difficult to attribute them to a single interaction event. Furthermore, fine structures such as spiral arms, if formed as a result of LMC-SMC interactions $1-3$ or $6-8$ Gyr cannot be clearly seen using VMC data as they were probably smoothed by the dynamical relaxation of the discs. Therefore most of these features can be attributed to the recent LMC-SMC interaction $\sim200$~Myr ago.

In the SMC, the young populations is limited to the Wing and bar (e.g.~\citealp{Harris2003,Rubele2015,Rubele2018}), in agreement with the highest concentration of stars in our morphology maps for these populations. Compared to the SFH maps of \cite{Rubele2018} the separation between the Wing and the bar is less evident with age and becomes inconspicuous for the oldest main sequence stars and yellow supergiant distributions. The asymmetric nature of the SMC is apparent until $\sim707$ Myr. It can be attributed to the $1-3$ Gyr interaction. The older populations show signatures of elongations in the south east towards the Magellanic Bridge and the trailing arm, as detected by \cite{Belokurov2017}, and extending to large distances from the centre of the galaxy. These elongations have also been detected in RR Lyrae stars \citep{Muraveva2018} and they can be attributed to the interaction that happened $\sim200$ Myr ago. It is difficult to discern these elongations in SFH studies, but the general circular structure in the distribution of old stars was found.
\subsection{Morphology and variable stars}
Variable stars have been used extensively to study the morphology as well as the three dimensional structure of the Magellanic Clouds.
\cite{Jacyszyn-Dobrzeniecka2016} used the OGLE survey to study Classical Cepheids in the Magellanic System. This population shows a peak at an age of $100$ Myr in the LMC, and can be compared with region A in terms of age and regions A, B, C in terms of spatial distribution. The classical bar of the LMC was redefined to include the western density as both the age and distance tomography show no clear separation between these structures. Our morphological maps shows two distinct overdensities at the western end of the classical bar as well as one in the western density itself in our three main sequence maps. This break is visible at $\Delta$RA=$2^\circ$. It is more enhanced in regions B and C. Region C has another distinct break in the bar at $\Delta$RA=$1^\circ$. \cite{Jacyszyn-Dobrzeniecka2016} also detect the connection of the SW arm to the bar visible in regions B and C, as well as other features visible in our maps. In the SMC, young and intermediate-age Cepheids have a heart-like distribution. Younger Cepheids are concentrated in the north ($10-140$ Myr) while older Cepheids are concentrated in the south west. This is also supported by \cite{Ripepi2017} who found a bimodal age distribution, with two
peaks at $120\pm10$ Myr and $220\pm10$ Myr showing a different spatial distribution, which supports the interaction between the Magellanic Clouds $\sim200$ Myr ago. This agrees well with our maps, since stars in region A are concentrated in the North while for the rest of the young populations the main overdensity is located in the south west.

Furthermore, \cite{Jacyszyn-Dobrzeniecka2017} and \cite{Muraveva2018}, the latter using VMC data, studied the structure of the Magellanic Clouds with RR Lyrae stars. RR Lyrae stars generally present a regular smooth structure similar to that of the old and intermediate-age stars in the LMC, although contrary to those populations, no hints of spiral arms are present in RR Lyrae stars. In the SMC, the spatial distribution shows no irregularities or substructures. The central region is composed of a unique nucleus, contrary to other old populations such as RGB and RC stars. These differences might be due to the fact that there is no pure old population among our investigated regions because even though RR Lyrae stars occupy predominantly region C \citep{Cioni2014}, their number is significantly lower than that of the main sequence stars in the same region. In the models (e.g.~Fig.~\ref{fig:SIMUCMD})), horizontal branch stars are present in both regions C and J. The eastern part of the SMC is closer than the western part, and a large number of RR Lyrae stars are found in the direction of the SMC's trailing arm \citep{Belokurov2017}.  

\begin{landscape}
	\begin{figure}
		\centering
		%		\includegraphics[scale=0.045]{Maps-LMC7/contour_a}
		%		\includegraphics[scale=0.045]{Maps-LMC7/contour_b}
		%		\includegraphics[scale=0.045]{Maps-LMC7/contour_c}
		%		\includegraphics[scale=0.045]{Maps-LMC7/contour_d}
		%		\includegraphics[scale=0.045]{Maps-LMC7/contour_e}
		%		%	\includegraphics[scale=0.33]{Maps-LMC41/contour_f}
		%		\includegraphics[scale=0.045]{Maps-LMC7/contour_g}
		%		\includegraphics[scale=0.045]{Maps-LMC7/contour_h}
		%		\includegraphics[scale=0.045]{Maps-LMC7/contour_i}
		%		\includegraphics[scale=0.045]{Maps-LMC7/contour_j}
		%		\includegraphics[scale=0.045]{Maps-LMC7/contour_k}
		%		\includegraphics[scale=0.045]{Maps-LMC7/contour_m}
		%		%	\includegraphics[scale=0.33]{Maps-LMC5/contour_n}
		\includegraphics[scale=0.48]{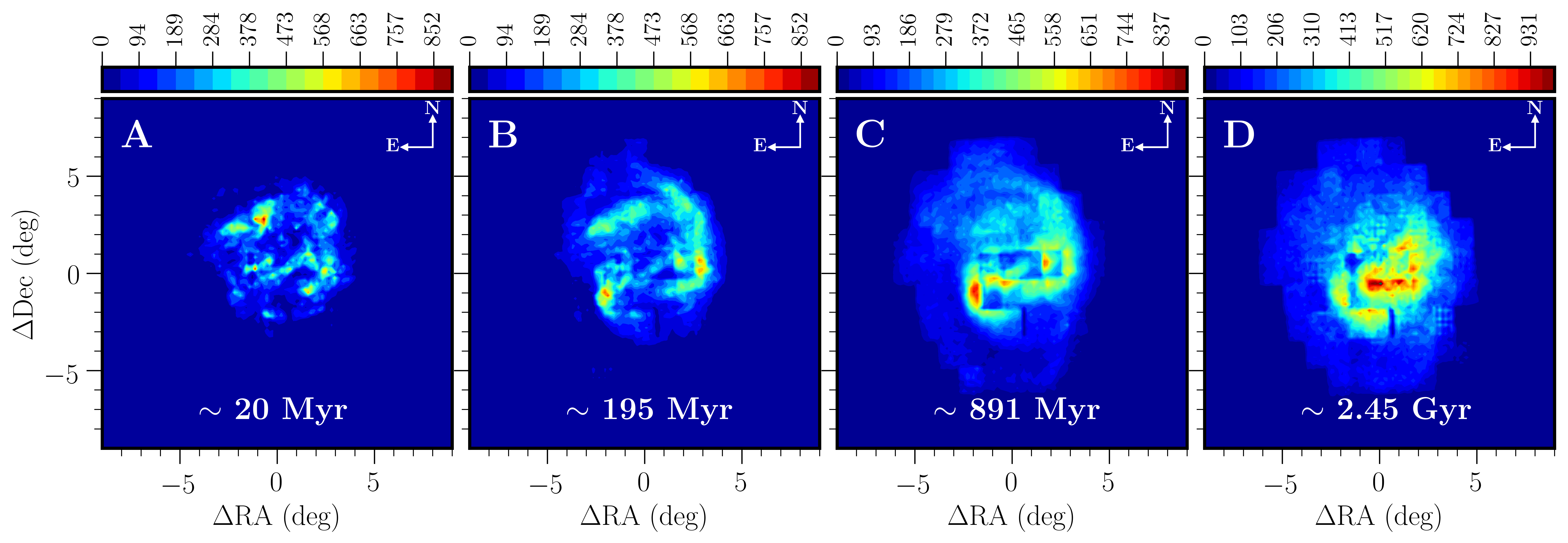}\\
		\includegraphics[scale=0.48]{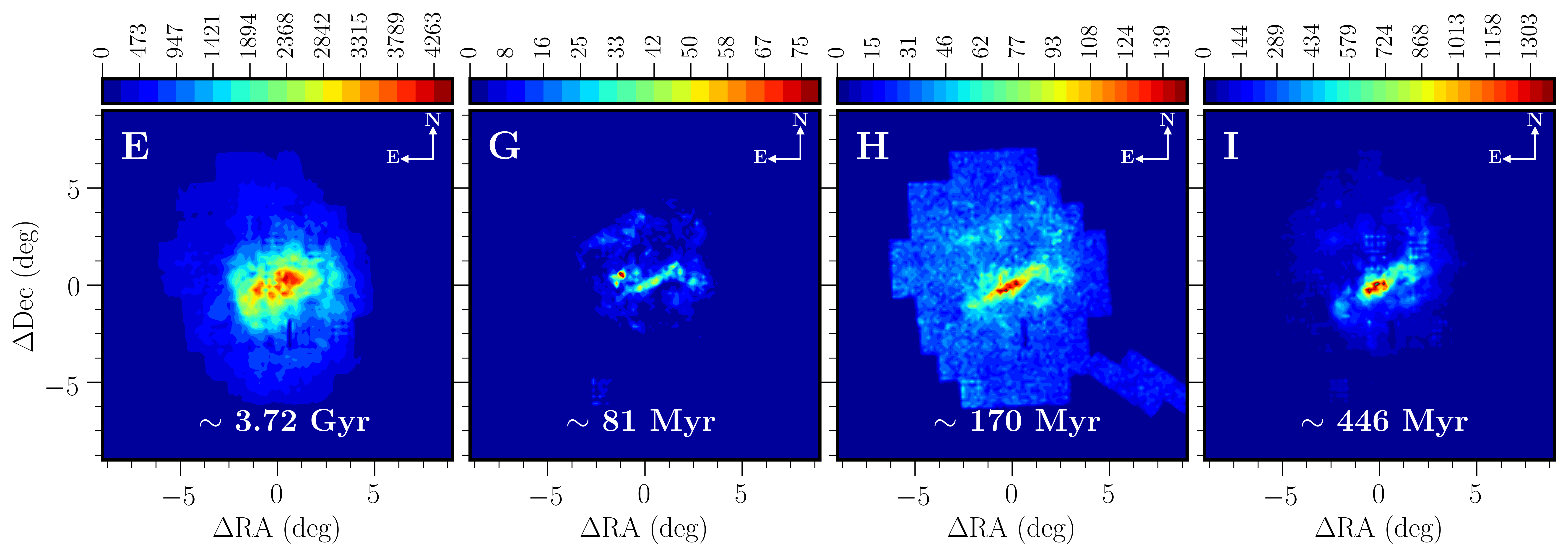}\\
		\hspace{0.2cm}\includegraphics[scale=0.48]{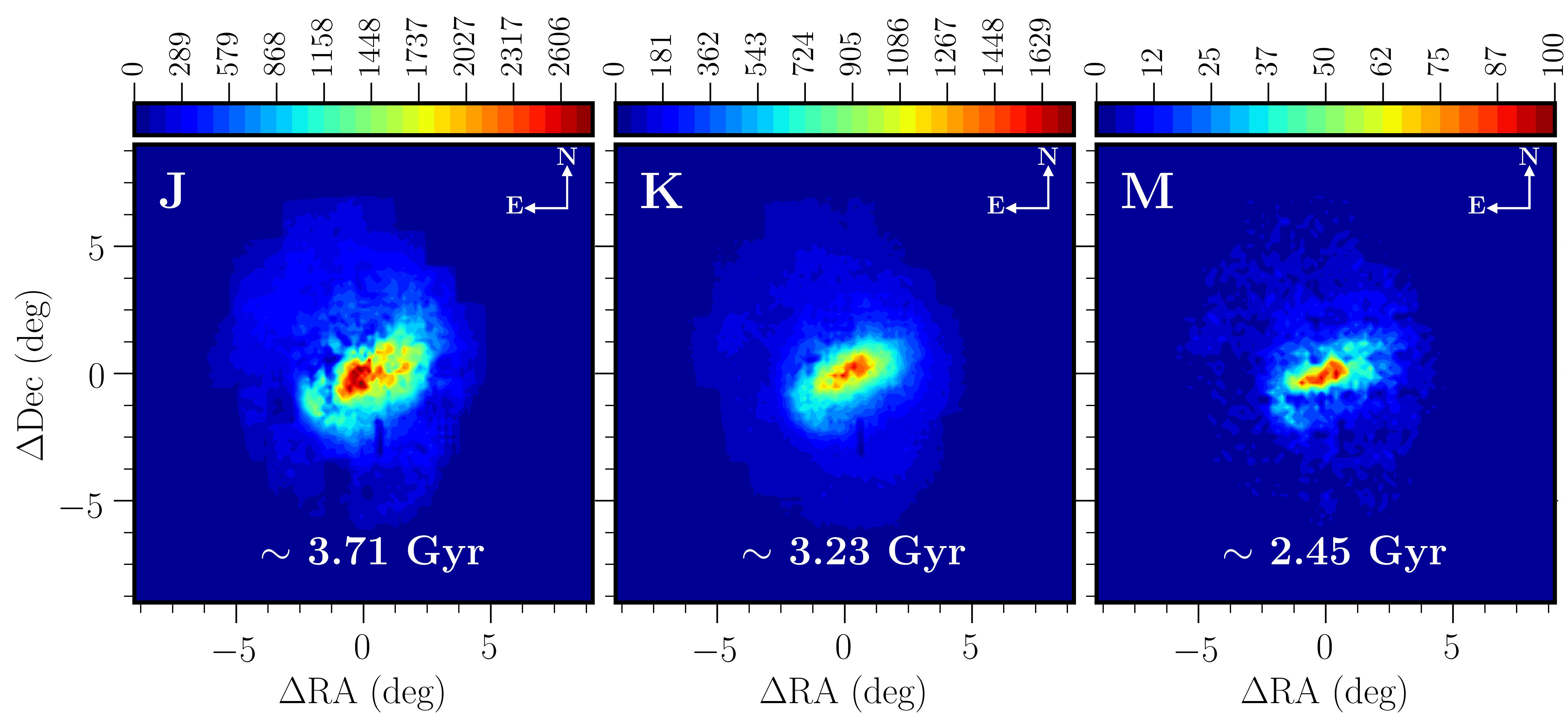}
		\caption{Stellar density/contour maps of the LMC's stellar populations extracted from the VMC survey. The bin size is $0.03$ deg$^2$ and the colour bar represents the number of stars per bin. Regions A, B and C refer to main sequence stars, D to main sequence/subgiant stars, G, H and I to supergiants and giant stars, J to RC stars, K to upper RGB stars, and M to thermally pulsing AGB stars, More details about the stellar populations can be found in Table \ref{table:lmc}.}
		\label{fig:morphlmc}
	\end{figure}
\end{landscape}
%\newpage
\begin{landscape}
	\begin{figure}
		\centering
		\includegraphics[scale=0.48]{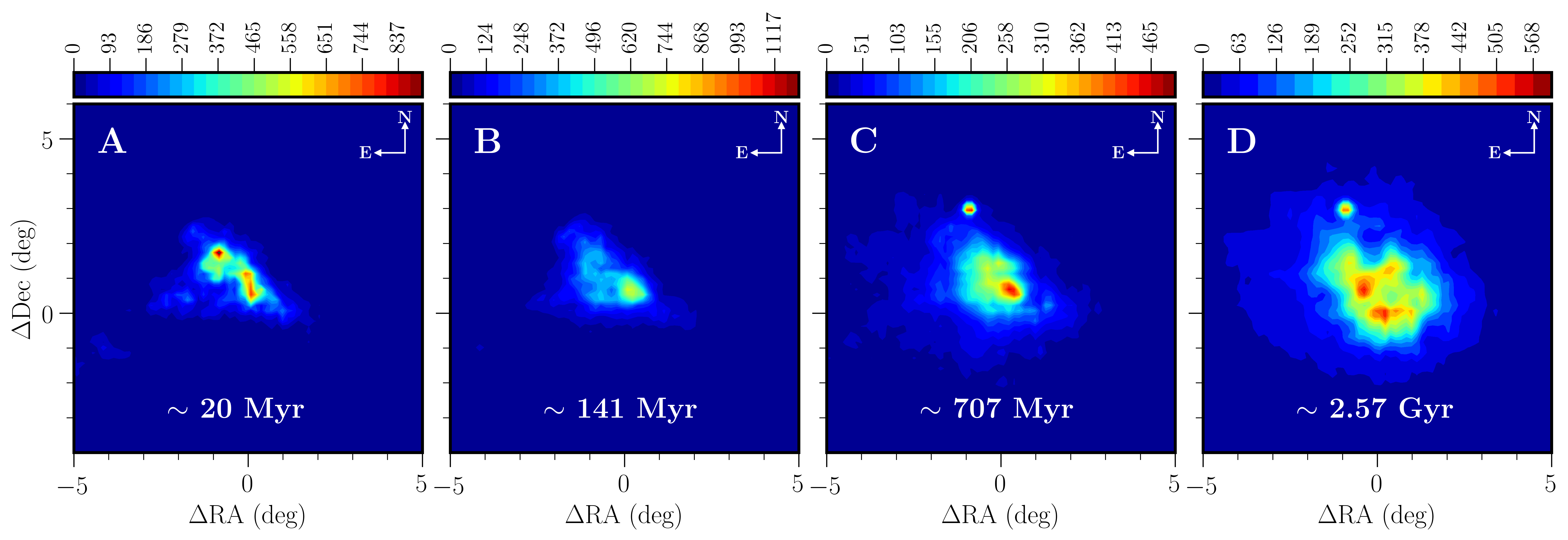}\\
		\includegraphics[scale=0.48]{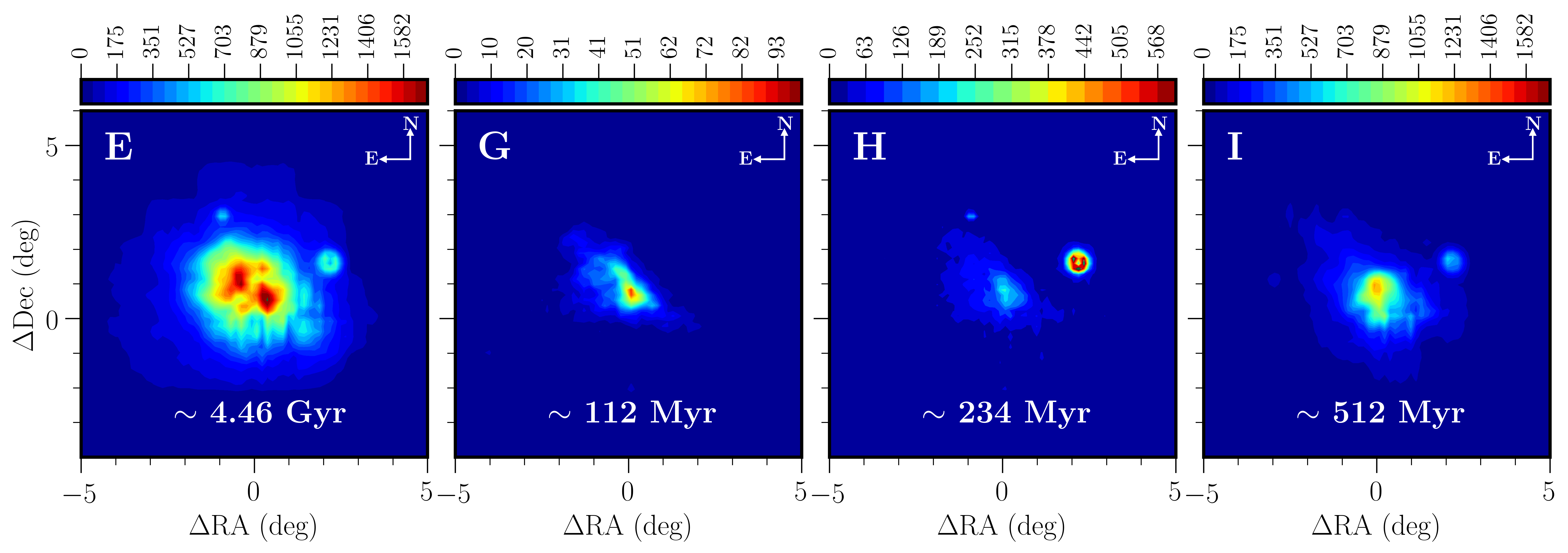}\\
		\hspace{0.2cm}\includegraphics[scale=0.48]{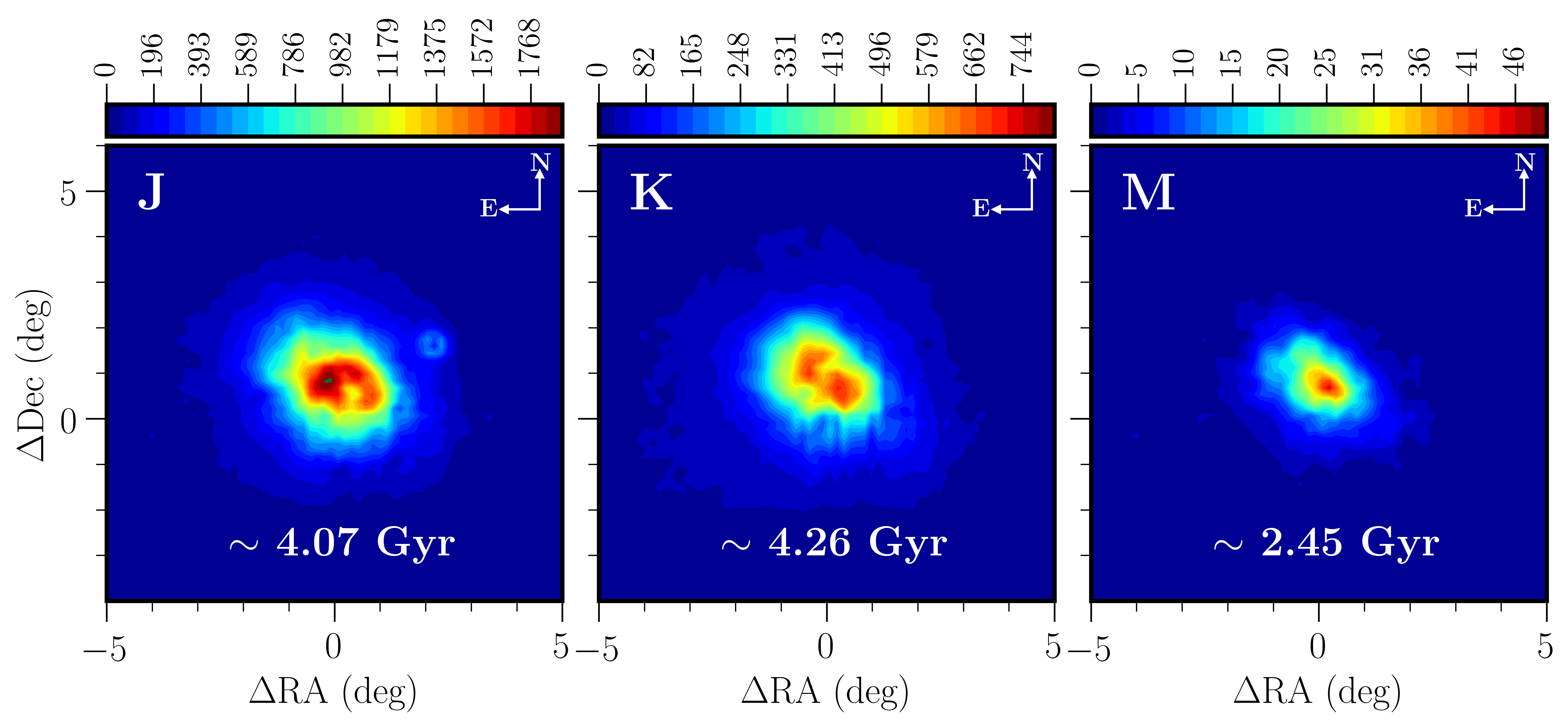}

		\caption{As Fig. \ref{fig:morphlmc} but for the SMC.}
		\label{fig:morphsmc}
	\end{figure}
\end{landscape}

%\\
\subsection{Multi-wavelength morphology}\label{sect44}

%The morphology of the Magellanic Clouds has been studied extensively using various wide field surveys across different wavelength bands. Being less sensitive to dust absorption than optical bands and offering the advantage of low Galactic extinction, both galaxies have been previously observed in the near-infrared by major surveys such as the DEep Near Infrared Survey of the Southern Sky (DENIS; \citealp{Cioni2000}), the Two Micron All Sky Survey (2MASS; \citealp{Skrutskie2006}), and the InfraRed Survey Facility (IRSF; \citealp{Kato2007}).

Galaxy morphology generally appears smoother in the near-infrared because the dominant stellar population traces the stellar mass distribution and gravitational potential of the galaxy. 2MASS was used by \cite{Nikolaev2000} to provide the most comprehensive morphological maps of the LMC to date. The sensitivity of the survey allowed them to only study stellar populations brighter than the RC. \cite{Cioni2000a} also studied only AGB stars, upper RGB stars and bright young stars, the latter including supergiants and blue loop stars. In the LMC, the young stars from 2MASS and DENIS have a mean age of $\sim 0.5$ Gyr, which can be compared with the young main sequence and supergiant populations (regions A, G, H and I) in our VMC maps. Similarly, this population shows a clumpy and irregular structure with elongations towards the extremities of the bar indicating the presence of spiral arms. The distribution of AGB stars shows a highly extended population with a broad and faint outer spiral arm in the north-west which in our study is better traced by RC stars. 
The upper RGB population is comparable to our map from region K, even though we sample a much larger portion of the branch itself. Their SE arm is undetected while it is clearly outlined by our map. The NW arm is very faint but it is emphasised by the dimming of the density in our map.\\
\cite{Gonidakis2008} investigated the spatial distributions of the SMC stellar component using also 2MASS data, providing isopleth contour maps of four age groups. Their K- and M-type giants show a central concentration, comparable to the morphology of RR Lyrae stars, according to their selection criteria, these stars should be regions H, I and K. Combining these stellar populations a double nucleus is present in our maps. \cite{Cioni2000a} found that the younger population is asymmetric with protuberances that can be due to tidal effects. These proturberances are visible in young main sequence and supergiant stars distribution, while the older populations show two main concentrations. The western concentration is dominated by RGB stars and the eastern concentration is populated by a significantly younger population. Overall, our maps agree well with theirs but provide a significant improvement in sensitivity and spatial resolution.

\ion{H}{i} imaging accounts for spiral structures in galaxies. It traces spiral arms and ring-like features. \ion{H}{i} maps can also trace enhanced surface brightness in star forming regions as well as extended gaseous discs compared to their optical and near-infrared counterparts.
\cite{Staveley-Smith2003} provided a new look at the large scale structure of the LMC. The \ion{H}{i} morphology of the LMC traces four main spiral arms. Arm `B' seems to be a tidal arm connecting the LMC to the Magellanic Bridge. The presence of this arm demonstrates the existence of LMC gas in the Bridge and it is composed of two filaments separated by $0.5$ deg. Arm `E' points towards the leading arm while arm 'W' leads to the north. Moreover, the main body of the LMC is bound by arm `S'. Based on our morphology maps, the spiral structure in the LMC is best traced by main sequence stars. The gaseous arms `E' and `W' have near-infrared stellar counterparts i.e. the horizontal part of the North-West arm and the South-East arm, the horizontal part of the South-East arm traces arm `S' although it is not as extended as in the \ion{H}{i} map. Furthermore, the main LMC body in \ion{H}{i} is also punctuated by holes. These holes show strong correlations with stellar associations, \ion{H}{ii} regions and supernova remnants (SNRs). \cite{Bozzetto2017} suggested an association between the spatial distribution of SNRs and the environmental density of the LMC, as well as their tendency to be located around supergiant shells. They also found a connection between the highest \ion{H}{i} density areas, tracing the spiral structure of \ion{H}{i}, and the location of SNRs. Comparing these objects with our main sequence stars, we find that they are also consistent with their near-infrared morphology tracing spiral arms not traced by \ion{H}{i} such as the South-West arm. Overdensities within the spiral arms shown in the map from the youngest main sequence stars correspond to the location of supershells described by \cite{Dawson2013}. \\
The morphology of the SMC in \ion{H}{i} is characterised by the presence of filaments, arcs and shells \citep{Staveley-Smith1995,Stanimirovic1999}. The \ion{H}{i} bar is more extended than its stellar counter part, and a bridge seems to connect the bar to the Wing. The smallest overdensity in the Wing, from stars in region A, corresponds to the supershell SMC1 while the larger one corresponds to the \ion{H}{ii} region N84A. The morphology of the youngest main sequence stars is closest to that of the \ion{H}{i}. \cite{McClure-Griffiths2018} showed that the SMC has an atomic outflow extending to $2$ kpc and that these cold outflows may have formed $25-60$ Myr ago, which corresponds to the age range of stars in our region A. The spatial distribution of SNRs also agrees well with its morphology. Most of the SNRs in the SMC are located in the bar \citep{Williams2009}, precisely in the northern and southern regions along the two main overdensities.

The distribution of ionised gas across the Magellanic Clouds, as traced by H$\alpha$ imaging \citep{Gaustard2001,Reid2012}, allows us to isolate star forming regions. In the LMC, H$\alpha$ evidently traces the spiral structure of the galaxy. Hence a correlation between H$\alpha$ and the distribution of young stars is expected, although the corresponding overdensities appear much larger and extended in H$\alpha$ than in our maps since a stellar association can lead to a bright \ion{H}{ii} region with only a few photoionising stars \citep{Harris2009}. The same is valid for the SMC where most of the activity is spread over the bar of the galaxy as in the youngest main sequence stars. UV imaging can also be used to estimate star formation rates and trace the distribution of young stars once dust extinction is estimated. Contrary to the near-infrared where the correction towards the Magellanic Clouds is small and where different methods yield results that agree within uncertainties, this is not the case for the UV. There is a strong correlation between UV and H$\alpha$ imaging as the UV is particularly used to trace recent star formation and decouple it from the overall SFH of the galaxy.

In the optical, \cite{Choi2018b} used the Survey of the MAgellanic Stellar History (SMASH; \citealp{Nidever2017}) and detected a ring like structure in the outskirts of the LMC's disc. They looked at the spatial distribution of main sequence stars in a range of magnitude bins where the age ranges from $100$ Myr to $1.8$ Gyr and found that stars <300 Myr old form clumpy structures while main sequence stars with ages between $300$~Myr and $1$~Gyr old show conspicuous structures such as the central bar and the NW arm. \cite{Zaritsky2000} compared the distribution of upper main sequence stars vs. red giant and RC stars using the Magellanic Clouds Photometric Survey (MCPS) and found that the asymmetric nature of the SMC is mostly made up of young main sequence stars while the older populations trace an extremely regular distribution. Their study was limited to two morphological maps and their spatial resolution had rendered it hard to characterise central features of the SMC. Furthermore \cite{Maragoudaki2001} used photometric plates in the optical domain to investigate the recent structural evolution of the SMC. Using main sequence stars, they found that the irregularity of the younger component offers evidence of the encounter with the LMC $0.2$ to $0.4$ Gyr ago. The recedence of the Wing as the population grows older is noted in their maps as well. They also detected the overdensities present in this feature as in our map of region A. 
Our data covers the trailing arm discovered by \cite{Belokurov2017} using \textit{Gaia} DR1 data \citep{Brown2016,Prusti2016}. However, we notice an elongation or extension of the contours in several populations. In the upper and lower RGB, it is directed towards the south-east, while in subgiant/main sequence stars (region D) it is directed towards the north-east. The distribution of RC stars does not show any elongations towards the trailing arm.
%The inner twist in the contours in region C can be related to the S shape referred to in \citealp{Belokurov2017} and \citealp{Mackey2018}, this phenomenon can be traced to the central 1.5$^\circ$ of the galaxy, while in previous studies was only traced to $\approx$ 4$^\circ$ - 5$^\circ$.\\

\begin{table*}
	\caption{Stellar populations towards the Large Magellanic Cloud using VMC data.}             
	\label{table:lmc}      
	\centering 
	\begin{tabular}{cccrccc}
		
		\hline
		Region & log(age) & [M/H] & \textit{N} & Milky Way & Milky Way & Dominant Stellar Population\\
		&  (yr) & (dex)& &Model (\%) & \textit{Gaia} (\%)$^{a}$\\
		\hline
		& \\
		
		A& 7.31 $\pm$ 0.32& --0.37 $\pm$ 0.02&\raggedright{328 034}&0&-& Main sequence stars \\
		B& 8.29 $\pm$ 0.46& --0.39 $\pm$ 0.02&547 577&0&-& Main sequence stars\\
		C& 8.95 $\pm$ 0.36& --0.39 $\pm$ 0.05&848 252&0.3&-& Main sequence stars\\
		D& 9.39 $\pm$ 0.27& --0.60 $\pm$ 0.24&1 337 683&1&-& Subgiants \& main sequence stars \\
		E& 9.57 $\pm$ 0.29& --0.96 $\pm$ 0.30& 973 616&1.4&-& Red giant branch stars\\
		F& 9.81 $\pm$ 0.19& --0.25 $\pm$ 0.47&1 053 240&100&50& Milky Way stars\\
		G& 7.91 $\pm$ 0.21& --0.43 $\pm$ 0.02&11 594&9.5&8.4& Supergiants \& giant stars\\
		H& 8.23 $\pm$ 0.15& --0.39 $\pm$ 0.02&145 785&87&78.7& Supergiants \& giant stars\\
		I& 8.65 $\pm$ 0.25& --0.39 $\pm$ 0.11&489 333&31&28.3& Supergiants \& giant stars\\
		J& 9.57 $\pm$ 0.40& --0.96 $\pm$ 0.34&1 901 035&1.6&-& Red clump stars\\
		K& 9.51 $\pm$ 0.29& --0.96 $\pm$ 0.01&918 552&5.6&6.4& Red giant branch stars\\
		M& 9.39 $\pm$ 0.28& --0.58 $\pm$ 0.23&20 451&1.2&4.6& Asymptotic-giant branch stars\\
		&&&\\
		\hline
		& \\
	\end{tabular} 

\vspace{1pt}
\begin{flushleft}
	$^{a}$ We only provide a lower limit of the Milky Way contamination \%  up to $K_\mathrm{s}\sim16$~mag (see Sect.~\ref{sect22} as well as Figs.~\ref{fig:ANNMW1} and \ref{fig:ANNMW2}).
\end{flushleft}
\end{table*}
\begin{table*}
	\caption{Stellar populations towards the Small Magellanic Cloud using VMC data.}                        
	\label{table:smc}      
	\centering  		
	\begin{tabular}{cccrccc}
		\hline
		Region & log(age) & [M/H] & \textit{N} & Milky Way & Milky Way& Dominant Stellar Population\\
		&  (yr) & (dex) & &Model (\%) & \textit{Gaia} (\%)$^{b}$ \\
		\hline
		& \\			
		
		A& 7.31 $\pm$ 0.33& --0.55 $\pm$ 0.01 &69 871&0&-& Main sequence stars\\
		B& 8.15 $\pm$ 0.46& --0.62 $\pm$ 0.05 &81 786&0&-& Main sequence stars\\
		C& 8.85 $\pm$ 0.41& --0.66 $\pm$ 0.05 &72 681&2&-& Main sequence stars\\
		D& 9.41 $\pm$ 0.28& --0.84 $\pm$ 0.19 &187 366&22&-& Subgiants \& main sequence stars \\
		E& 9.65 $\pm$ 0.25& --1.14 $\pm$ 0.22 &565 019&5.2&-& Red giant branch stars\\
		F& 9.81 $\pm$ 0.19& --0.35 $\pm$ 0.54 &354 022&100&90& Milky Way stars\\
		G& 8.05 $\pm$ 0.31& --0.60 $\pm$ 0.03 &5830&7.4&3.7& Supergiants \& giant stars\\
		H& 8.37 $\pm$ 0.19& --0.68 $\pm$ 0.03 &61 529&82&72& Supergiants \& giant stars\\
		I& 8.71 $\pm$ 0.75& --0.66 $\pm$ 0.07 &209 664&47&31.4& Supergiants \& giant stars\\
		J& 9.61 $\pm$ 0.35& --1.12 $\pm$ 0.23 &475 076&4&-& Red clump stars\\
		K& 9.63 $\pm$ 0.27& --1.12 $\pm$ 0.22 &192 047&16&14.8& Red giant branch stars\\
		M& 9.39 $\pm$ 0.33& --0.86 $\pm$ 0.24 &5788&1&3.8& Asymptotic-giant branch stars\\
		&&&\\
		\hline
	\end{tabular} 
\vspace{1pt}
\begin{flushleft}
	$^{b}$ As Table \ref{table:lmc} but for the SMC.
\end{flushleft}
\end{table*}

\section{Summary and Conclusions}\label{section5}
We have presented a morphological study of the Magellanic Clouds using the VMC survey, the most sensitive and higher resolution near-infrared imaging survey of the Magellanic system to date. We used ($J - K_\mathrm{s}$, $K_\mathrm{s}$) CMDs along with stellar population models to select different stellar populations of different median ages. We estimated the influence of Milky Way stars using these models and from the \textit{Gaia} DR2 data. Our main results of the paper are high-resolution maps of the Magellanic Clouds that trace the morphological evolution of the galaxies with age as well as characterisation of the central features for the first time at bin-size resolutions of 0.13 kpc and 0.16 kpc. Below we summarise our main findings:\\

\noindent
In the LMC:
\begin{itemize}
	\item young main sequence stars exhibit coherent arms and trace the multi-distinct spiral structure of the galaxy. Along with the bar these structures are more enhanced at a median age of $\sim$738 Myr;
	\item a clear connection between the SW arm and the bar is detected in main sequence stars, and its density also correlates with age;
	\item supergiant stars also trace the spiral structure of the galaxy but to a lesser extent. Most of the stars in these populations are concentrated in the bar of the galaxy;
	\item using main sequence stars for age tomography, a break in the bar becomes relevant as the population becomes younger;
	\item lower RGB stars are leaning towards a radially symmetric morphology but still show irregular clumps in the centre;
	\item unlike RR Lyrae stars, RGB and RC stars still show signatures of tidal arms despite their regular morphology.
\end{itemize}
In the SMC:
\begin{itemize}
	\item the bar in the youngest main sequence stars is more prominent than the central density in the rest of the young populations and its broken appearence may be the result of tidal interactions;
	\item upper and lower RGB stars show elongations and extensions towards the trailing arm of the SMC;
	\item intermediate-age populations show irregular central features that are characterised for the first time. These features suggest that tidal interactions influenced the inner SMC.
\end{itemize}
The VMC survey shows a great capability to trace the morphology of the Magellanic Clouds due to its sensitivity and high spatial resolution, its large area and its reduced sensitivity to dust. In the future, we will explore the outer morphology of the Magellanic Clouds using the VISTA Hemisphere Survey (VHS; \citealp{McMahon2013}) while upcoming large-scale multi-fibre spectroscopic facilities (e.g. 4MOST and MOONs) will be able to characterise substructures with kinematics and chemistry.

\section*{acknowledgements}
	We thank the Cambridge Astronomy Survey Unit (CASU) and the Wide Field Astronomy Unit (WFAU) in Edinburgh for providing calibrated data products under the support of the Science and Technology Facility Council (STFC). This project has received funding from the European Research Council (ERC) under European Union's Horizon 2020 research and innovation programme (project INTERCLOUDS, grant agreement no. 682115). This study is based on observations obtained with VISTA at the Paranal Observatory under program ID 179.B-2003. SR and LG acknowledge support from the ERC Consolidator Grant funding scheme (project STARKEY, grant agreement no. 615604).
\bibliographystyle{mnras} % style aa.bst
\bibliography{ref}

%-------------------------------------------------------------------

\appendix

	\section{Additional maps}
	\begin{figure}
		\centering
		\includegraphics[scale=0.4]{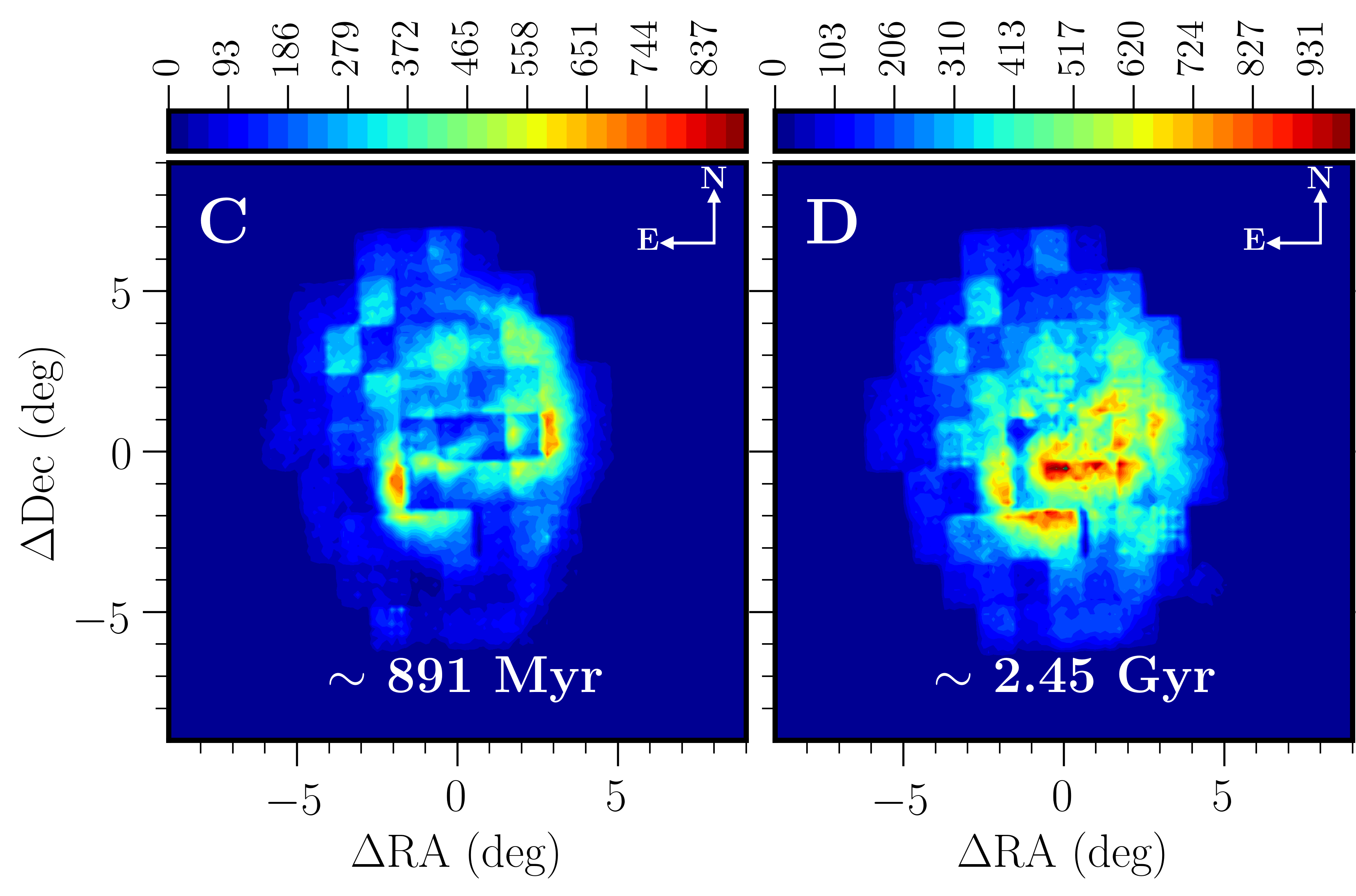}
		
		\caption{{Stellar density/contour maps of the LMC's stellar populations within regions C and D obtained from CMD boxes limited to $K_\mathrm{682115s}<19.8$ mag. The bin size is $0.03$ deg$^2$ and the colour bars represents the number of stars per bin.}}
		\label{fig:mapscd}
	\end{figure}
	
	Maps in Fig.~\ref{fig:mapscd} were made to show the tiling pattern apparent in regions C and D when we probe these populations to $K_\mathrm{s}<19.8$ mag. This pattern might be due to the different completeness levels among tiles. It appears to affect only these two CMD regions. In order to take possible incompleteness effects into account, we limited the morphology of the stellar populations within these two regions to $K_\mathrm{s}\leq19.4$ mag (Fig.~\ref{fig:morphlmc}).
	
	\section{Using Gaia to disentangle the Milky Way population}
\begin{figure*}
		\centering
		\includegraphics[scale=0.065]{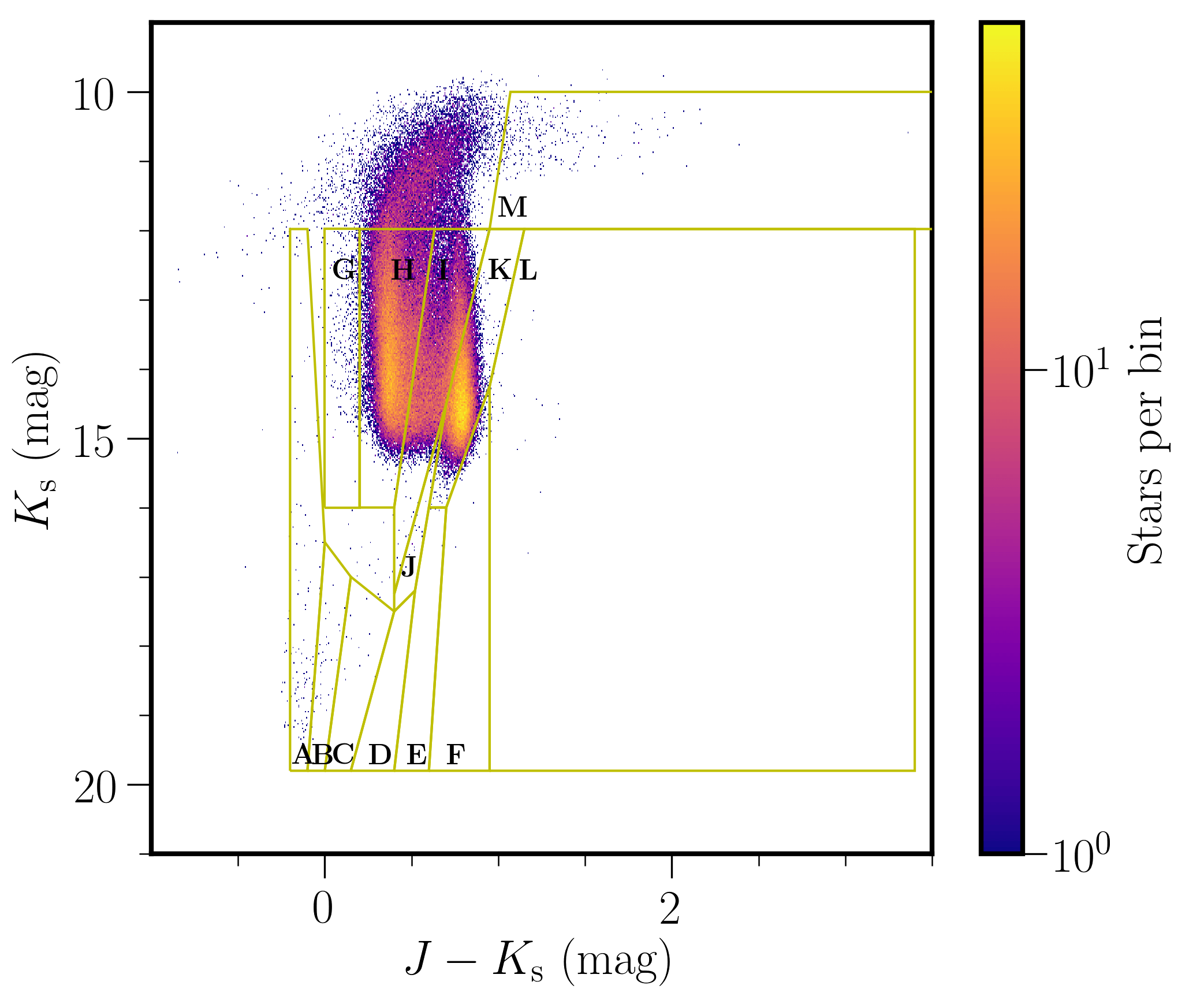}
		\includegraphics[scale=0.065]{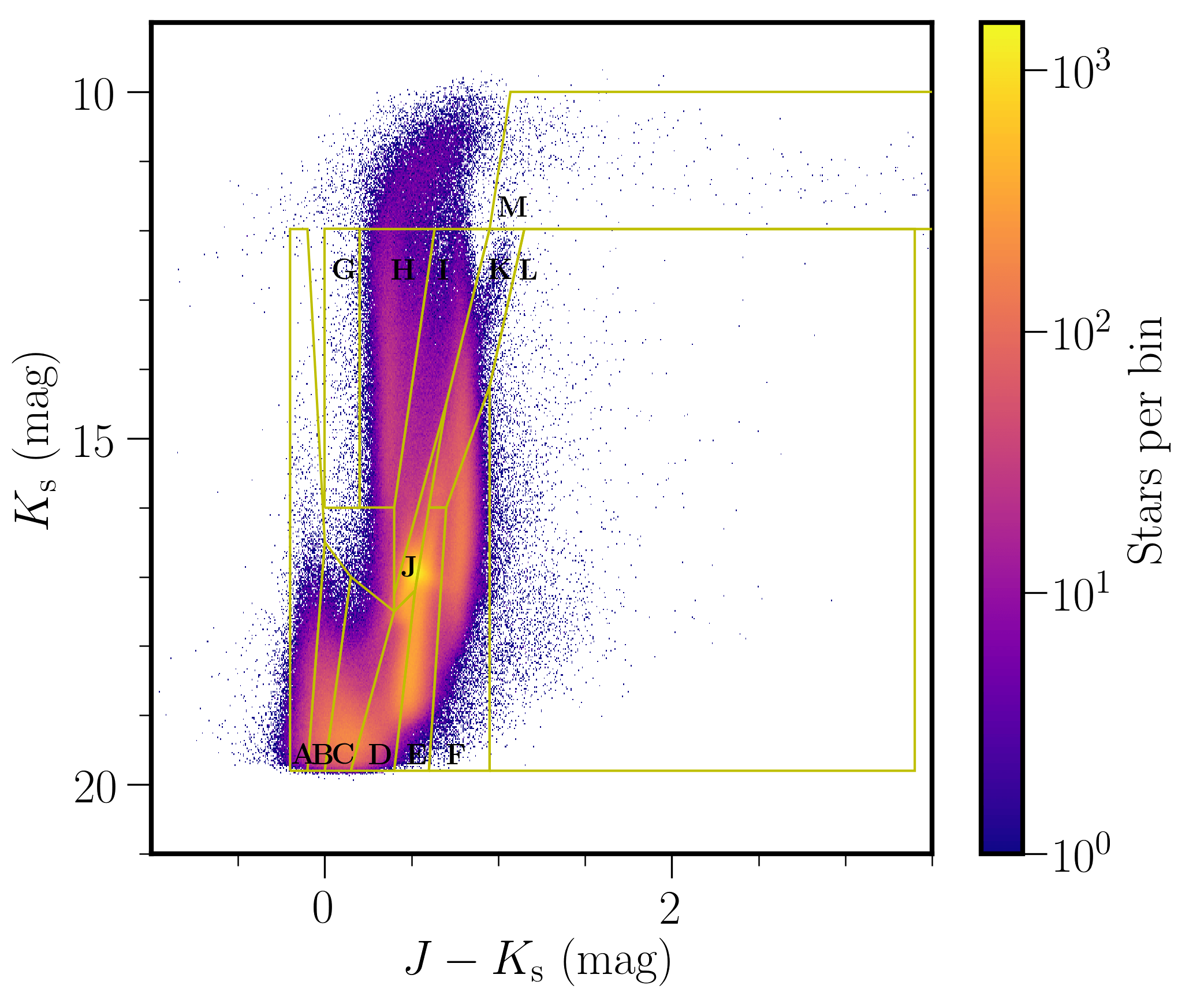}
		\includegraphics[scale=0.065]{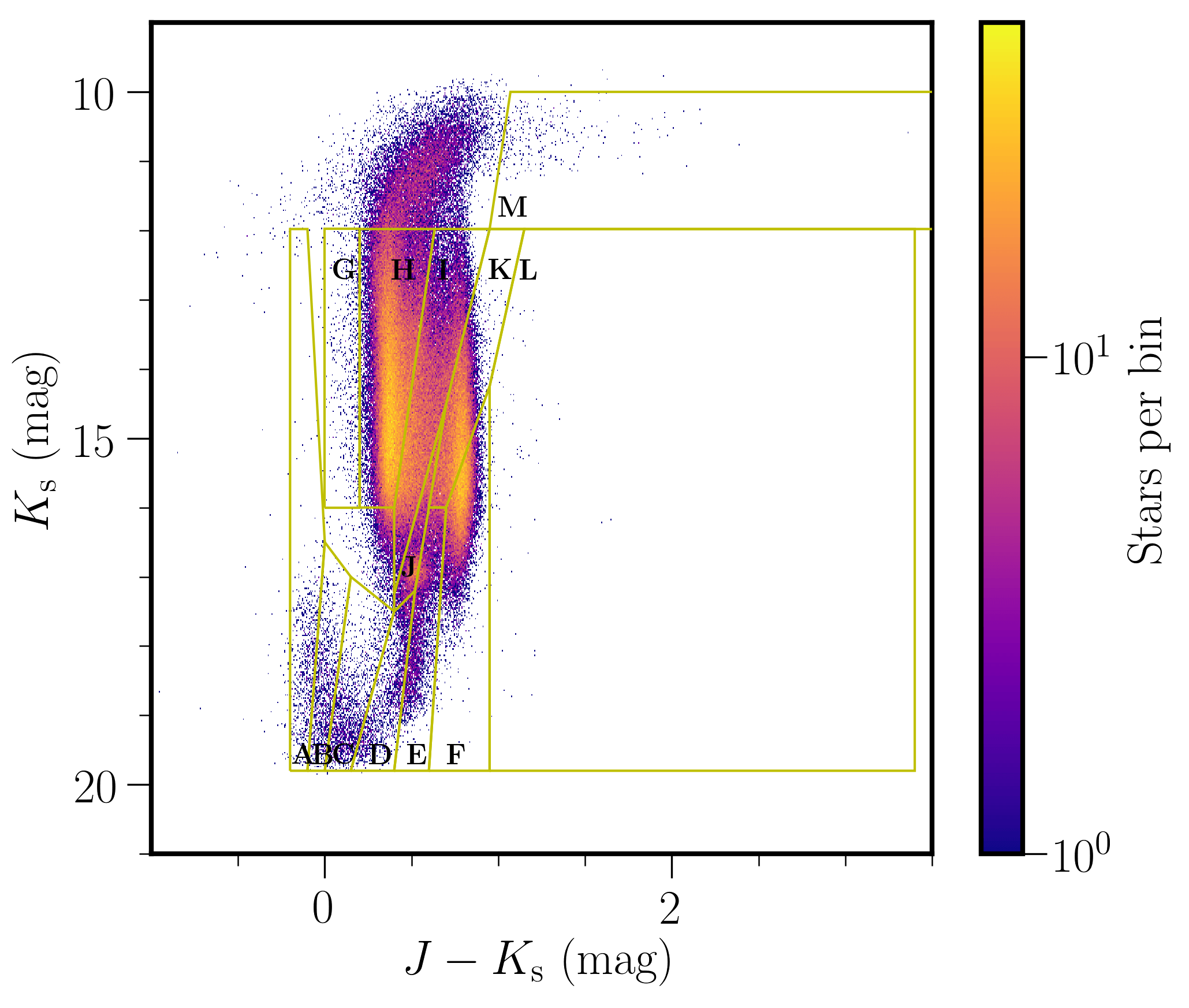}
		\caption{Near-infrared ($J - K_\mathrm{s}$,$K_\mathrm{s}$) CMDs illustrating the distribution of Milky Way stars across the CMD regions in the LMC using different selection criteria: \protect\cite{Helmi2018A} (left), \protect\cite{Vasiliev2018} (centre), $\omega$>0.2 mas (right).}
		\label{fig:ANNMW1}
	\end{figure*}

	The CMDs shown in Fig.~\ref{fig:ANNMW1} refer to the Milky Way stars in the direction of the LMC, distinguished using different selection criteria. These criteria are discussed in Sect. \ref{sect22} and the set adopted here was proposed by \cite{Vasiliev2018}.

	\begin{figure*}
		\centering
		\includegraphics[scale=0.065]{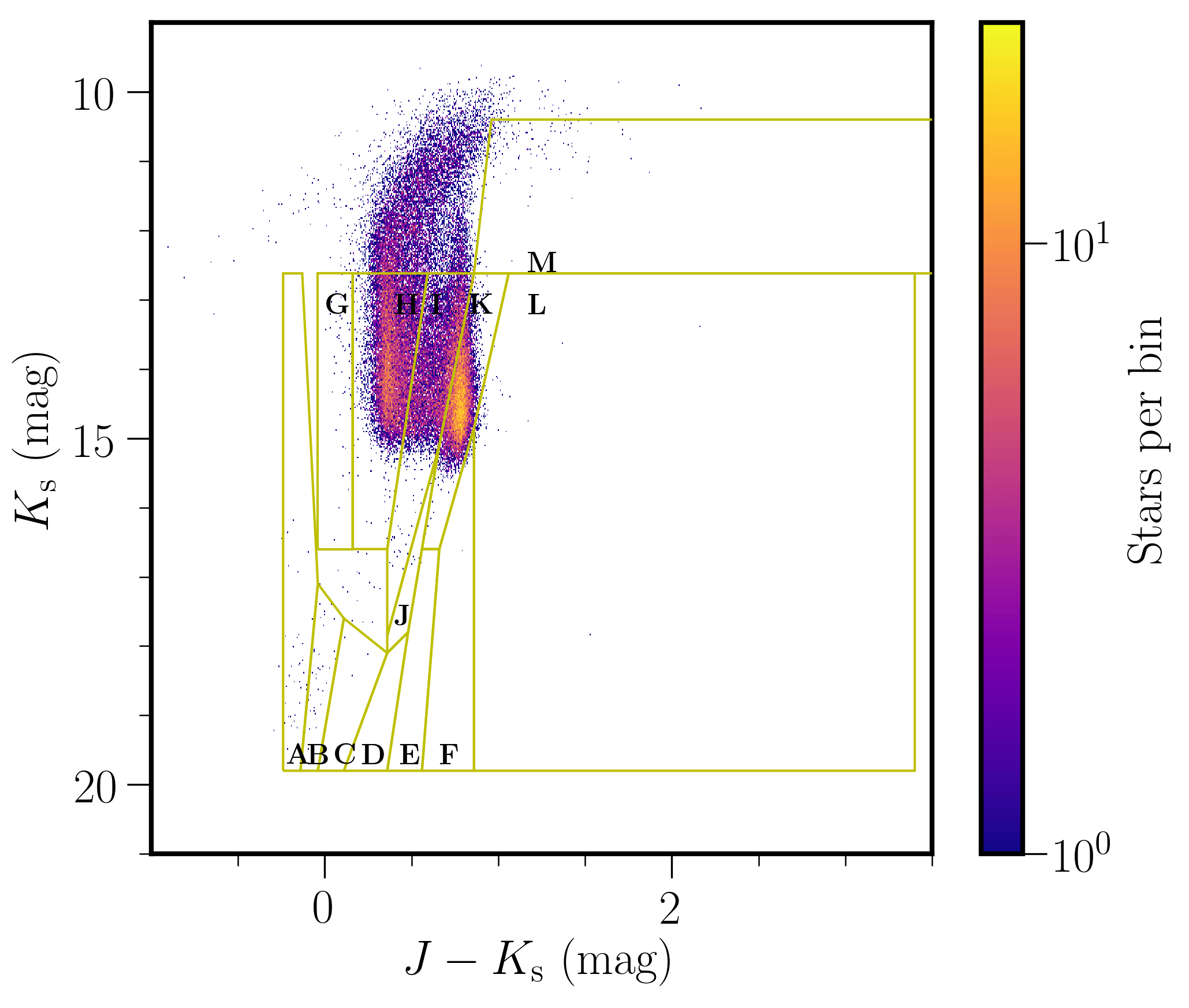}
		\includegraphics[scale=0.065]{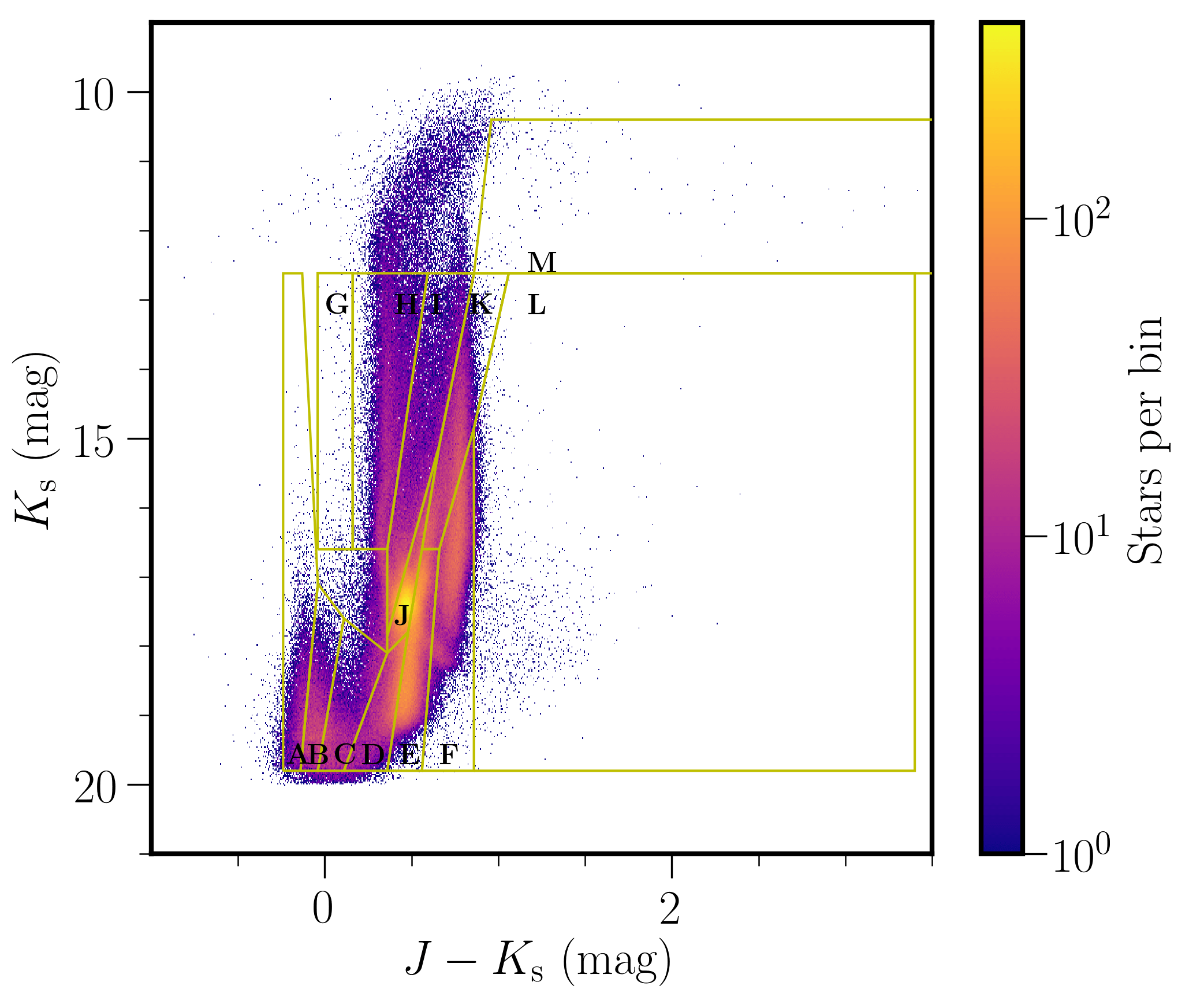}
		\includegraphics[scale=0.065]{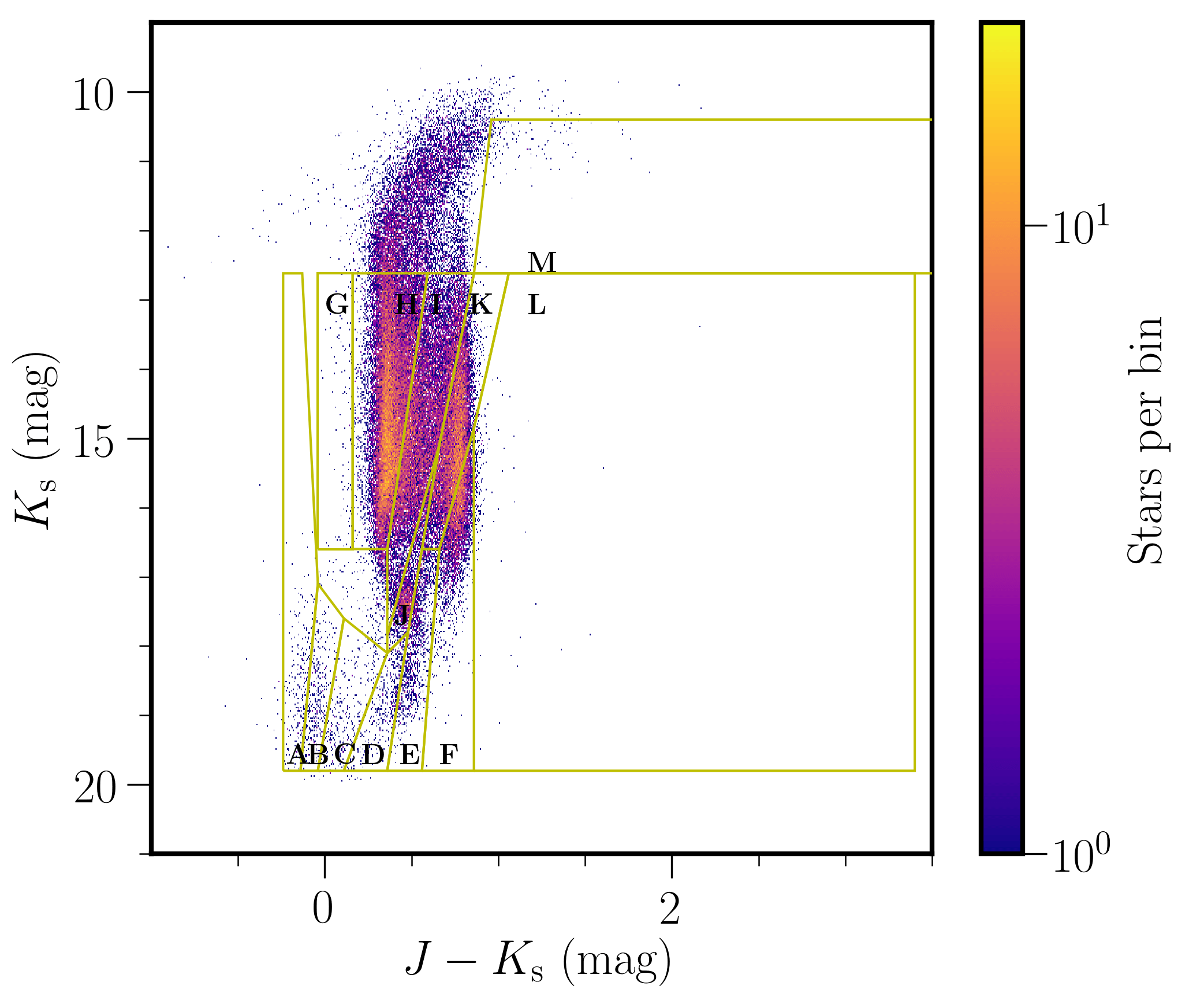}
		
		\caption{As Fig. \ref{fig:ANNMW2} but for the SMC.}

		\label{fig:ANNMW2}
	\end{figure*}
	
	\section{Stellar population models of the LMC}
\begin{figure*}
		\centering
		\includegraphics[scale=0.1]{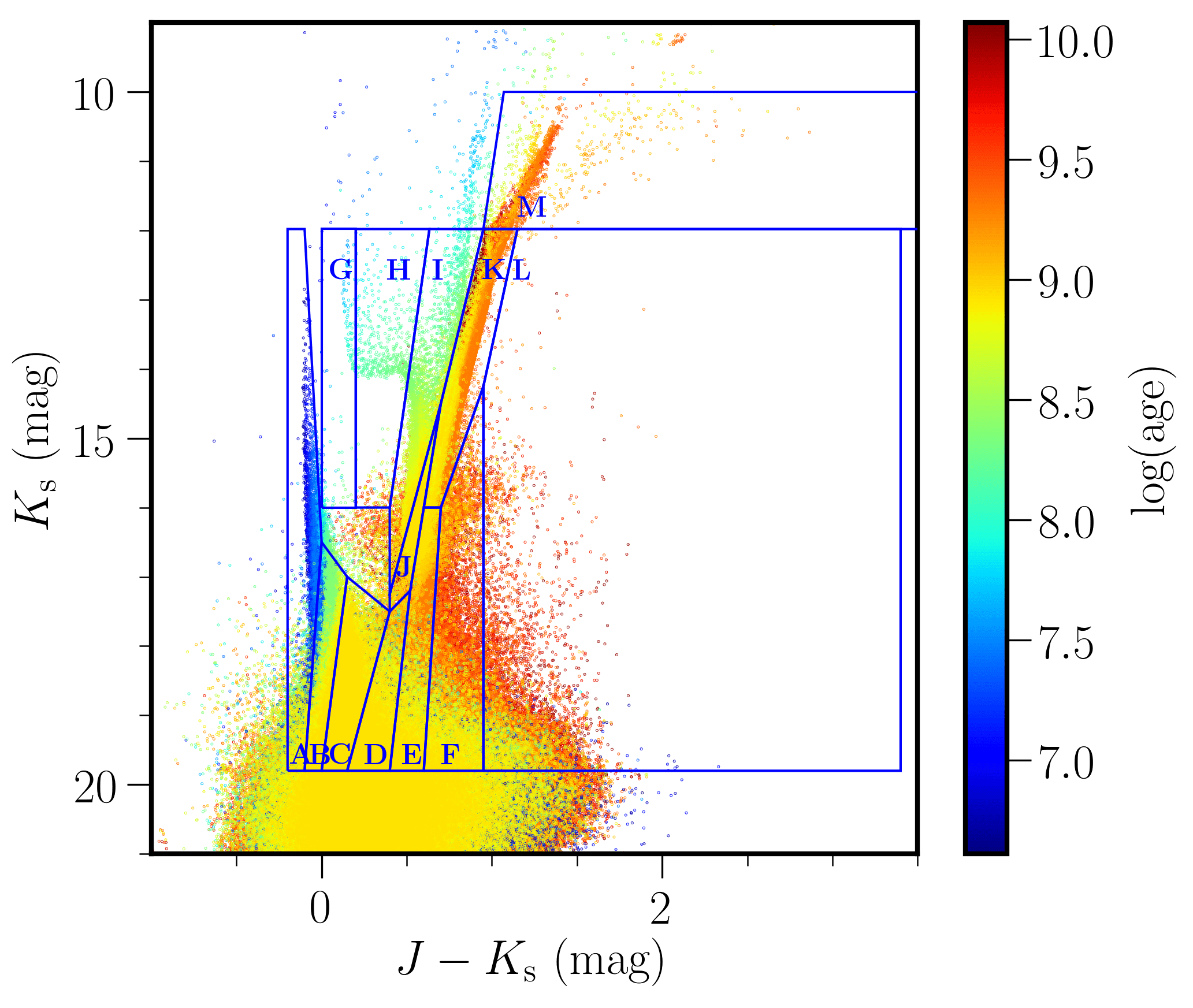}
		\includegraphics[scale=0.1]{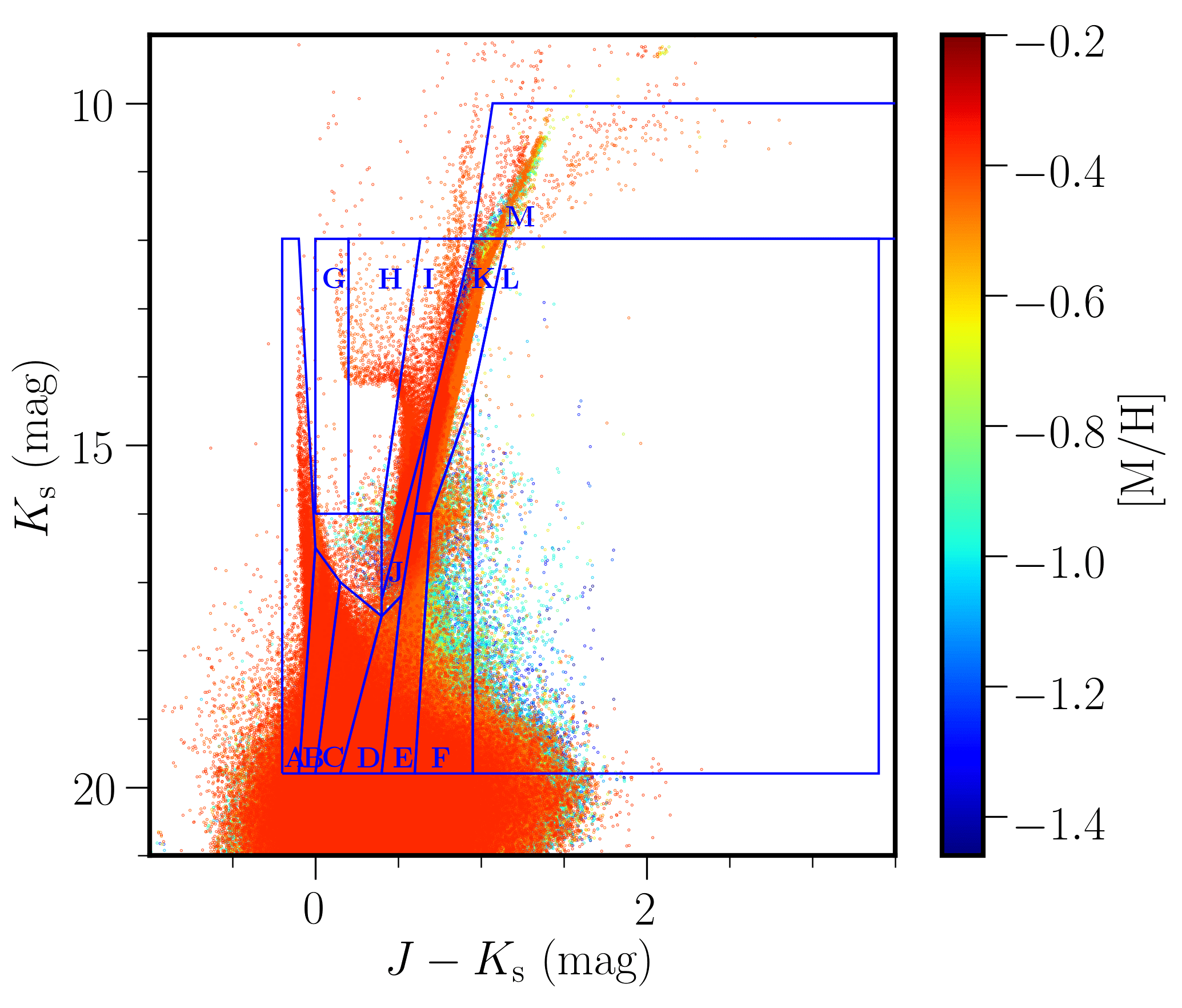}
		\includegraphics[scale=0.065]{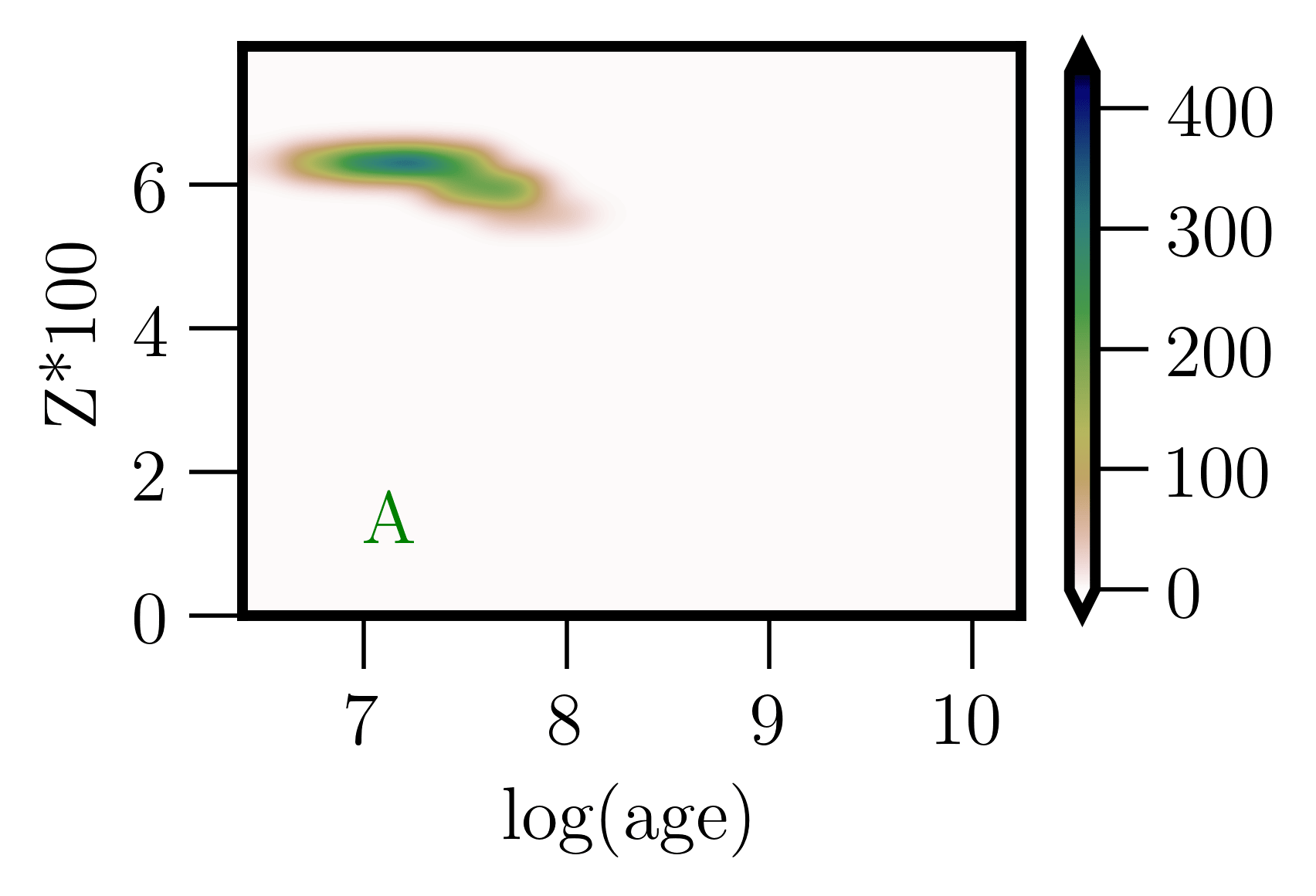}
		\includegraphics[scale=0.065]{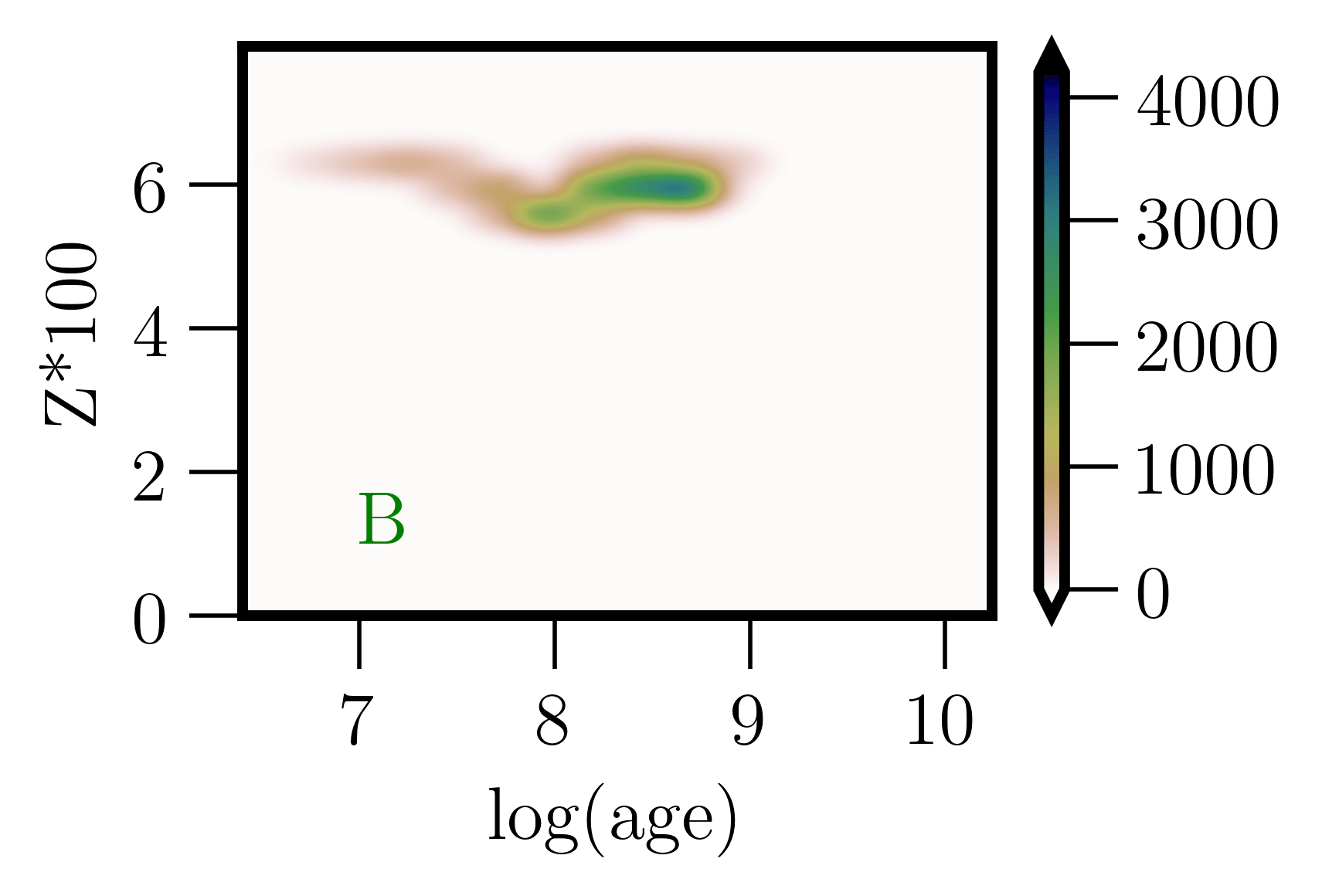}
		\includegraphics[scale=0.065]{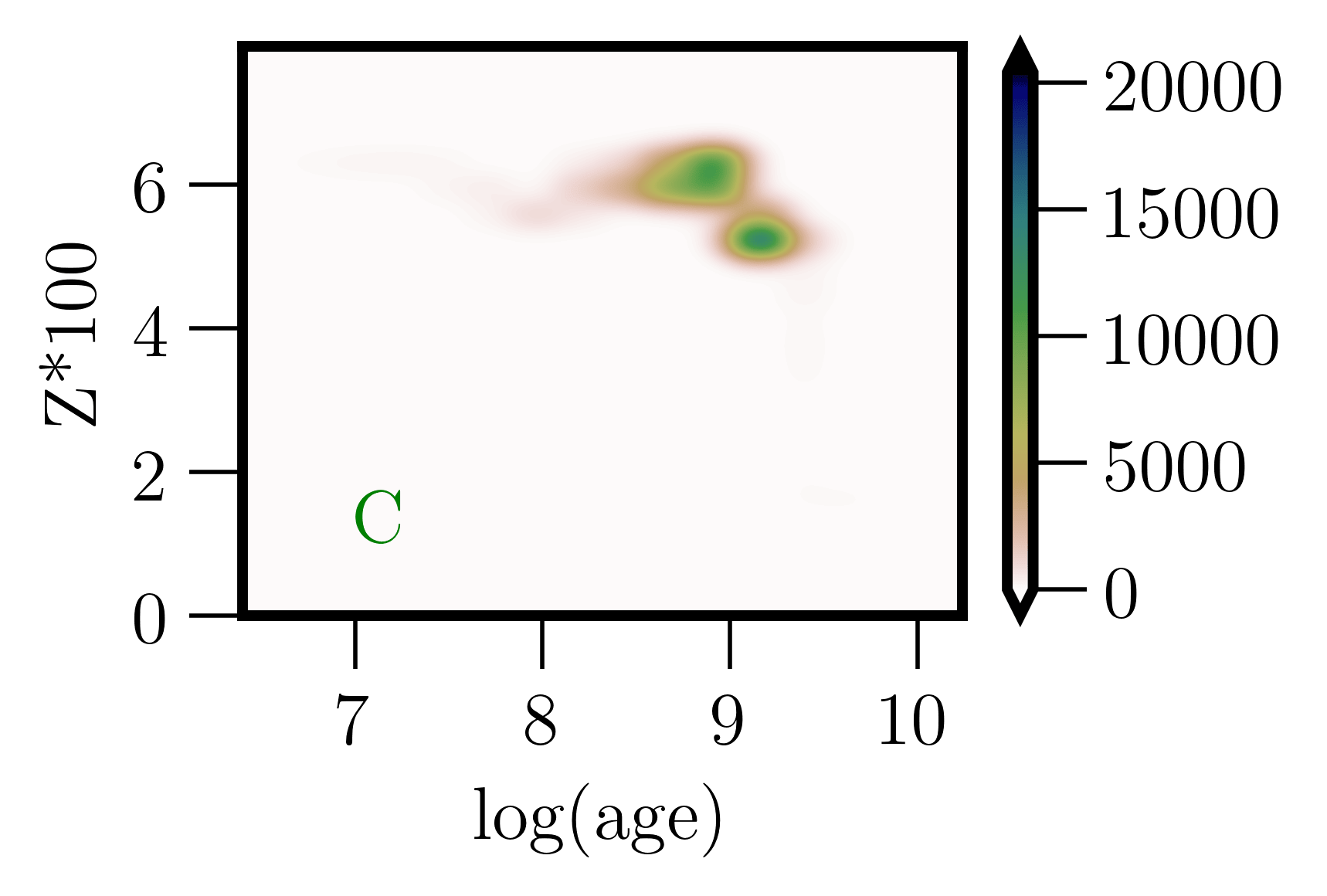}
		\includegraphics[scale=0.065]{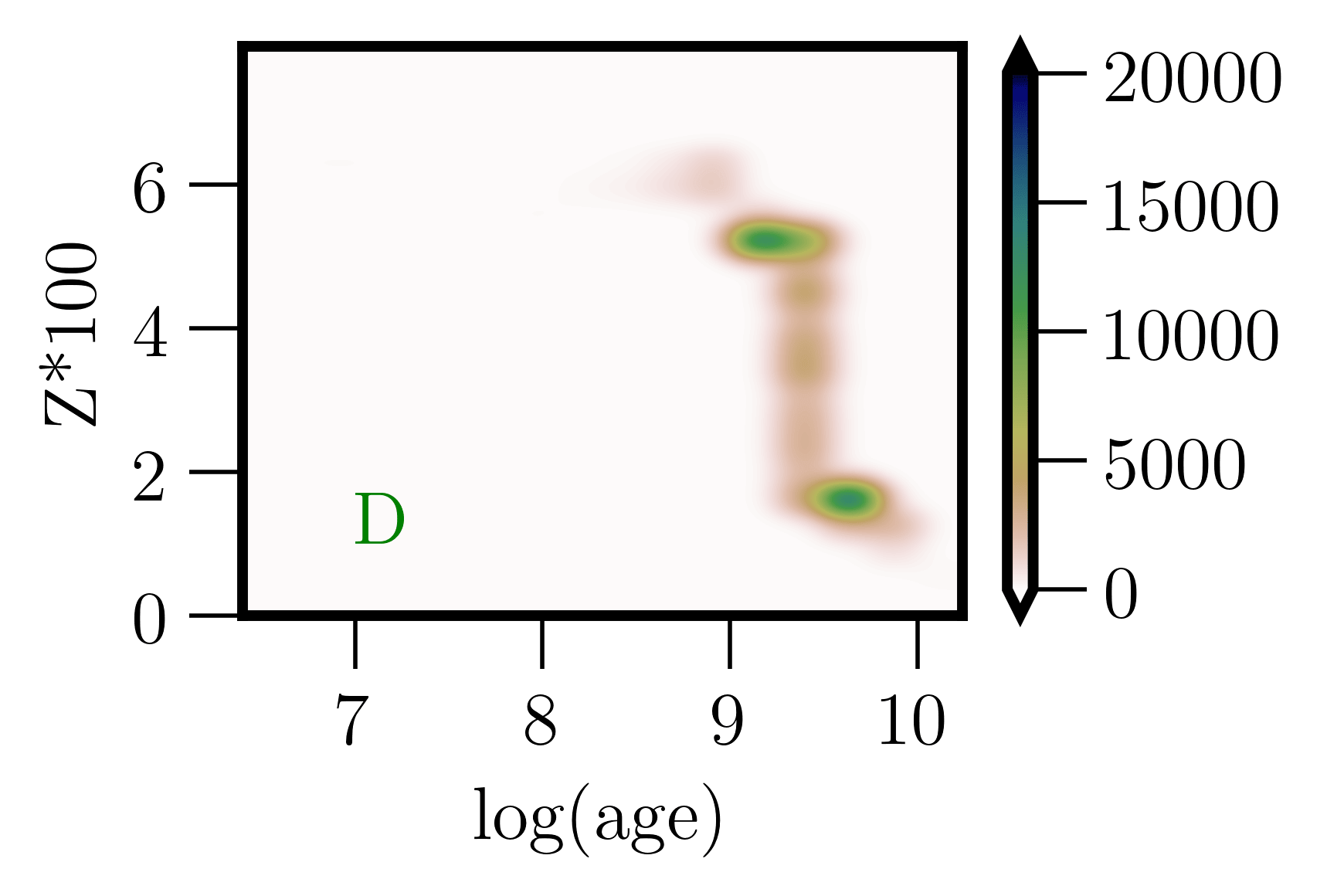}\\
		\includegraphics[scale=0.065]{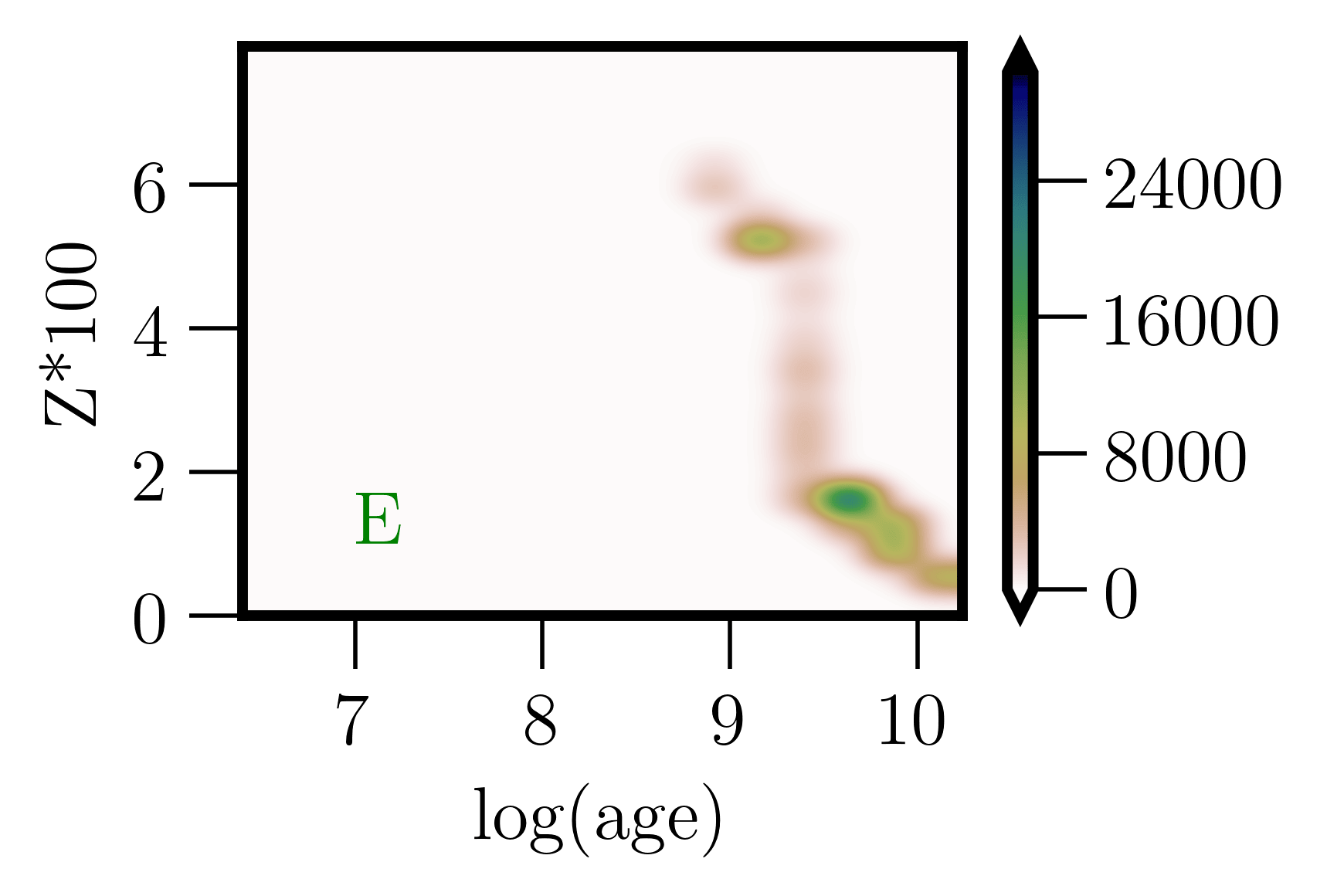}
		\includegraphics[scale=0.065]{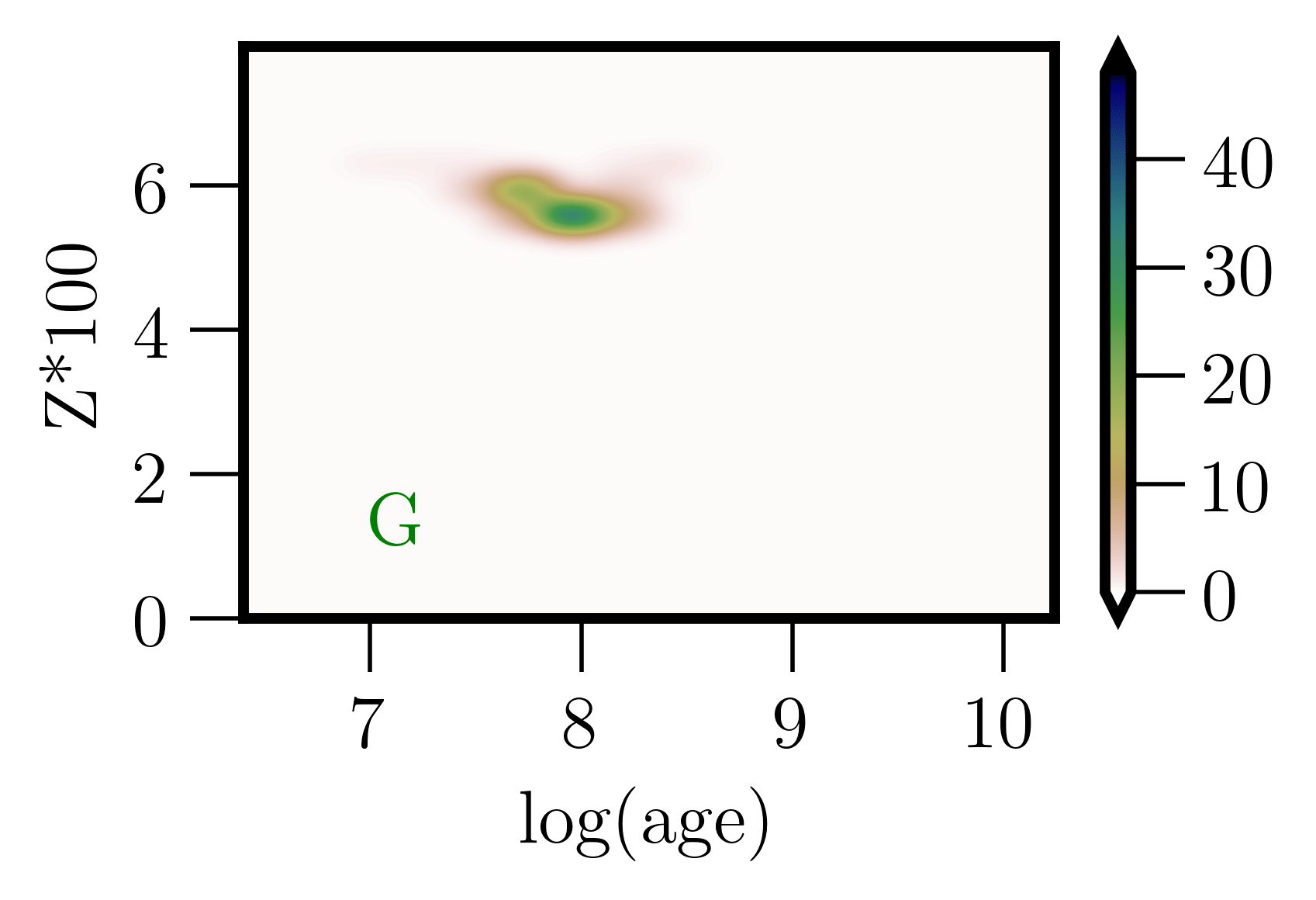}
		\includegraphics[scale=0.065]{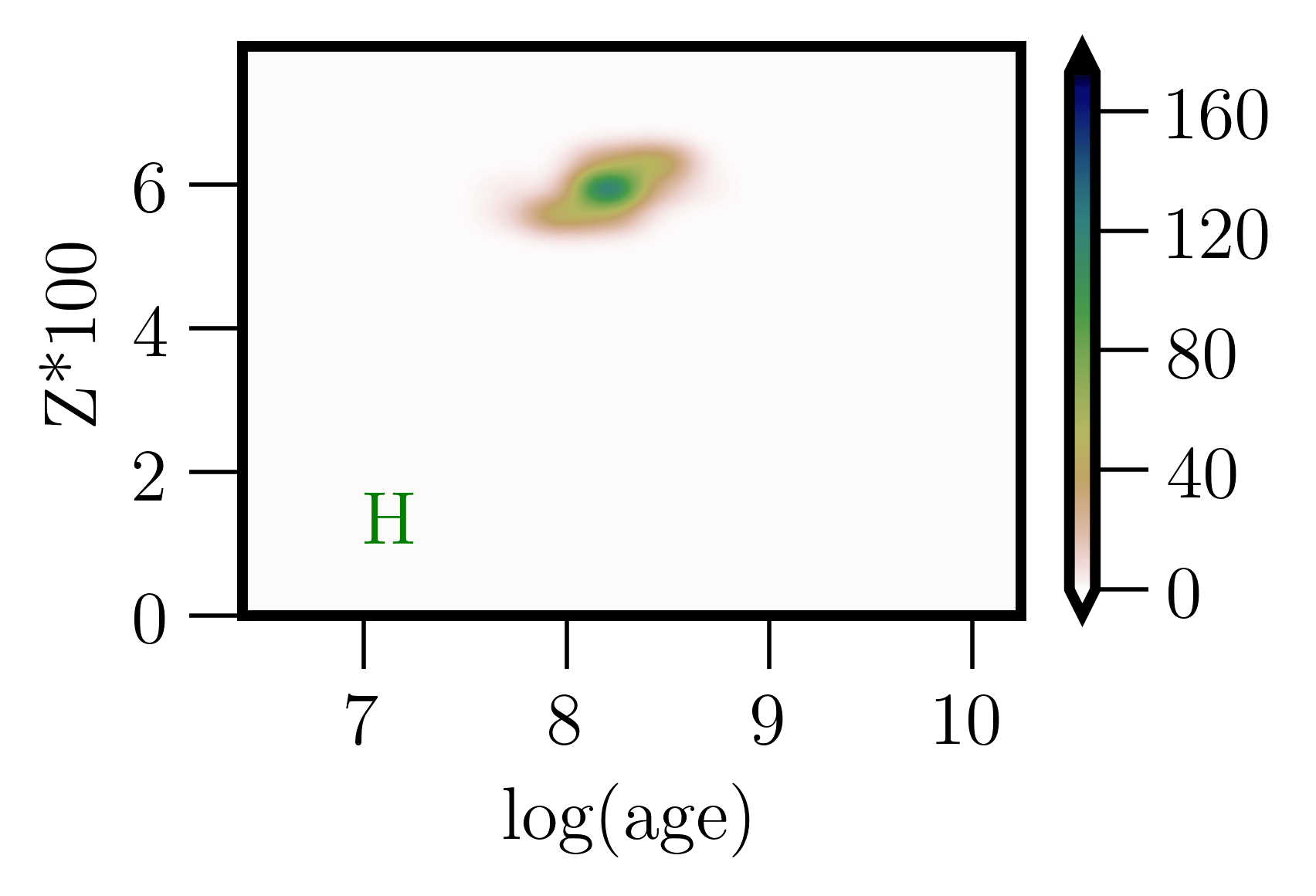}
		\includegraphics[scale=0.065]{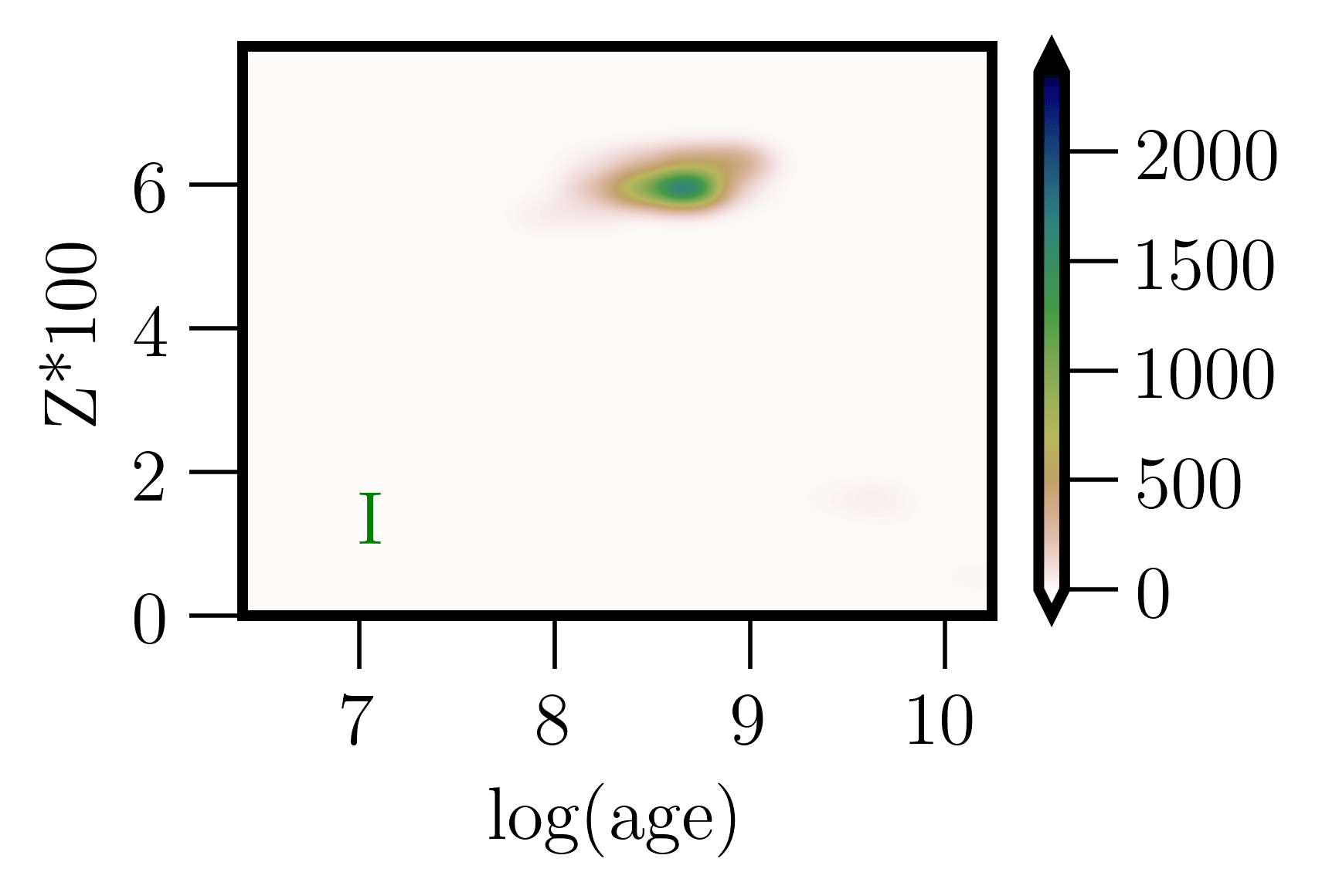}\\
		\includegraphics[scale=0.065]{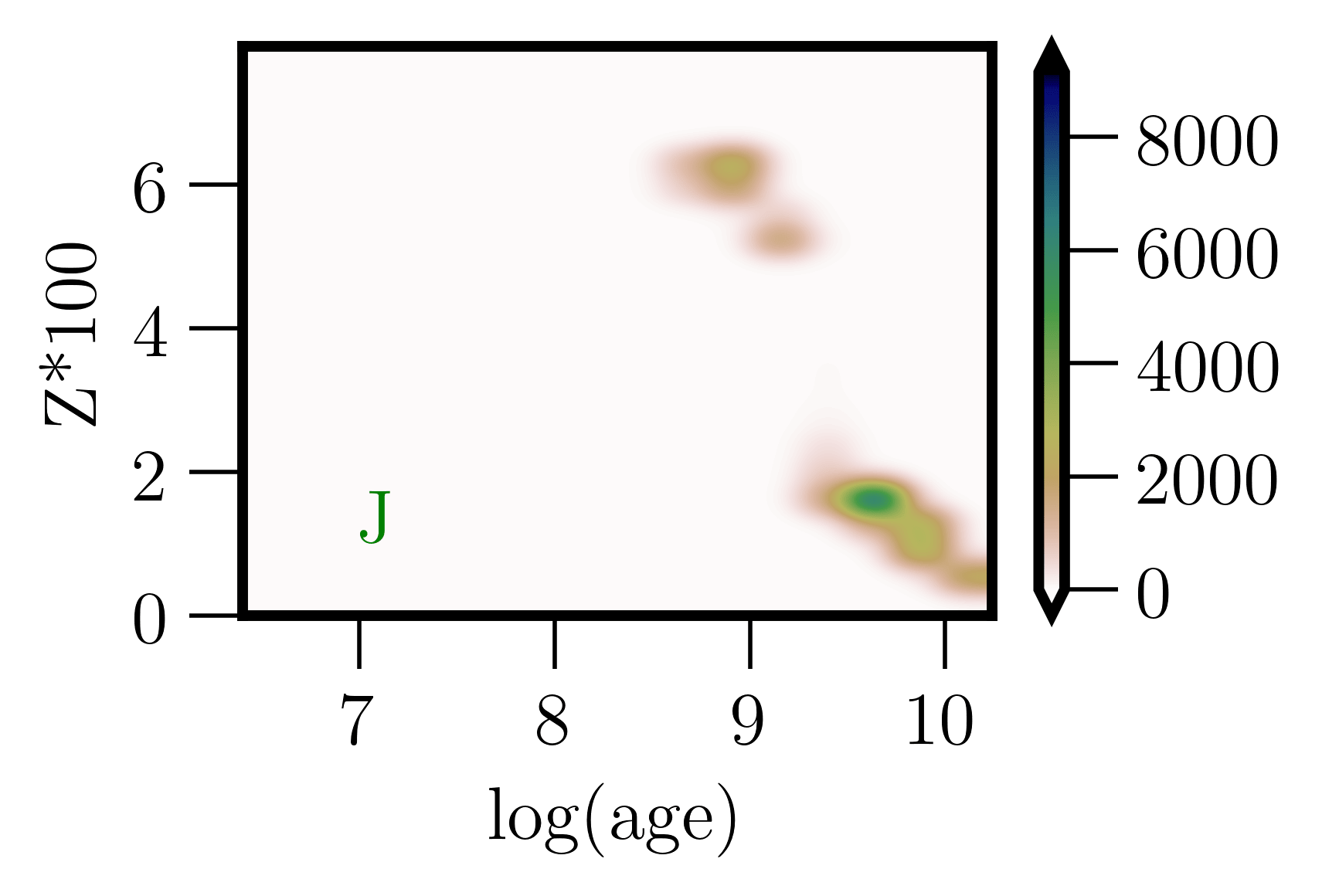}
		\includegraphics[scale=0.065]{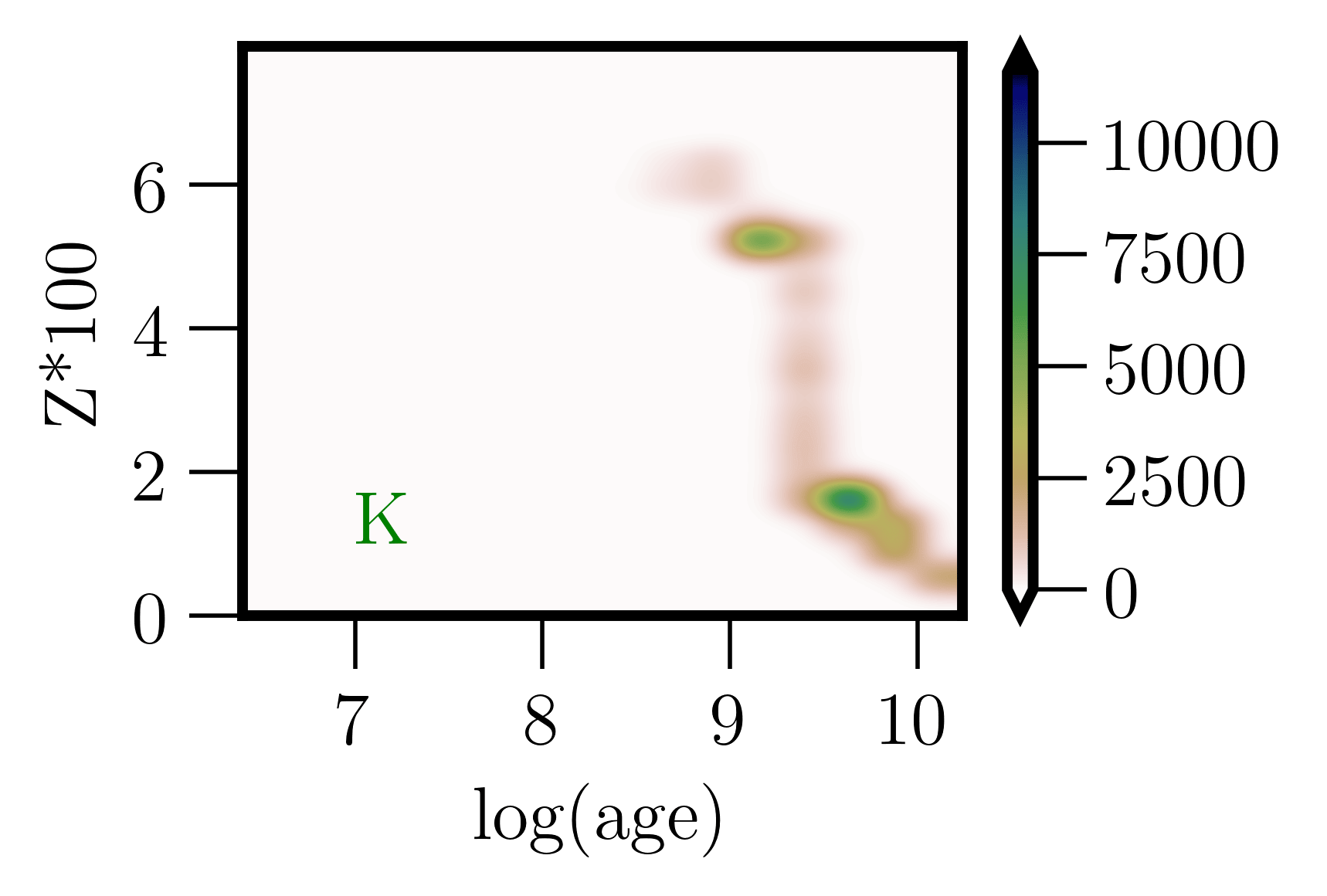}
		\includegraphics[scale=0.065]{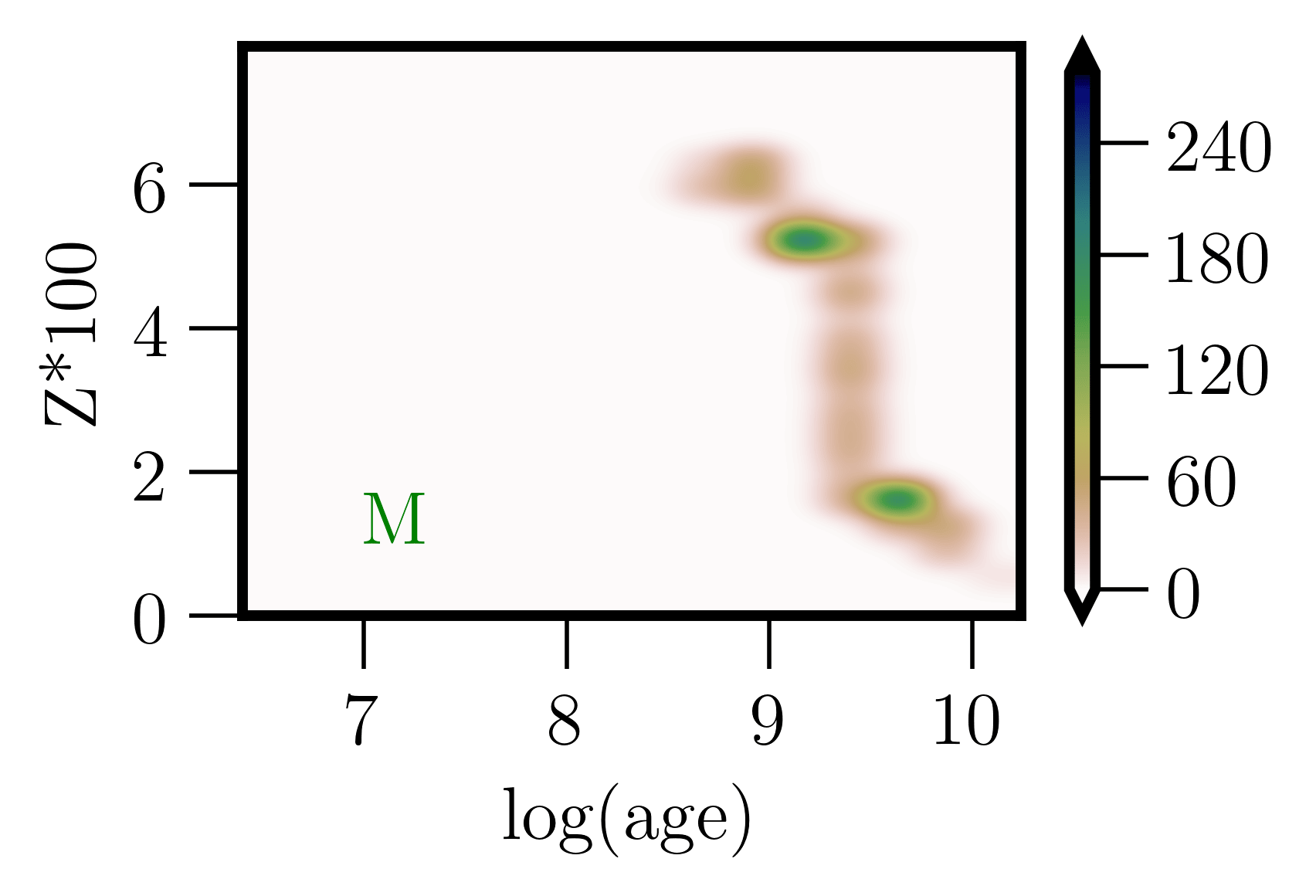}

		\caption{(top) Simulated ($J-K_\mathrm{s}$,$K_\mathrm{s}$) CMDs illustrating  stellar populations of the LMC. The colours correspond to a range of ages (left) and metallicities (right). The boxes refer to the regions used to disentangle different stellar populations. (bottom) Age-metallicity diagrams showing the distribution of ages and metallicities for stars inside each CMD region. The bin size is $0.08$ dex$^2$. The colour bar reflects the number of objects per bin.}
		
		\label{fig:SIMUCMDLMC}
	\end{figure*}

	Similarly to Fig. \ref{fig:SIMUCMD}, Fig. \ref{fig:SIMUCMDLMC} shows the theoretical models that have been derived from the analysis of the star formation history within several LMC tiles. These models were obtained from an ongoing SFH study by S. Rubele et al. (in preparation) and are used to update region boxes boundaries, stellar population ages as well as the Milky Way percentages published by \cite{Cioni2014}.
	
	\section{Photometric errors}
\begin{figure*}
				\centering
		\includegraphics[scale=0.1]{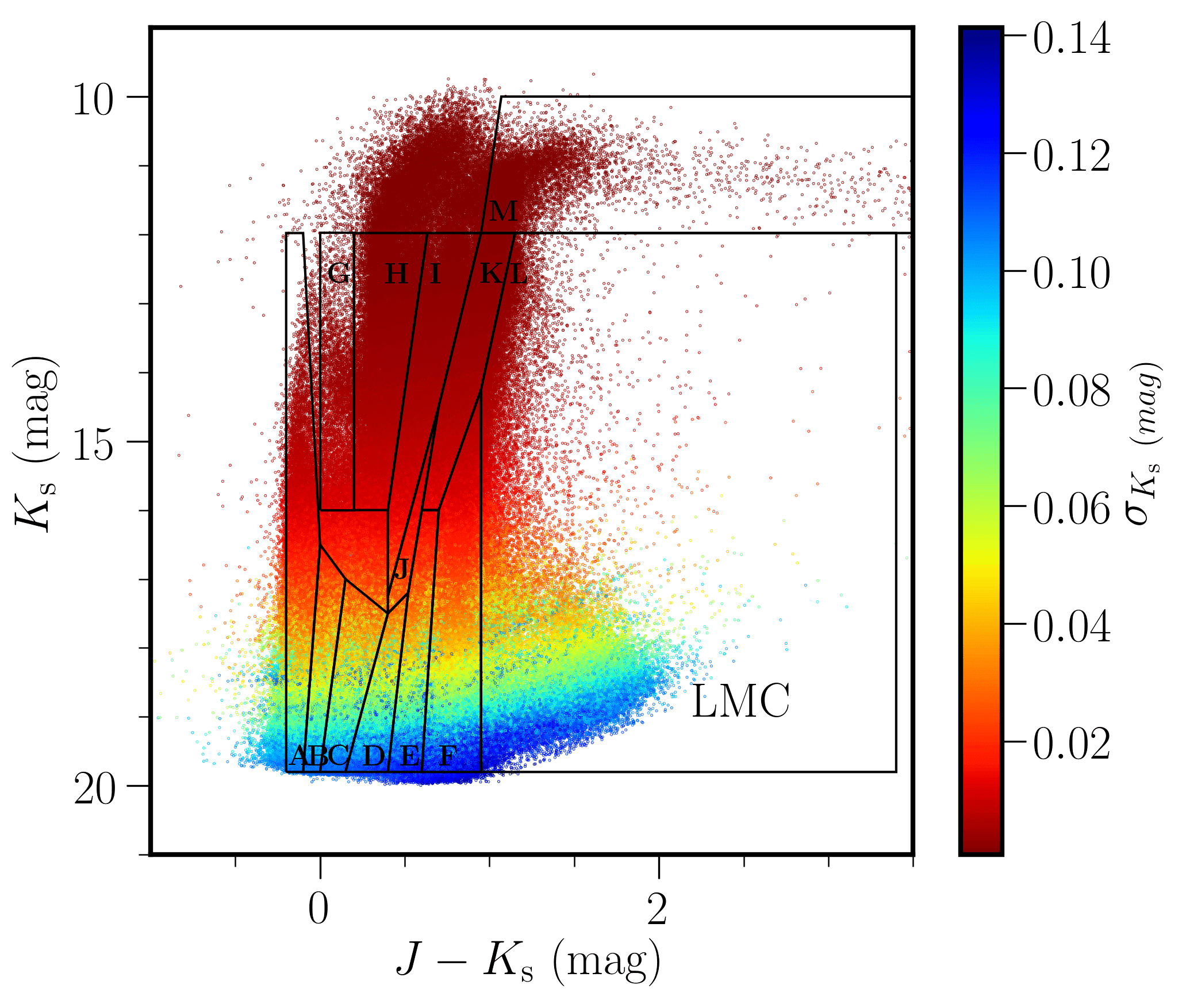}
		\includegraphics[scale=0.1]{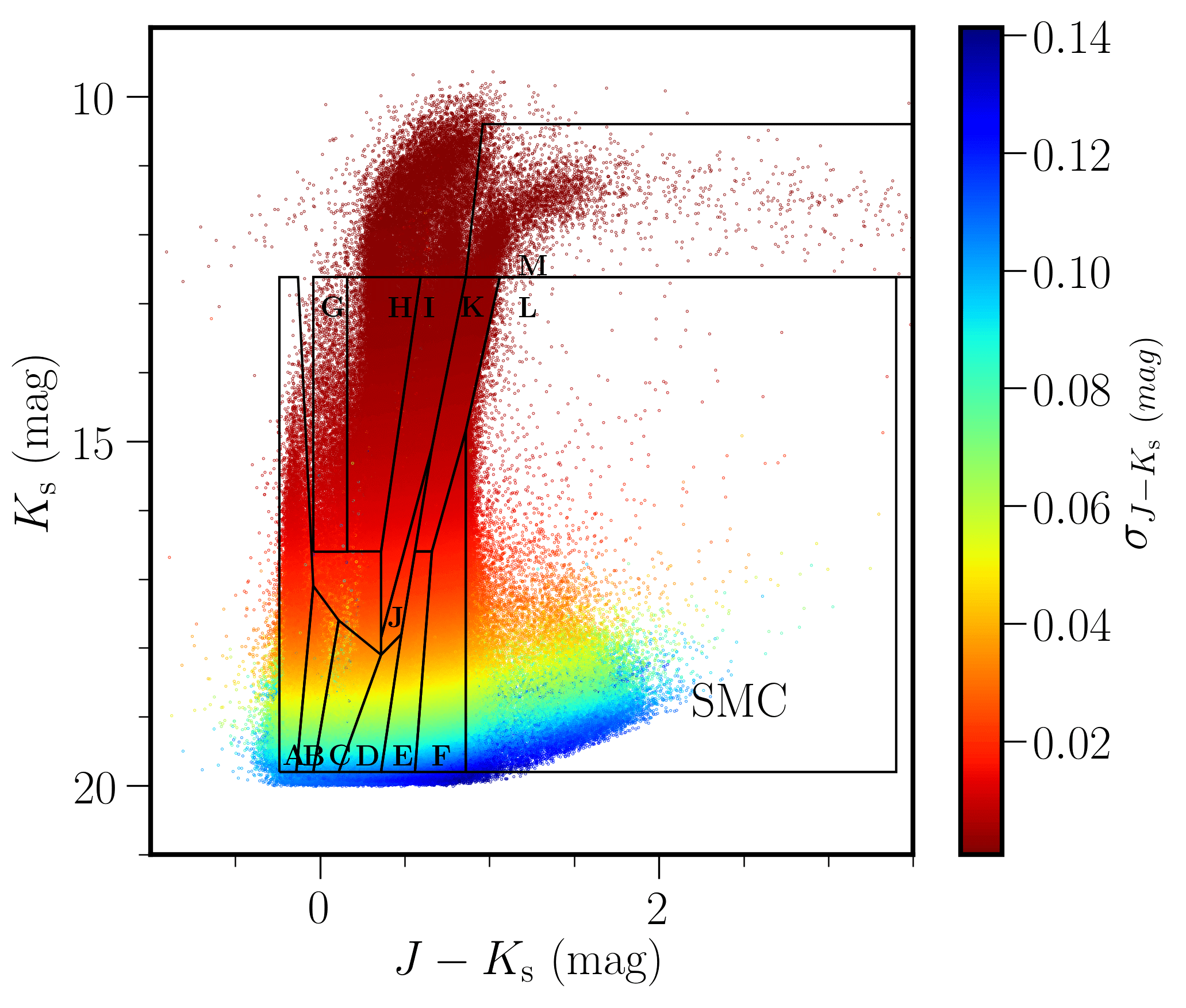}\\
		\includegraphics[scale=0.1]{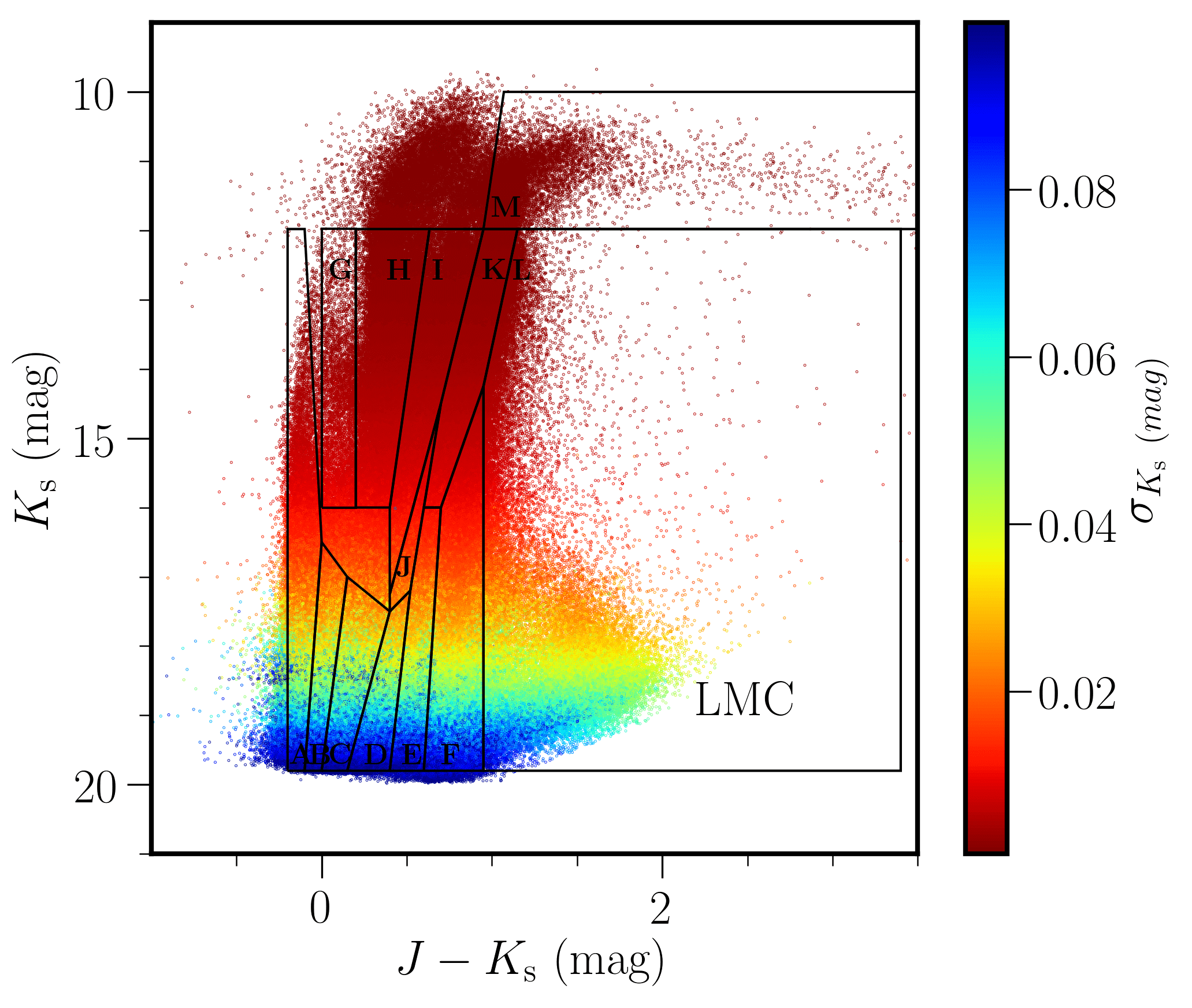}
		\includegraphics[scale=0.1]{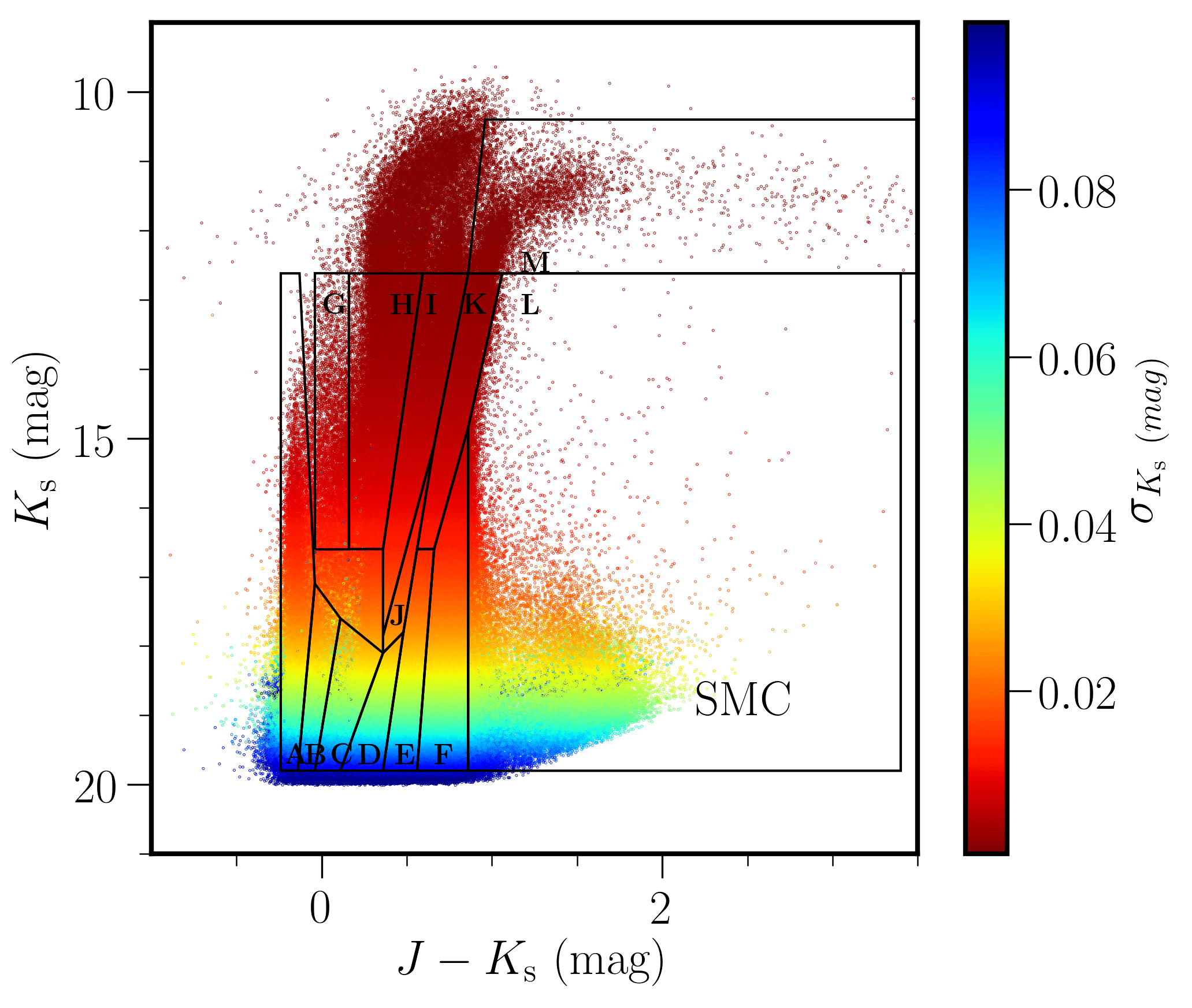}
		
		\caption{Colour-magnitude diagrams of photometric uncertainties in colour $\sigma_{J-K_\mathrm{s}}$ (top) and magnitude $\sigma_{K_\mathrm{s}}$ (bottom) for LMC (left) and SMC (right) stars. For clarity, we only plotted half the number of stars for the LMC.}

		\label{fig:errors}
	\end{figure*}

	The CMDs shown in Fig.~\ref{fig:errors} were constructed to show the distribution of $\sigma_{J-K_\mathrm{s}}$ and $\sigma_{K_\mathrm{s}}$ across the stellar population boxes of the LMC and SMC. The boxes widths range from 0.1 to 0.3 mag in colour. We note that we reach half the width of the lower limit at $\sim$ 18.5 mag.

%%%%%%%%%%%%%%%%%%%%%%%%%%%%%%%%%%%%%%%%%%%%%%%%%%



%\bibliographystyle{mnras}


%%%%%%%%%%%%%%%%%%%%%%%%%%%%%%%%%%%%%%%%%%%%%%%%%%

%%%%%%%%%%%%%%%%% APPENDICES %%%%%%%%%%%%%%%%%%%%%

%\appendix

%%%%%%%%%%%%%%%%%%%%%%%%%%%%%%%%%%%%%%%%%%%%%%%%%%

% Don't change these lines
\bsp	% typesetting comment
\label{lastpage}
\end{document}